\documentclass[12pt,twocolumn]{aastex63}
\usepackage{graphics}
\usepackage{color}
\usepackage{amsmath}
\usepackage{empheq} 
\newcommand{\etal}{et\,al.}
\newcommand{\ltsimeq}{\la}
\newcommand{\gtsimeq}{\ga}

\newcommand{\msun}{M$_{\odot}$}

\newcommand{\hi}{H{\sc i}}

\newcommand{\CH}[1]{\colhead{#1}}

\newcommand{\kms}{km~s$^{-1}$}

\shortauthors{McQuinn et al.}
\shorttitle{Galaxy Properties at the Faint End of the HI Mass Function}

\begin{document}
\title{GALAXY PROPERTIES AT THE FAINT END OF THE HI MASS FUNCTION}

\author{Kristen B.~W. McQuinn}
\affiliation{Rutgers University, Department of Physics and Astronomy, 136 Frelinghuysen Road, Piscataway, NJ 08854, USA} 
\email{kristen.mcquinn@rutgers.edu}

\author{Anjana K. Telidevara}
\affiliation{University of Texas at Austin, McDonald Observatory, 2515 Speedway, Stop C1400, Austin, Texas 78712, USA}
\affiliation{University of Minnesota, Minnesota Institute for Astrophysics, School of Physics and Astronomy, 116 Church Street, S.E., Minneapolis, MN 55455, USA} 

\author{Jackson Fuson}
\affiliation{Department of Physics and Astronomy, Macalester College, Saint Paul, MN 55105, USA}

\author{Elizabeth~A.~K. Adams}
\affiliation{ASTRON, The Netherlands Institute for Radio Astronomy, Oude Hoogeveensedijk 4, 7991 PD, Dwingeloo, The Netherlands}
\affiliation{Kapteyn Astronomical Institute, University of Groningen Postbus 800, 9700 AV Groningen, The Netherlands}

\author{John M. Cannon}
\affiliation{Department of Physics and Astronomy, Macalester College, Saint Paul, MN 55105, USA}

\author{Evan D. Skillman}
\affiliation{University of Minnesota, Minnesota Institute for Astrophysics, School of Physics and Astronomy, 116 Church Street, S.E., Minneapolis, MN 55455, USA} 

\author{Andrew E.~Dolphin}
\affiliation{Raytheon Company, 1151 E. Hermans Road, Tucson, AZ 85756, USA}
\affiliation{University of Arizona, Steward Observatory, 933 North Cherry Avenue, Tucson, AZ 85721, USA}

\author{Martha P.~Haynes}
\affiliation{Center for Astrophysics and Planetary Science, Space Sciences Building, Cornell University, Ithaca,
NY 14853, USA}

\author{Katherine L.~Rhode}
\affiliation{Department of Astronomy, Indiana University, 727 East Third Street, Bloomington, IN 47405, USA}

\author{John.~J.~Salzer}
\affiliation{Department of Astronomy, Indiana University, 727 East Third Street, Bloomington, IN 47405, USA}

\author{Riccardo Giovanelli}
\affiliation{Center for Astrophysics and Planetary Science, Space Sciences Building, Cornell University, Ithaca,
NY 14853, USA}

\author{Alex J.~R. Gordon}
\affiliation{Department of Physics and Astronomy, Macalester College, Saint Paul, MN 55105, USA}

\begin{abstract}
The Survey of \hi\ in Extremely Low-mass Dwarfs (SHIELD) includes a volumetrically complete sample of 82 gas-rich dwarfs with M$_{HI}\ltsimeq10^{7.2}$ \msun\ selected from the ALFALFA survey. We are obtaining extensive follow-up observations of the SHIELD galaxies to study their gas, stellar, and chemical content, and to better understand galaxy evolution at the faint end of the \hi\ mass function.  Here, we investigate the properties of 30 SHIELD galaxies using Hubble Space Telescope imaging of their resolved stars and Westerbork Synthesis Radio Telescope observations of their neutral hydrogen. We measure tip of the red giant branch (TRGB) distances, star formation activity, and gas properties. The TRGB distances are up to $4\times$ greater than estimates from flow models, highlighting the importance of velocity-independent distance indicators in the nearby universe. The SHIELD galaxies are in under-dense regions, with 23\% located in voids; one galaxy appears paired with a more massive dwarf. We quantify galaxy properties at low masses including stellar and \hi\ masses, SFRs, sSFRs, SFEs, birthrate parameters, and gas fractions. The lowest mass systems lie below the mass thresholds where stellar mass assembly is predicted to be impacted by reionization. Even so, we find the star formation properties  follow the same trends as higher mass gas-rich systems, albeit with a different normalization. The \hi\ disks are small ($\langle$r$\rangle<$0.7 kpc) making it difficult to measure the \hi\ rotation using standard techniques; we develop a new methodology and report the velocity extent, and its associated spatial extent, with robust uncertainties.
\end{abstract} 


\section{Introduction}\label{sec:intro}
\subsection{On the Cosmological Importance of Extremely Low-mass Galaxies}
In the current paradigm, galaxies are formed hierarchically, and the numbers of galaxies increase as galaxy stellar mass decreases. Large surveys support this framework and the galaxy mass function (as traced by stellar luminosity) continues to rise at lower masses. At small enough masses, galaxy counts should eventually decline (e.g., due to the combined loss of baryons from blow-out, ram pressure stripping, tidal interactions, and reionization at the earliest epochs). Identifying this turnover in the galaxy luminosity function and its environmental dependence would place strong constraints on the limits of structure formation in the early universe. 

The present-day masses of `turnover galaxies', while a subject of debate, are expected to be low (M$_* \ltsimeq 10^{7.5}$ \msun) with shallow potentials and correspondingly low  rotation velocities \citep[V$_{\rm rot} \ltsimeq 30$ km s$^{-1}$; e.g.,][and references therein]{Rees1986, Gnedin2000, Hoeft2006, Okamoto2008a}. The existence of {\em gas-rich} galaxies with such very shallow potential wells poses interesting puzzles for the $\Lambda$CDM paradigm and for our understanding of baryon physics \citep[][and references therein]{Bullock2017}. Star formation activates feedback mechanisms that result in gas loss via superwinds \citep{McQuinn2019b}. Metagalactic UV radiation inhibits gas accretion and cooling. A hot IGM should vaporize a small, unshielded cold gas mass within less than a Hubble time \citep[e.g.,][]{Ikeuchi1986, Rees1986, MacLow1999, Ferrara2000, Hoeft2006}. How gas is retained in these very low mass systems is under debate; it might depend on the protection provided by a shielding envelope of warm, ionized gas, as shown in the models of \citet{Sternberg2002}, and/or environment, i.e., low-mass galaxies can be found outside the immediate vicinity of dense coronal gas. There is also speculation about whether such systems will have different properties than more massive dwarfs due to the impacts of the same internal and external processes, or if their properties will simply extend well-established galaxy scaling relations. 

In sum, there are numerous open questions about galaxies at the faint end of the luminosity function, particularly for gas-rich systems, including (i) their number counts, (ii) their overall properties, and (iii) the ability for at least some systems with such shallow potentials to retain their gas until the present day. 

\subsection{Finding Extremely Low-Mass Gas-Rich Galaxies}
Observationally, a main challenge to counting and characterizing very low-mass galaxies and, subsequently, testing the various predictions, has been finding large enough samples of these systems. As one moves to lower and lower masses, the systems host fewer stars and thus are intrinsically faint in the optical, infrared, and ultraviolet regimes with correspondingly low surface brightnesses. There are numerous, on-going efforts to search for low-mass galaxies including, for example, searches for ultra-faint dwarfs within the Local Group \citep[e.g., the Dark Energy Survey;][]{Drlica-Wagner2020}, searches for low surface brightness and satellite galaxies outside the Local Group but within the Local Volume \citep[e.g.,][]{Smercina2018, Greco2018, Carlsten2020}, and searches for satellites of more massive galaxies outside the Local Volume in a statistical sample of galaxies \citep[the SAGA Survey;][]{Geha2017, Mao2020}. 

As our interests lie in the faint end of the galaxy luminosity function populated with {\em gas-rich, star-forming galaxies}, rather than searching for the faint emission from their stellar populations, we take a different approach and search for galaxies via their neutral hydrogen. The Arecibo Legacy Fast ALFA (ALFALFA) survey \citep{Giovanelli2005, Haynes2011} is an extragalactic survey that mapped neutral hydrogen (\hi) over $\sim$7000 deg$^2$ of high Galactic latitude sky in the nearby universe. ALFALFA was designed to populate the faint end of the \hi\ mass function with an \hi\ mass detection limit of $\sim10^6$ \msun\ in the local universe and 10$^{9.5}$ at the edge of the survey volume at $z\sim0.06$. Because of the higher sensitivity of the ALFALFA survey, low-mass galaxies can be detected over a volume 4$\times$ larger than the \hi\ Parkes All Sky Survey \citep[HIPASS][]{Barnes2001, Meyer2004}, despite the smaller areal coverage \citep{Jones2018}. The full ALFALFA catalog includes over 30,000 extragalactic \hi\ sources \citep{Haynes2018}.

From the rich ALFALFA catalog of \hi\ sources, we selected low-mass systems in the Local Volume (D$\ltsimeq 4-11$ Mpc; see \S\ref{sec:sample} for specific criteria) that also have stellar counterparts identified in optical imaging of SDSS for detailed follow-up in the Survey of \hi\ in Extremely Low-mass Dwarfs \citep[SHIELD;][]{Cannon2011c}. The full SHIELD sample includes 82 galaxies, many of which were discovered by the ALFALFA survey, and span the range of halo masses over which the interesting transition from the cosmic baryon fraction (f$_b$) value of $\sim$ 0.16 to $< 0.01$ takes place \citep[see, e.g.,][]{Hoeft2006, McGaugh2010}. Here, we present results for 30 galaxies in the SHIELD sample, including incorporating results from the initial study of 12 systems (hereafter the SHIELD~I galaxies) and new results for an additional 18 systems (hereafter the SHIELD~II galaxies). 

As the SHIELD galaxies are an \hi\ selected sample, the expectation is that the majority of the systems will be located in a field environment based on the low frequency of gas-rich satellites of the Milky Way and M~31. While a full exploration of the environment of the SHIELD sample requires accurate distances to all 82 galaxies, the present results for the first 30 systems allow us to produce initial statistics regarding whether the systems are in under-dense environments and the distances to their nearest neighbors.

Using the SHIELD galaxies, we work toward quantifying the physical properties in very low-mass galaxies at the faint end of the \hi\ mass function, understanding their environments, and testing galaxy formation and evolution theories. We begin in \S\ref{sec:sample} with a comparison to some of the existing surveys of low-mass galaxies, including an exploration of \hi\ mass detections as a function of distance in the Local Volume. We present new HST optical imaging for 18 galaxies and WSRT \hi\ observations for 16 galaxies, including introducing a new approach for measuring the velocity and extent of \hi\ from velocity fields with limited spatial sampling in \S\ref{sec:obs}, measurements of the TRGB distances to the galaxies in \S\ref{sec:trgb}, and measurements of their star formation properties in \S\ref{sec:sf}. Using the TRGB distances, we explore the galaxies' surrounding neighborhoods in \S\ref{sec:neighbors}. Galaxy properties at the faint end of the \hi\ mass function, including stellar and gas content and star formation properties of the SHIELD galaxies are discussed in \S\ref{sec:properties}, with comparisons to results in the literature on other very low-mass galaxies. Our conclusions are presented in \S\ref{sec:conclusions}. Finally, in the Appendix, we include an atlas of the HST imaging (Appendix~\ref{app:HST_atlas}), an atlas of the WSRT \hi\ data (Appendix~\ref{app:HI_atlas}), and details on our new methodology for measuring the rotational motion and spatial extent of the \hi\ in very low-mass galaxies (Appendix~\ref{app:pvvel}).

\begin{figure}
\begin{center}
\includegraphics[height=0.73\textheight]{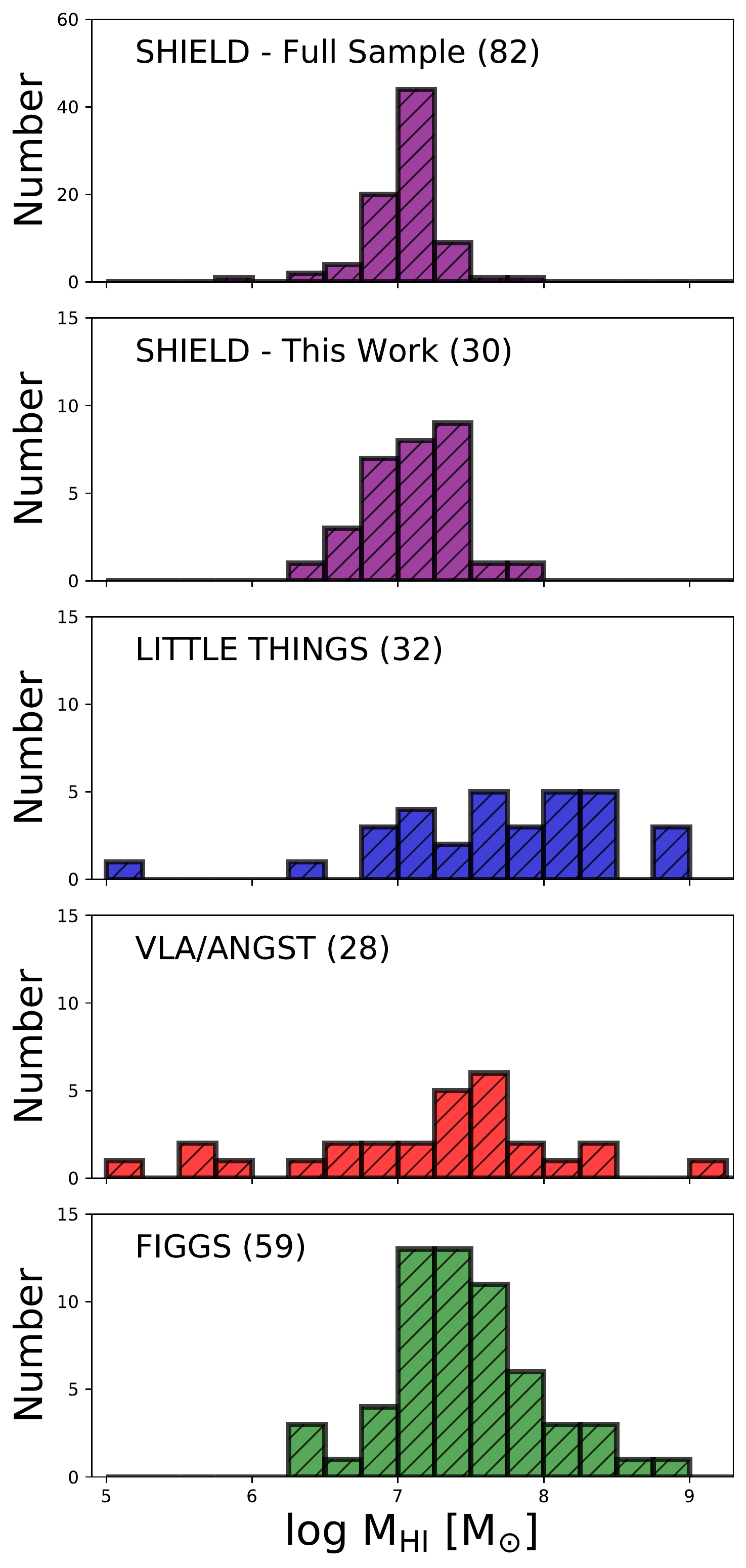}
\end{center}
\caption{From top to bottom: The \hi\ mass distribution of galaxies in the full SHIELD survey, the subset of SHIELD galaxies included in this work (the SHIELD~I and II samples), and three other surveys of low-mass galaxies including LITTLE THINGS galaxies, the VLA/ANGST survey, and the FIGGS survey. Both LITTLE THINGS and FIGGS are \hi\ selected surveys, while VLA/ANGST is an \hi\ follow-up on the optically selected galaxies in ANGST \citep{Dalcanton2009}. We include the galaxies for which we have compiled stellar masses; the parenthetical numbers represent the total galaxy counts in each histogram. Collectively, these surveys provide a basis to characterize galaxy properties at the faint end of the \hi\ mass function. See, also, Figure~\ref{fig:D_MHI}.}
\label{fig:MHI_compare}
\end{figure}

\section{Overview of SHIELD and Comparison with other Surveys}\label{sec:sample}
\subsection{The SHIELD program}
The SHIELD sample consist of 82 galaxies selected from the ALFALFA catalog that met the following criteria: (i) low gas masses (log(M$_{HI}$/\msun) $<$ 7.2) based on the \hi\ line fluxes and distance estimates from the flow model of \citet{Masters2005}, (ii) narrow \hi\ line widths (\hi\ FWHM $<$65 km s$^{-1}$), which selected against  massive but \hi-deficient galaxies, (iii) optically identified counterparts in SDSS imaging, and (iv) distance estimates within $\sim11$ Mpc, which ensured the galaxies are sufficiently close for detailed analysis while eliminating the closest and most well-studied gas-rich dwarfs. Note that a few galaxies within $\sim4$ Mpc met these criteria but were not included in the SHIELD sample as they were part of existing \hi\ selected surveys. 

Results for the first 12 SHIELD galaxies \citep[SHIELD~I galaxies;][]{Cannon2011c} include (i) measuring tip of the red giant branch (TRGB) distances \citep{McQuinn2014}, (ii) characterizing their star formation properties from color-magnitude diagrams \citep[CMD;][]{McQuinn2015a}, (iii) deriving their gas kinematics from a combination of B, C, and D configuration data from the Very Large Array \citep[VLA; see][for details]{McNichols2016}, (iv) comparing their star formation and gas properties \citep{Teich2016}, and (v) measuring their oxygen abundances \citep{Haurberg2015}. 

Here, we expand the analysis from 12 to 30 galaxies, adding new observations of the resolved stellar populations from HST and of the \hi\ from WSRT of 18 galaxies (SHIELD~II galaxies). With the combined 30 galaxies from the SHIELD~I and II samples, and expanding on the work of previous surveys \citep[e.g., FIGGS, VLA/ANGST, LITTLE THINGS;][respectively]{Begum2008, Ott2012, Hunter2012}, we begin to quantify the properties of galaxies at the faint end of the galaxy \hi\ mass function with statistical confidence. Follow-up analysis will place the SHIELD galaxies in the context of the Baryonic Tully-Fisher Relation (BTFR; K.~McQuinn et al.\ in preparation).

\subsection{Comparison of SHIELD with Other Surveys of Gas-Rich Low Mass Galaxies}
Figure~\ref{fig:MHI_compare} shows the number distribution of the \hi\ masses for the full SHIELD sample (top panel) based on the ALFALFA \hi\ fluxes and adopting the Virgo-centric flow model distances from \citet{Masters2005} (see \S\ref{sec:flow}). The number shown parenthetically is the number of galaxies in the histogram. The second panel shows the distribution of the SHIELD galaxies presented in this work (i.e., SHIELD~I and II samples), after adopting the more robust TRGB distance measured in the present work (see \S\ref{sec:trgb}). Note that the ordinate range shown for the SHIELD~I and II galaxies is a factor of four smaller than for the full SHELD sample shown in the top panel. The majority of the galaxies span a narrow \hi\ mass range of $10^{6.5} - 10^{7.5}$ \msun, which is mainly due to the upper mass limit imposed in our selection criteria and the growing incompleteness of the ALFALFA catalog at lower masses.

As our main goal is to characterize the properties of galaxies in the nearby universe at the faint end of the \hi\ mass function, we also present samples from three existing surveys focused on dwarfs galaxies in the subsequent panels in Figure~\ref{fig:MHI_compare}. Specifically, we include the data from the LITTLE THINGS survey \citep{Hunter2012}, the VLA/ANGST survey \citep{Ott2012}, and the FIGGS survey \citep{Begum2008}. We excluded galaxies from VLA/ANGST that are classified as early-type galaxies or were not detected in  \hi. Note that the numbers for each survey, shown parenthetically, represent the galaxies for which we have compiled stellar masses; the total number of galaxies in each survey is slightly higher. While not a complete census of Local Volume gas-rich dwarfs,\footnote{Note that in particular the LITTLE THINGS galaxies were selected to sample a broad range of galaxies properties and did not attempt to be volumetrically complete.} these surveys are the largest detailed studies of \hi\ bearing low-mass galaxies in the Northern Hemisphere and, thus, help characterize galaxy properties over a larger mass range and with improved statistics, complementing the SHIELD results. 

\subsection{\hi\ Masses as a Function of Distance}\label{sec:survey}
\begin{figure}
\includegraphics[width=0.48\textwidth]{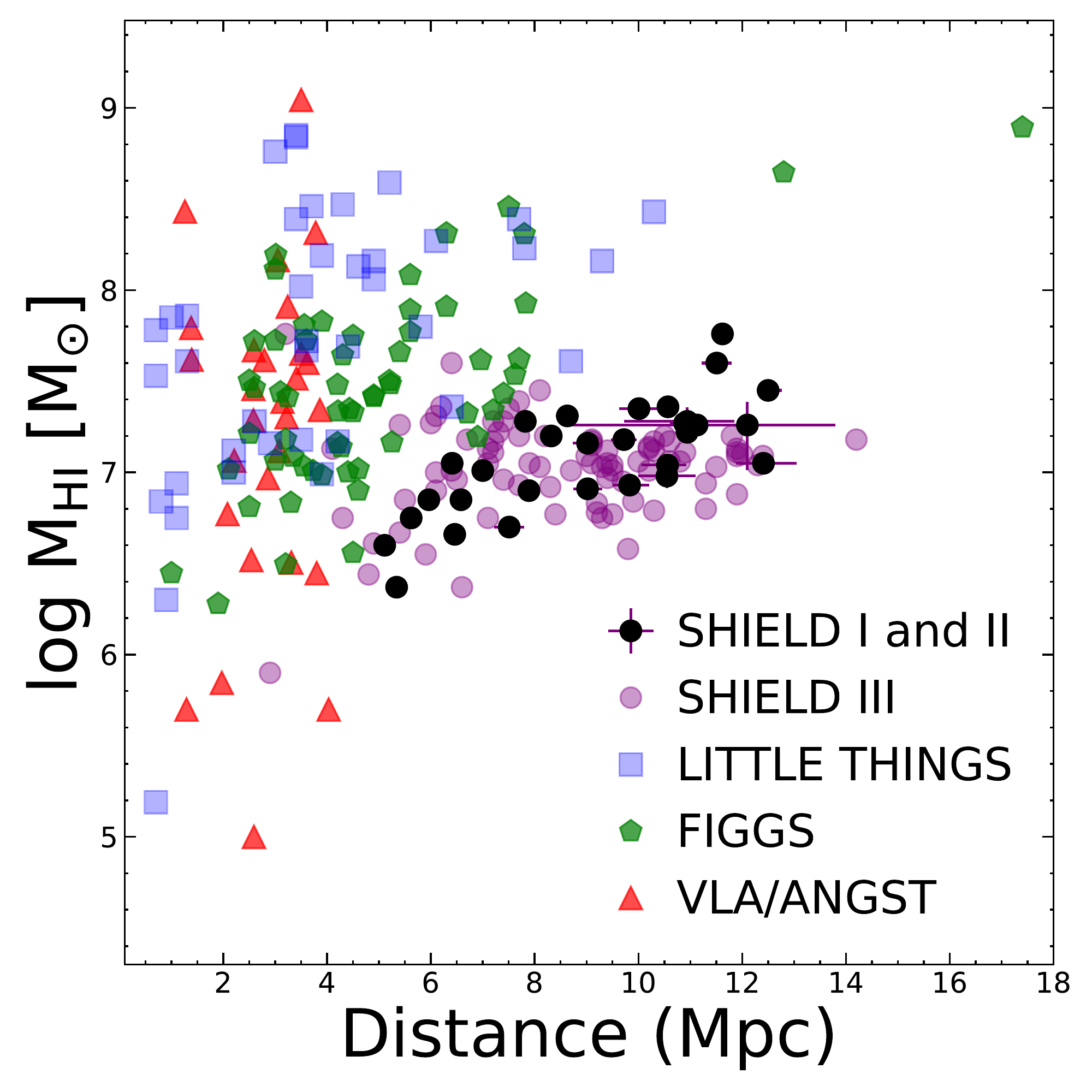}
\caption{Log(M$_{\rm HI}$/\msun) vs.\ Distance for SHIELD and three other surveys of low-mass galaxies. The SHIELD~I and II samples studied in detail in the present work are shown with black circles and purple uncertainties; the remainder of the SHIELD sample (labelled as SHIELD~III) are shown with transparent purple circles. VLA/ANGST includes optically selected galaxies within $\sim$4 Mpc by design. LITTLE THINGS extends to larger \hi\ masses than VLA/ANGST and spans a larger range in distance. The FIGGS galaxies are on average slightly more massive than the SHIELD galaxies (see also Figure~\ref{fig:MHI_compare}) and less distant.}
\label{fig:D_MHI}
\end{figure}

The distribution of \hi\ masses as a function of distance for SHIELD and the three other surveys described in \S\ref{sec:sample} is presented in Figure~\ref{fig:D_MHI}. We include not only the SHIELD~I and II samples, but the remaining 52 galaxies from SHIELD (labelled SHIELD~III which includes the all SHIELD galaxies not presented in detail in this work). Collectively, these galaxies provide a statistical population of gas-rich dwarfs down to M$_{\rm HI} \sim 10^{6.5}$ \msun, distributed over $\sim12$ Mpc. Galaxies below this mass threshold have almost exclusively been detected within a few Mpc. 

\begin{deluxetable*}{lcLLLcLLLccCc}
\colnumbers
\tabletypesize{\scriptsize}
\tablecaption{Stellar Properties \label{tab:stars}}
\tablehead{
\CH{} 		& \CH{Galactic}	& \CH{F814W$_{TRGB}$}& \CH{Distance}	& \CH{Distance} 	& \CH{Internal} &  \CH{log} &  \CH{$\langle$SFR$\rangle_{life}$} & \CH{$\langle$SFR$\rangle_{\rm 200~Myr}$} 	& \CH{a}	& \CH{1-{b/a}}	& \CH{PA}	& \CH{Optical} \\
\CH{Galaxy}	&  \CH{A$_{F606W}$}	& \CH{ML } & \CH{Modulus} & \CH{} 	&  \CH{A$_V$}	& \CH{(M$_*$/\msun)}&  \CH{$\times10^{-3}$} 			&  \CH{$\times10^{-3}$ } 		& \CH{} 		& \CH{} 	& \CH{}	&\CH{Diameter} \\
\CH{} 		& \CH{(mag)} 	& \CH{(mag)} & \CH{(mag)}& \CH{(Mpc)} 		& \CH{(mag)} 	& \CH{} 	& \CH{(\msun\ yr$^{-1})$} & \CH{(\msun\ yr$^{-1}$)}	& \CH{(\arcsec)} & \CH{}	& \CH{($^{\circ}$)} & \CH{(kpc)}}
\startdata
\\
\multicolumn{12}{c}{SHIELD I Galaxies} \\
\hline \\
AGC 110482	& 0.23	&25.38\pm0.05 		& 29.47\pm0.05		&7.82\pm0.21 		& 0.00	& 7.40_{-0.21}^{+0.25} 	&5.17_{-2.47}^{+2.99} 	&5.72_{-3.66}^{+2.69} 	&37 	&0.30	& 15		& 2.8\\ 
AGC 111164	& 0.14	&24.45\pm0.02 		& 28.54\pm0.03		&5.11\pm0.07 		& 0.00	& 6.54_{-0.14}^{+0.12} 	&0.72_{-0.23}^{+0.20} 	&0.39_{-0.21}^{+0.21} 	&60 	&0.40	& 6		&3.0\\ 
AGC 111946	& 0.20	&25.69^{+0.04}_{-0.06} 	& 29.78^{+0.05}_{-0.06}	&9.02^{+0.20}_{-0.29}	& 0.00	& 6.81_{-0.24}^{+0.20} 	&1.34_{-0.73}^{+0.61}	&2.17_{-1.20}^{+0.66} 	&27 &0.56 	& 6		&2.4\\
AGC 111977	& 0.17	&24.81^{+0.03}_{-0.02} 	& 28.88^{+0.04}_{-0.03}	&5.96^{+0.11}_{-0.09} 	& 0.00	& 7.12_{-0.17}^{+0.16} 	&2.74_{-1.09}^{+0.99}  	&1.81_{-0.93}^{+1.04} 	&70 &0.47 	& 30		&4.0 \\
AGC 112521	& 0.15	&25.00\pm0.05 		& 29.09\pm0.05		&6.58\pm0.18   		& 0.00	& 6.47_{-0.19}^{+0.18}	&0.61_{-0.27}^{+0.26} 	&0.52_{-0.38}^{+0.37} 	&32 &0.38 	& 18		&2.0\\
AGC 174585	& 0.10	&25.39^{+0.05}_{-0.04} 	& 29.49\pm0.05		&7.89^{+0.21}_{-0.17} 	& 0.00	& 6.56_{-0.16}^{+0.09}	&0.75_{-0.27}^{+0.16} 	&1.57_{-0.76}^{+0.52} 	&23  &0.26	& 347	&1.8\\
AGC 174605	& 0.06	&26.09\pm0.05 		& 30.19\pm0.05		&10.89\pm0.28 		& 0.00	& 7.10_{-0.20}^{+0.63}	&2.60_{-1.21}^{+3.75} 	&3.09_{-1.35}^{+1.96} 	&26 &0.19		& 0		&2.7\\ 
AGC 182595	& 0.10	&25.68\pm0.06		& 29.78\pm0.06		&9.02\pm0.28 	 	& 0.00	& 7.41_{-0.25}^{+0.15}	&5.28_{-3.01}^{+1.88} 	&3.95_{-1.81}^{+1.65} 	&52	&0.19 	& 0		&4.5 \\
AGC 731457 	& 0.07	&26.13^{+0.03}_{-0.02}	& 30.23^{+0.04}_{-0.03}	&11.13^{+0.20}_{-0.16} 	& 0.00	& 7.82_{-0.33}^{+0.03}	&13.5_{-10.3}^{+0.88} 	&13.1_{-6.23}^{+0.48} 	&37	&0.38 	& 0		&4.0\\
AGC 748778	& 0.16	&24.95^{+0.04}_{-0.05}	& 29.05\pm0.05 		&6.46^{+0.14}_{-0.17} 	& 0.00	& 6.13_{-0.21}^{+0.11}	&0.28_{-0.14}^{+0.07} 	&0.62_{-0.39}^{+0.24} 	&27	&0.11 	& 0		&1.7\\
AGC 749237	& 0.05	&26.24^{+0.03}_{-0.02} 	& 30.33^{+0.04}_{-0.03}	&11.62^{+0.20}_{-0.16}	& 0.00	& 7.40_{-0.31}^{+0.39} 	&5.26_{-3.78}^{+4.75} 	&8.29_{-2.99}^{+3.99} 	&38	&0.45 	& 44		&4.3 \\
AGC 749241 	& 0.04	& 24.64^{+0.06}_{-0.05}	&28.75^{+0.06}_{-0.05}	&5.62^{+0.17}_{-0.14} 	& 0.00	& 6.26_{-0.13}^{+0.15}	&0.38_{-0.11}^{+0.13} 	&0.50_{-0.27}^{+0.24}	&27	&0.07 	& 0		&1.5 \\
\hline
\\
\multicolumn{12}{c}{SHIELD II Galaxies} \\
\hline \\
AGC 102728	& 0.11	&26.45^{+0.11}_{-0.09}	& 30.47^{+0.11}_{-0.09}	&12.41^{+0.64}_{-0.53}	& 0.00	& 6.58_{-0.32}^{+0.15}  	&0.79_{-0.58}^{+0.28} 	&1.07_{-0.44}^{+0.37} 	&9.5	&0.48 	& 283	&1.1\\ 
AGC 123352	& 0.61	&26.17\pm0.05		& 29.94\pm0.05		&9.72\pm0.25		& 0.00	& 6.71_{-0.23}^{+0.11}	&1.06_{-0.56}^{+0.26} 	&1.62_{-0.82}^{+0.35} 	&8.2	&0.43 	& 137	&1.3\\
AGC 198507	& 0.08	&26.11^{+0.18}_{-0.24}	& 30.20^{+0.18}_{-0.24}	&10.94^{+0.91}_{-1.22}	& 0.00	& 6.27_{-0.25}^{+0.09} 	&0.38_{-0.22}^{+0.08} 	&1.95_{-0.71}^{+0.54} 	&6.9	&0.11 	& 97		&0.7\\ 
AGC 198508	& 0.09	&25.91^{+0.08}_{-0.07}	& 29.96^{+0.08}_{-0.07}	&9.83^{+0.38}_{-0.33}	& 0.00	& 6.74_{-0.21}^{+0.14} 	&1.14_{-0.56}^{+0.37} 	&2.39_{-1.20}^{+0.55} 	&17 	&0.39	& 67		&1.6\\ 
AGC 198691 	& 0.03	&\nodata			& 30.4^{+0.31}_{-0.60}	&12.1^{+1.7}_{-3.4}		& \nodata	& 5.74_{-0.26}^{+0.13}	&0.04_{-0.02}^{+0.01}	& \nodata				&6.8 &0.61 	& 80		&0.8\\
AGC 200232	& 0.07	&26.06^{+0.01}_{-0.02}	& 30.12^{+0.02}_{-0.03}	&10.57^{+0.12}_{-0.15}	& 0.00	& 7.65_{-0.26}^{+0.19}	&9.30_{-5.58}^{+4.01}	& 4.51_{-1.91}^{+1.90}	&28 	&0.18 	& 95		&2.9\\ 
AGC 205590	& 0.0	5	&26.04\pm0.11	& 30.12\pm0.11		&10.55\pm0.55		& 0.00	& 7.08_{-0.36}^{+0.37}	&2.50_{-2.05}^{+0.21} 	&3.20_{-2.39}^{+0.17}	&16	&0.14 	& 330	&1.3\\
AGC 223231	& 0.03	&25.5\pm0.05		& 29.60\pm0.05		&8.32\pm0.21		& 0.00	& 6.81_{-0.24}^{+0.05}	&1.33_{-0.75}^{+0.15} 	&2.77_{-1.16}^{+1.07} 	&19	&0.18 	& 220	&1.5\\ 
AGC 223254	& 0.05	&24.98^{+0.02}_{-0.01}	& 29.03^{+0.03}_{-0.02}	&6.41^{+0.09}_{-0.07}	& 0.00	& 6.84_{-0.15}^{+0.09}	&1.42_{-0.49}^{+0.28} 	&2.41_{-1.11}^{+0.54} 	&43	&0.10	& 13		&2.7\\ 
AGC 229053	& 0.07	&26.42^{+0.04}_{-0.02}	& 30.48^{+0.05}_{-0.03}	&12.50^{+0.26}_{-0.17}	& 0.00	& 7.39_{-0.29}^{+0.21}	&5.12_{-3.42}^{+2.47} 	&2.19_{-0.74}^{+0.83} 	&41	&0.54 	& 104	&5.0 \\ 
AGC 229379	& 0.04	&25.28\pm0.08		& 29.38\pm0.08		&7.51\pm0.29		& 0.00	& 6.42_{-0.20}^{+0.14}	&0.54_{-0.24}^{+0.17} 	&0.44_{-0.35}^{+0.19}  	&15	&0.04 	& 151	&1.1 \\
AGC 238890	& 0.03	&24.58\pm0.02		& 28.64\pm0.03		&5.34\pm0.07		& 0.15	& 7.12_{-0.19}^{+0.07}	&2.73_{-1.22}^{+0.47} 	&0.57_{-0.30}^{+0.17} 	&53	&0.33 	& 85		&2.7\\
AGC 731448	& 0.07	&25.92\pm0.08		& 30.00\pm0.08		&10.01\pm0.38		& 0.00	& 7.44_{-0.26}^{+0.15}	&4.45_{-2.70}^{+1.57} 	&4.13_{-1.49}^{+1.31} 	&19	&0.38 	& 145	&1.8 \\
AGC 731921	& 0.05	&26.23\pm0.05		& 30.31\pm0.05		&11.51\pm0.29		& 0.00	& 7.66_{-0.32}^{+0.06}	&9.45_{-7.06}^{+1.32} 	&5.07_{-2.46}^{+0.41} 	&24	&0.15 	& 12		&2.7 \\ 
AGC 739005	& 0.11	&25.64^{+0.04}_{-0.05}	& 29.68\pm0.05		&8.63^{+0.18}_{-0.22}	& 0.00	& 7.11_{-0.26}^{+0.06}	&2.64_{-1.60}^{+0.38} 	&3.23_{-1.96}^{+1.19} 	&27	&0.43 	& 125	&2.3\\ 
AGC 740112	& 0.11	&26.08^{+0.07}_{-0.10}	& 30.12^{+0.07}_{-0.10}	&10.56^{+0.36}_{-0.50}	& 0.00	& 7.60_{-0.34}^{+0.10}	&8.16_{-6.39}^{+1.88} 	&0.57_{-0.56}^{+0.27} 	&36	&0.55 	& 102	&3.6 \\ 
AGC 742601	& 0.09	&25.18\pm0.05		& 29.22\pm0.05		&7.00\pm0.18		& 0.10	& 6.46_{-0.22}^{+0.05}	&0.60_{-0.30}^{+0.08} 	&0.71_{-0.45}^{+0.20} 	&17	&0.34 	& 193	&1.1 \\ 
AGC 747826	& 0.05	&26.13^{+0.04}_{-0.05}	& 30.20\pm0.05		&10.94^{+0.28}_{-0.23}	& 0.00	& 7.44_{-0.35}^{+0.11}	&5.74_{-4.64}^{+1.44}	&2.01_{-1.39}^{+0.28} 	&23	&0.32 	& 5		&2.4 \\
\enddata
\tablecomments{Col. 2 lists the Galactic extinction from the dust maps of \citet{Schlegel1998} with recalibration from \citet{Schlafly2011}; values for SHIELD I galaxies are updated from \citet{McQuinn2014}. Cols. 3, 4, and 5 list the TRGB identified in the F814W luminosity function of the resolved stars, the distance modulus calculating from the TRGB mag using the calibration of \citet{Rizzi2007a}, and the corresponding distance respectively. Col. 6 is the internal extinction estimating by fitting the CMD with stellar evolution libraries. Col. 7, 8, and 9 are the present-day stellar mass, average lifetime SFR, and average recent SFR over the past 200 Myr derived from the CMDs using the PARSEC stellar library. Cols. 10, 11, and 12 are the semi-major axis, ellipticity, and position angle determined from the resolved stars. Col. 13 is the optical diameter of the stellar disk estimated from the semi-major axis and adopting our distances. See text for details. The distance to AGC~198691 is from \citet{McQuinn2020}; due to the spareness of the CMD, no recent SFR for this galaxy was measured from the CMD-fitting technique.}
\end{deluxetable*}

All the surveys shown are Northern hemisphere surveys, which allows us to calculate an approximate number density. Selecting systems in the mass range $6.5 \leq$ log(M$_{\rm HI}$/\msun)$ \leq 7.5$, the number density within 4, 8 and 12 Mpc is 0.25, 0.06, and 0.04 Mpc$^{-3}$ respectively. Focusing on just the SHIELD sample, the volume probed is based on a solid angle of 6630 square degrees \citep{Jones2018}, much smaller than the Northern hemisphere, over a distance range of 4-12 Mpc. The resulting number density is 0.02 Mpc$^{-3}$. 

As one moves out in distance, the number counts of galaxies at a given mass should increase as the cube of the distance. While this is not necessarily true over the volume probed by the ALFALFA footprint within $\sim12$ Mpc due to cosmic variance, the decreasing number density from 0.25, 0.06, 0.04, to 0.02 Mpc$^{-3}$ as a function of distance is, in large part, a result of the growing incompleteness below M$_{\rm HI} \sim 10^7$ \msun\ due to the sensitivity limits of the various surveys. The distribution and number density of galaxies in Figure~\ref{fig:D_MHI} highlights the incomplete nature of our current studies of gas-rich, low-mass galaxies, and the rich opportunity for future \hi\ surveys with even greater sensitivity using new facilities such as the Square Kilometer Array (SKA) and its precursors.

\section{The SHIELD Observations and Data Processing}\label{sec:obs}
Our new observations consist of HST images of the resolved stars, which are used to measure distances, stellar masses, and recent SFRs, and WSRT images of the neutral hydrogen, which are used to characterize the gas distributions and kinematics. The HST data processing and analysis follow the same approach used for the SHIELD~I galaxies. For more detailed descriptions, we refer the reader to \citet{McQuinn2014} for HST data processing and TRGB distance methodology,  and to \citet{McQuinn2015a} for measurements of the star formation rates from CMD-fitting. The WSRT data processing and estimates of the \hi\ kinematics are described in full below. 

\subsection{Hubble Space Telescope Observations of the Stars}\label{sec:stars}
Table~\ref{tab:stars} lists the 18 SHIELD~II galaxies for which we have obtained new observations from HST (PID HST$-$GO$-$13750). All galaxies were observed for a single orbit using the Advanced Camera for Surveys (ACS) Wide Field Channel \citep{Ford1998}. Integration times were divided between the F606W (1000 s) and F814W filters ($\sim$1200 s) with \textsc{crsplit} $=$ 2 to allow for the removal of cosmic-rays. The images were processed by the standard HST pipeline. The data from each filter were cosmic-ray cleaned, aligned, and median combined at the native resolution of the ACS instrument using the HST Drizzlepac v2.0 software \citep{Gonzaga2012}. 

Color images of four galaxies are shown in the left panels of Figure~\ref{fig:image1}, created with the F606W (blue), F814W (red), and an average of the two images (green). An atlas of the HST data for the remainder of the SHIELD~II sample is presented in Appendix~\ref{app:HST_atlas}. In order to highlight the low surface brightness features in the galaxies, we apply an arcsinh stretch to the images. The fields of view encompass thrice the optical major diameter or $6\times${\it a}, where {\it a} is the semi-major axis listed in Table~\ref{tab:stars}; the exception is AGC~238890 which has a smaller field of view of twice the optical diameter. Using the same approach adopted for SHIELD~I galaxies studied in \citet{McQuinn2014}, we determined the optical semi-major axes of the galaxies by iteratively examining color-magnitude diagrams (CMDs) of the stars with different elliptical parameters and in concentric annuli. We describe this in more detail in \S\ref{sec:photometry}. 

Similarly to the SHIELD~I sample, the galaxies have irregular morphologies and their stellar concentrations range from compact (AGC~189691) to more distributed (AGC~731921). The physical scales, gauged by the markers in the lower right of each panel, are small, with nearly all of the stellar disks less than 2 kpc in diameter. Knots of recent star formation are apparent in much of the sample, although only 12 of the 18 galaxies are detected in H$\alpha$ (M.~Shepley et al.\ in preparation).

\subsection{Photometry of the Resolved Stars}\label{sec:photometry}
We analyzed the resolved stars in the HST images using the same methodology applied in \citet{McQuinn2014}. Briefly, point-spread function (PSF) photometry was performed with the software package HSTphot using the ACS-specific module  \citep{Dolphin2000}. As input to the photometry, we used the pipeline processed, cosmic-ray rejected {\sc crj.fits} images. These files were also pipeline corrected for charge transfer efficiency (CTE) non-linearities caused by space radiation damage to the ACS instrument \citep[e.g.,][]{Anderson2010, Massey2010}. 

The photometry results were filtered to exclude: objects with large errors (error flag $\ge$ 4); cosmic rays and background galaxies (V$_{\rm sharp} +$ I$_{\rm sharp})^2 > 0.75$ and objects in crowded regions (V$_{\rm crowd} +$ I$_{\rm crowd}) > 0.8$, which have higher photometric uncertainties. The F814W photometry was filtered for sources with a signal-to-noise ratio (S/N) $>$ 4, ensuring only sources with high fidelity measurements are used as input to the TRGB and star formation measurements. The F606W photometry was filtered twice. First, for the purposes of creating clean CMDs and measuring the star formation rates in the galaxies, we made a cut on sources with the more stringent F606W S/N $>$ 4. Second, for the purpose of measuring the TRGB from the F814W luminosity function, we made a more liberal cut with F606W S/N $>$ 2. This more generous S/N cut in F606W avoids excluding sources with a low S/N in the F606W filter but higher S/N in the F814W photometry. This is particularly important given the photometric depth of the data and prevents completeness effects in the F606W data from removing high quality point sources in the F814W data needed for the TRGB measurements.

Finally, we applied spatial cuts to the photometry. The SHIELD galaxies typically subtend only a small fraction of the ACS field of view. Therefore, spatial cuts are used to reduce contamination of background galaxies and foreground stars that were not already rejected by our quality cuts. We used the same approach employed for the SHIELD~I galaxies, given in detail in \citet{McQuinn2014}. Briefly, the spatial extent of the galaxies was determined by plotting the CMD of well-recovered point sources within concentric ellipses centered on the galaxy with position angles and ellipticities that approximately matched the distribution of point sources. Starting with the inner ellipse, the CMD is dominated by stars in the galaxies. As the axes of the ellipses are increased, the stellar density in the galaxy begins to drop until more contaminating point sources are added to the CMD than bona fide stars in the galaxy. We determine that we have reached the edge of the main stellar population detected when the CMD of sources from the outer annulus matches the approximate number and distribution of sources from a field region in the image. Table~\ref{tab:stars} lists these empirically determined ellipticities, position angles, and semi-major axes. 

\begin{figure*}
\begin{center}
\vspace{-0.12in}
\includegraphics[width=0.9\textwidth]{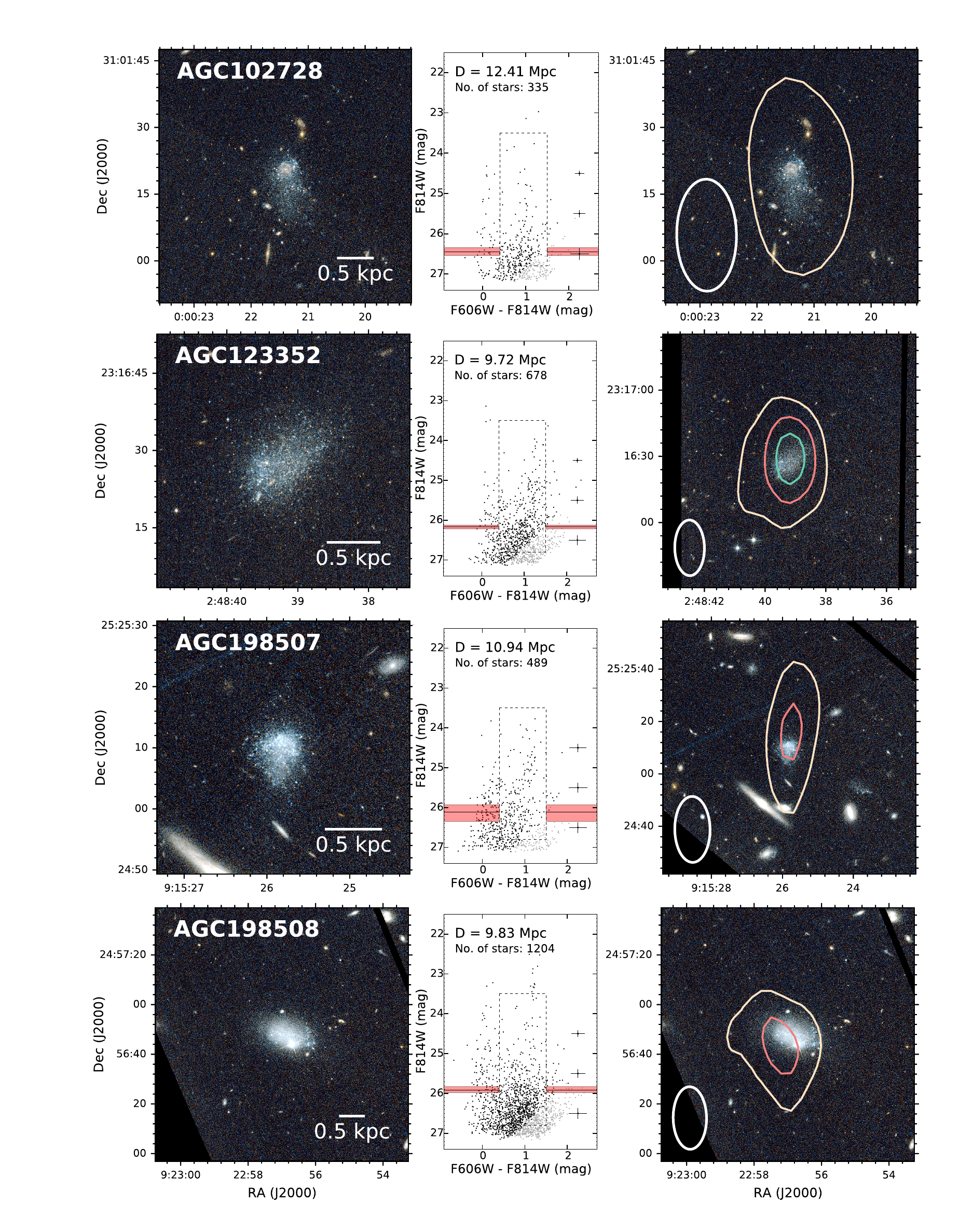}
\end{center}
\vspace{-0.35in}
\caption{HST optical images, CMDs, and HST optical images with WSRT \hi\ contours overlaid for AGC~102728, AGC~123352, AGC~198507, AGC~198508. Left: HST optical images of the galaxies with North up and East left. The physical scales of the galaxies are marked in the lower right. The galaxies range from highly compact (e.g., AGC~198691; see Figure~\ref{fig:image2}) to more extended (e.g., AGC~123352). Middle: CMDs of the resolved stars from the HST imaging. Grey points at the bottom of CMDs are point sources with low S/N in the F606W filter that were used in the distance determinations. The dashed boxes highlight the region of the CMD used for the TRGB measurement. The black horizontal lines mark the measured TRGBs and the shaded red regions represent the measurement uncertainties. Distances to the galaxies and total number of stars in each CMD are listed at the top for each galaxy. Right: HST optical images with WSRT \hi\ contours overlaid at \hi\ column densities corresponding to 4, 8, and 12$\sigma$ in tan, red, and green, respectively. The fields of view in the right panels are the same as in the left panel except for AGC~123352, AGC~198507, and AGC~198691 (Figure~\ref{fig:image2}) which encompass fields $7\times$ the optical diameter to show the more extended \hi\ distributions. The WSRT beam size is shown in the lower left for each galaxy. Note the elongated beam results in somewhat elongated \hi\ contours and, thus, care must be taken when interpreting the apparent \hi\ morphologies of the galaxies. An atlas of the HST data for the remainder of the SHIELD~II sample is presented in Appendix~\ref{app:HST_atlas}.}
\label{fig:image1}
\end{figure*}

Artificial star tests were performed on the images to measure the completeness limits of the data using the same photometry package. The stars were injected in an area encompassing each galaxy using the geometric parameters in Table~\ref{tab:stars}, but over a slightly larger area. We filtered the artificial star tests using the same quality cuts applied to the photometry. 

The CMDs for 4 of the SHIELD~II galaxies are shown in the middle panels of Figure~\ref{fig:image1}, based on the photometry with S/N cuts of 4-$\sigma$ in both filters. We plot in grey the point sources with lower S/N in the F606W image that were used in distance determinations. Representative uncertainties per magnitude are shown which include uncertainties from the photometry and the completeness tests. The depths of the CMDs approximately correspond to the 50\% completeness limits measured by the artificial star tests. An atlas of the CMDs for the remainder of the SHIELD~II sample is presented in Appendix~\ref{app:HST_atlas}.

The CMDs have well-populated red giant branch (RGB) sequences, which allows for a robust distance determination. The exception is AGC~198691 \citep[a.k.a. the Leoncino dwarf;][]{Hirschauer2016}; for this galaxy, deeper HST observations were obtained separately in HST-GO-15423. We adopt the distance and galaxy properties reported from this program in \citet{McQuinn2020} and discuss these results in more detail below. The CMDs of the galaxies also ubiquitously host upper main sequence and helium burning stars, indicative of recent star formation. There are, however, notable differences in the number of these young stars between systems. For example, AGC~223231 has well-defined upper main and helium burning sequences, whereas in AGC~747826 these sequences are comparatively under-populated. We investigate the differences in recent star formation properties in more detail in \S\ref{sec:sf}.

\begin{deluxetable*}{ccccccrCc}
\tablecaption{Summary of WSRT Observations \label{tab:obs}}
\tablehead{
\CH{Galaxy} 	& \CH{RA} 	& \CH{Dec }	& \CH{No. of} 	& \CH{rms noise} 		& \CH{Beam size} 	 & \CH{$B_{PA}$}	& \CH{S$_{\rm HI}$}	& \CH{Resolution}  \\
\CH{} 		& \CH{(J2000)} & \CH{(J2000)}	& \CH{Antennas} 	& \CH{(mJy Bm$^{-1}$)}	& \CH{(\arcsec~$\times$~\arcsec)} & \CH{($^{\circ}$)} & \CH{Jy km s$^{-1}$} & \CH{(kpc)} } 
\colnumbers
\startdata
\\
\multicolumn{9}{c}{SHIELD II Galaxies} \\
\hline \\
AGC~102728	& 00:00:21.4	& +31:01:19	& 8	& 1.64& 43.17$\times$19.41 & 0.5	& 0.20\pm0.02	& 1.74 \\
AGC~123352	& 02:48:39.2	& +23:16:28	& 9	& 1.23& 35.30$\times$12.85 & 0.1	& 0.66\pm0.07	& 1.00 \\ 
AGC~198507	& 09:15:25.8	& +25:25:10	& 10	& 1.48& 33.90$\times$13.32 & 0.3	& 0.58\pm0.06	& 1.13 \\ 
AGC~198508	& 09:22:57.0	& +24:56:48	& 9	& 1.18& 35.64$\times$13.40 & 0.3	& 0.46\pm0.05	& 1.04 \\ 
AGC~198691	& 09:43:32.4	& +33:26:58	& 8	& 1.43& 22.86$\times$11.97 & -1.2& 0.32\pm0.03	& 1.13 \\ 
AGC~200232	& 10:17:26.4	& +29:22:11	& 11	& 1.17& 27.32$\times$13.81 & 0.3	& 0.82\pm0.08	& 1.00 \\ 
AGC~205590	& 10:00:36.5	& +30:32:10	& 11	& 1.16& 26.37$\times$13.76 & 0.4	& 0.36\pm0.04	& 0.97\\ 
AGC~223231	& 12:22:52.7	& +33:49:43	& 9	& 1.33& 23.51$\times$13.49 & 0.6	& 0.83\pm0.08	& 0.72\\ 
AGC~223254	& 12:28:05.0	& +22:17:27	& 9	& 1.18& 39.60$\times$13.78 & 0.3	& 0.89\pm0.08	& 0.73 \\ 
AGC~229053	& 12:18:15.5	& +25:34:05	& 10	& 1.20& 31.36$\times$13.56 & -1.0& 0.81\pm0.08	& 1.25 \\ 
AGC~229379	& 12:30:34.3	& +23:12:19	& 10	& 2.42& 37.08$\times$13.91 & -0.1& 0.25\pm0.03	& 0.83 \\ 
AGC~238890	& 13:32:30.3	& +25:07:24	& 10	& 1.36& 29.77$\times$12.35 & -0.2& 0.25\pm0.03	& 0.50 \\ 
AGC~731448	& 10:23:45.0	& +27:06:39	& \nodata	& \nodata & \nodata &\nodata& \nodata & \nodata\\ 
AGC~731921	& 12:05:34.3	& +28:13:56	& 10	& 1.34& 26.64$\times$12.65 & 0.4	& 1.14\pm0.12	& 0.89 \\
AGC~739005	& 09:13:39.0	& +19:37:07	& 10	& 1.32& 39.29$\times$12.16 & -0.1& 1.02\pm0.10	& 1.22 \\
AGC~740112	& 10:49:55.4	& +23:04:06	& 8  	& 1.37& 30.03$\times$11.85 & 1.2	& \nodata & \nodata\\ 
AGC~742601	& 12:49:36.9	& +21:55:05	& 10	& 1.40& 34.02$\times$12.35 & 0.0	& 0.73\pm0.07	& 0.86 \\
AGC~747826	& 12:07:50.0	& +31:33:07	& 9	& 1.24& 25.19$\times$13.46 & 0.7	& 0.61\pm0.06	& 0.94 \\
\enddata
\tablecomments{Coordinates are based on the optical emission. From the WSRT data, we list the number of dishes available for each observation, rms noise per channel, angular beam size, position angle of the restoring beam ($B_{PA}$) used in calculating the effective beam, total integrated flux density, and the resolution based on the effective radius of a circle with the same area as an ellipse with semi-major and semi-minor axes defined by the beam size and adopting the distances in Table~\ref{tab:stars}. Each \hi\ data cube has a velocity resolution of 4.1 km s$^{-1}$ per channel. Observations for AGC~731448 were interrupted; AGC~740112 is a non-detection.}

\end{deluxetable*}

\subsection{WSRT Observations of the \hi}\label{sec:gas}
The \hi\ data for SHIELD~II were obtained using the WSRT (Program Code 14A$-$018). Table~\ref{tab:obs} lists the number of antennas available for the observations, rms noise per channel, beam sizes, position angles of the restoring beams, and physical resolutions of the \hi\ observations. WSRT acquisition began in December 2013, and was completed in September 2014. The observing campaign coincided with the WSRT APERTIF project upgrade which removed three of the fourteen antennas from use. Additional antennas were unavailable for various maintenance reasons for many of the observations. As a result, the observations were obtained with a smaller collecting area than WSRT has when it is fully operational, and consequently have lower sensitivity. All WSRT observations are 12-hour tracks. For two sources, we were unable to detect the system in the WSRT interferometric data. For AGC~731448, the observations were truncated due to a high priority transient, and, thus, the beam is heavily skewed and the data uninterpretable. For AGC~740112 the source was not detected at a level above the background noise\footnote{The observations were obtained with only eight of fourteen total antennas, which may account for the non-detection. AGC~740112 is detected in the new VLA observations; J.~M.~Cannon et al.\ in preparation.}.

The WSRT \hi\ data reduction was performed using {\sc Miriad}. Radio Frequency Interference (RFI) was excised by hand. Bad baselines and broken antennas were likewise flagged and the data were corrected for system temperature. A combination of 3C48, 3C147, and 3C286 calibrators were observed at the beginning and end of the WSRT observations. As the telescope has excellent phase stability, targets are followed from the moment they peek above the horizon until the moment they set. A phase-only self-calibration was performed, using the {\tt self-cal} task, to correct for changes in the phases during the observation.

\begin{deluxetable*}{lLLLRLCCRRL}
\colnumbers
\tablecaption{HI Properties\label{tab:gas}}
\tablehead{
\CH{Galaxy}     & \CH{V$_{21}$} & \CH{W$_{50}$}      & \CH{S$_{\rm HI}$}     & \CH{PA} 	& \CH{$i$}   	& \CH{D$_{\rm PV}$}	& \CH{R$_{\rm PV}$}   & \CH{V$_{\rm PV}$}    &  \CH{V$_{\rm rot}$} & \CH{log (M$_{\rm HI}$/\msun)}  \\
\CH{}           & \CH{(km s$^{-1}$)} & \CH{(km s$^{-1}$)}       & \CH{(Jy km s$^{-1}$)} & \CH{($^{\circ}$)} & \CH{($^{\circ}$)} & \CH{($\arcsec$)}  & \CH{(kpc)}  & \CH{(km s$^{-1}$)} & \CH{(km s$^{-1}$)} & \CH{}  
}
\startdata
 \\
\multicolumn{11}{c}{SHIELD I Galaxies}  \\
\hline  \\
AGC~110482	& 357\pm1	& 30\pm2	& 1.33\pm0.04	& 84	& 55\pm5	& 21\pm4 	& 0.4\pm0.1	& 15\pm4 	& 9\pm3	& 7.28\pm0.03	\\
AGC~111164	& 163\pm3	& 27\pm6	& 0.65\pm0.04	&326	& 50\pm5	& 20\pm1 	& 0.3\pm0.1	& 20\pm2	& 13\pm2	& 6.60\pm0.03 	\\ 
AGC~111946	& 367\pm2	& 21\pm3	& 0.76\pm0.03	&285	& 62\pm5	& \nodata	& \nodata	& \nodata	& \nodata 	& 7.16\pm0.03	\\ 
AGC~111977	& 207\pm2	& 26\pm4 	& 0.85\pm0.05	& 29	& 59\pm5	& \nodata	& \nodata	& \nodata	& \nodata	& 6.85\pm0.03	\\ 
AGC~112521	& 274\pm1	& 26\pm1 	& 0.69\pm0.04	&180	& 55\pm5	& 23\pm2 	& 0.4\pm0.1 	& 14\pm1 	& 9\pm1 	& 6.85\pm0.03 	\\ 
AGC~174585	& 356\pm3	& 21\pm6 	& 0.54\pm0.04	&290	& 42\pm5	& \nodata	& \nodata 	& \nodata	& \nodata 	& 6.90\pm0.04	\\ 
AGC~174605	& 351\pm1	& 24\pm2	& 0.66\pm0.04	& 90	& 19\pm10 	& 10\pm7	& 0.3\pm0.2	& 10\pm7	& 15\pm13	& 7.27\pm0.03 	\\
AGC~182595	& 398\pm2	& 20\pm4	& 0.42\pm0.03 	& 74	& 39\pm10 	& \nodata 	& \nodata	& \nodata	& \nodata 	& 6.91\pm0.04 	\\
AGC~731457 & 454\pm3	& 36\pm6	& 0.62\pm0.04	& 18	& 34\pm10 	& \nodata 	& \nodata 	& \nodata	& \nodata 	& 7.26\pm0.03	\\
AGC~748778	& 258\pm2	& 16\pm3	& 0.46\pm0.04 	& 21	& 40\pm15 	& \nodata 	& \nodata 	& \nodata	& \nodata 	& 6.66\pm0.04	\\ 
AGC~749237	& 372\pm1	& 65\pm2	& 1.80\pm0.05	&254	& 54\pm5	& 38\pm4 	& 1.1\pm0.1	& 47\pm6	& 29\pm4	& 7.76\pm0.02	\\
AGC~749241& 451\pm1	& 18\pm2	& 0.76\pm0.03	&301	& 45\pm20 	& \nodata 	& \nodata 	& \nodata	& \nodata 	& 6.75\pm0.03	\\ 
\hline  \\
\multicolumn{11}{c}{SHIELD II Galaxies}  \\
\hline  \\
AGC~102728	& 566\pm3	& 21\pm6	& 0.31\pm0.03	&36	& 30\pm15 	& \nodata 	& \nodata 	& \nodata	& \nodata	& 7.05\pm0.06 	\\ 
AGC~123352	& 467\pm3	& 25\pm5	& 0.68\pm0.03	&64	& 50\pm5	& \nodata 	& \nodata 	& \nodata	& \nodata	& 7.18\pm0.03	\\ 
AGC~198507	& 502\pm2	& 37\pm3	& 0.68\pm0.04	&246	& 35\pm10 	& \nodata 	& \nodata 	& \nodata	& \nodata	& 7.28^{+0.08}_{-0.10}\\
AGC~198508	& 519\pm4	& 29\pm7	& 0.37\pm0.04	&223	&50\pm5	& \nodata 	& \nodata 	& \nodata	& \nodata	& 6.93\pm0.06	\\ 
AGC~198691	& 514\pm3	& 33\pm5	& 0.53\pm0.04	&10	& 45\pm5	& \nodata 	& \nodata 	& \nodata	& \nodata 	& 7.26^{+0.13}_{-0.25}\\
AGC~200232	& 450\pm6	& 49\pm11 	& 0.86\pm0.05	&0	& 40\pm5	& \nodata 	& \nodata 	& \nodata	& \nodata	& 7.36\pm0.03	\\
AGC~205590	& 494\pm4	& 29\pm7	& 0.36\pm0.04	&345	& 40\pm5	& \nodata 	& \nodata 	& \nodata	& \nodata	& 6.98\pm0.07	\\ 
AGC~223231	& 571\pm2	& 19\pm3	& 0.97\pm0.04	&104	& 50\pm5	& \nodata 	& \nodata 	& \nodata	& \nodata  	& 7.20\pm0.03	\\ 
AGC~223254	& 603\pm2	& 19\pm3	& 1.16\pm0.04	&80	& 45\pm10 	& \nodata 	& \nodata 	& \nodata	& \nodata	& 7.05\pm0.02	\\ 
AGC~229053	& 425\pm2	& 40\pm4	& 0.77\pm0.04	&190	& 50\pm5	& 34\pm3 	& 1.0\pm0.1 	& 29\pm3	& 19\pm2 	& 7.45\pm0.03	\\
AGC~229379	& 624\pm3	& 22\pm6	& 0.38\pm0.04	&40	& 20\pm10 	& \nodata 	& \nodata 	& \nodata	& \nodata 	& 6.70\pm0.06	\\
AGC~238890	& 360\pm3	& 20\pm6	& 0.35\pm0.04	&180	& 45\pm5	& \nodata 	& \nodata 	& \nodata	& \nodata 	& 6.37\pm0.06	 \\ 
AGC~731448	& 540\pm2	& 39\pm4	& 0.94\pm0.04	& \nodata& 55\pm5	& \nodata	& \nodata	& \nodata	& \nodata 	& 7.35\pm0.04	\\
AGC~731921	& 504\pm2	& 62\pm3	& 1.26\pm0.04	&110	& 40\pm5	& 53\pm4 	& 1.5\pm0.1	& 46\pm3	& 36\pm4 	& 7.60\pm0.03	 \\
AGC~739005	& 433\pm2	& 46\pm3	& 1.16\pm0.05	&308	& 55\pm5	& 38\pm2 	& 0.8\pm0.1 	& 37\pm2	& 23\pm2	& 7.31\pm0.03	 \\
AGC~740112	& 609\pm5	& 37\pm9	& 0.42\pm0.04	& \nodata& 55\pm10 & \nodata	& \nodata	& \nodata	& \nodata	& 7.04^{+0.05}_{-0.06}\\
AGC~742601	& 539\pm2	& 27\pm3	& 0.88\pm0.06	& 266	& 45\pm5	& 24\pm2 	& 0.4\pm0.1	& 13\pm1	& 9\pm1	& 7.01\pm0.04	 \\ 
AGC~747826	& 558\pm2	& 31\pm4	& 0.59\pm0.03	& 204	& 50\pm5	& 25\pm2 	& 0.7\pm0.1	& 10\pm3	& 7\pm2	& 7.22\pm0.03	 \\
\enddata
\tablecomments{Heliocentric velocity of the \hi\ (V$_{21}$), the velocity width of the \hi\ line profile measured at 50\% of its peak value (W$_{50}$), and integrated \hi\ flux density (S$_{\rm HI}$) values are from the ALFALFA catalog \citep{Haynes2018}. The statistical uncertainty on V$_{21}$ is adopted as half the error on the width W$_{50}$ tabulated in Col. 3 \citep{Haynes2018}. $i$ is the inclination angle of the source determined from the $HST$ optical imaging. D$_{\rm PV}$ is the maximum spatial extent in arcsec determined from the PV diagrams using with our new methodology, and R$_{\rm PV}$ is the radius in kpc based on $\frac{1}{2}$D$_{\rm PV}$ and adopting the TRGB distances. V$_{\rm PV}$ is full maximum velocity extent measured from the PV diagrams, also determined using our new methodology, and V$_{\rm rot}$ is $\frac{1}{2}$V$_{\rm PV} / sin~i$ (i.e., half the full velocity extent corrected for inclination). \hi\ masses are based on adopting the TRGB distance in Table~\ref{tab:stars}. See text for more details on all parameters.}
\end{deluxetable*}

A first-order continuum subtraction was performed for all datasets using the task {\tt uvlin}. The ALFALFA spectra were used to determine the velocity range that should be excluded from fitting for the continuum. Dirty image cubes were created with a 4 channel binning, corresponding to a velocity resolution of 4.1 km s$^{-1}$. These cubes, along with the ALFALFA spectra, were used to determine the velocity extent of each source. A single channel image over this entire velocity extent was created. The emission from the galaxy was selected and cleaned in an iterative fashion, with the mask growing as more emission was cleaned. A final mask was created by smoothing to a resolution of 60\arcsec\ $\times$ 30\arcsec, clipping at the 2-$\sigma$ level, and masking to only consider central emission associated with the source. This global clean mask was then applied to all channels in the original cube determined to have emission and a deep cleaning to 0.5-$\sigma$ was performed. A spectrum was then extracted using this global clean mask and the channel range considered for cleaning was compared to the extent of emission seen in this spectrum. If needed, the channel range considered for cleaning was changed to match the extent of emission seen in the spectrum. This approach optimizes the detection and cleaning of low S/N emission, following the approach of \citet{Adams2018}.

\subsection{HI Moment Zero and Moment One Maps}
Once the clean cubes were finalized, moment zero and moment one maps were created over the channel range used for cleaning. Each moment zero map, or total \hi\ intensity map, was clipped at the 2-$\sigma$ level to derive a spectrum over the region where emission was directly detected. These moment zero maps were then used to derive \hi\ column density maps assuming optically thin emission. The moment zero maps were also clipped at the 3-$\sigma$ level and used as a mask on the moment one maps, or velocity fields, so that only areas of significant emission were included.

An example \hi\ intensity map and \hi\ velocity field for AGC~731921 are shown in the right panels of the Figure~\ref{fig:hi}; an atlas of the \hi\ data for the remainder of the SHIELD~II sample is presented in Appendix~\ref{app:HI_atlas}. In the top right panel of Figure~\ref{fig:hi}, we show the \hi\ column density maps with contours corresponding to the 4, 8, and 12-$\sigma$ levels of emission in tan, red, and green colors respectively, matching the \hi\ column density contours overplotted in the HST images in Figure~\ref{fig:image1} and in Appendix~\ref{app:HST_atlas}. In the bottom right panel, we show the velocity field. Also shown are the beam size, physical scale of the image, and the location of the position-velocity (PV) slice used in the kinematics analysis below. The position angle of the PV slice is provided in Table~\ref{tab:gas}. Note that, while the elongated elliptical beam of WSRT observations prevents a detailed analysis of the \hi\ morphology, the overall extent of the \hi\ is accurately represented. The neutral gas is more extended than the detected stellar disks in all cases. A few galaxies do have East-West velocity gradients that are resolved by the beam sizes; we make note of this for the individual systems in the \hi\ image atlas in Appendix~\ref{app:HI_atlas}. 

Figure~\ref{fig:hi} also shows example spectra in the top left panel, along with the channel range used for cleaning, for AGC~731921. Three spectra are shown for comparison: a WSRT spectrum based on the global masks used for cleaning, a WSRT spectrum based on clipping the moment zero map at the 2-$\sigma$ level, and the ALFALFA spectrum; similar comparisons for the remaining galaxies with WSRT detections are presented in Appendix~\ref{app:HI_atlas}. Both the spectrum from the global clean mask and the 2-$\sigma$ moment zero mask were used to derive the flux of the galaxies. The fluxes derived from spectra using the clean masks are significantly larger than those derived using the moment zero 2-$\sigma$ mask as the global clean mask was spatially smoothed and includes real emission that is at the level of the noise. We report the fluxes from the global clean mask in Table~\ref{tab:obs}, assuming a 10\% uncertainty from the flux scale accuracy. The fluxes  are generally smaller than but consistent with the fluxes reported from the ALFALFA survey using the single-dish measurements; we list these values in Table~\ref{tab:gas}. The exception is AGC~198508 where noise peaks in the global WSRT spectrum artificially inflate the flux.

\begin{figure*}
\begin{centering}
\includegraphics[width=0.98\textwidth]{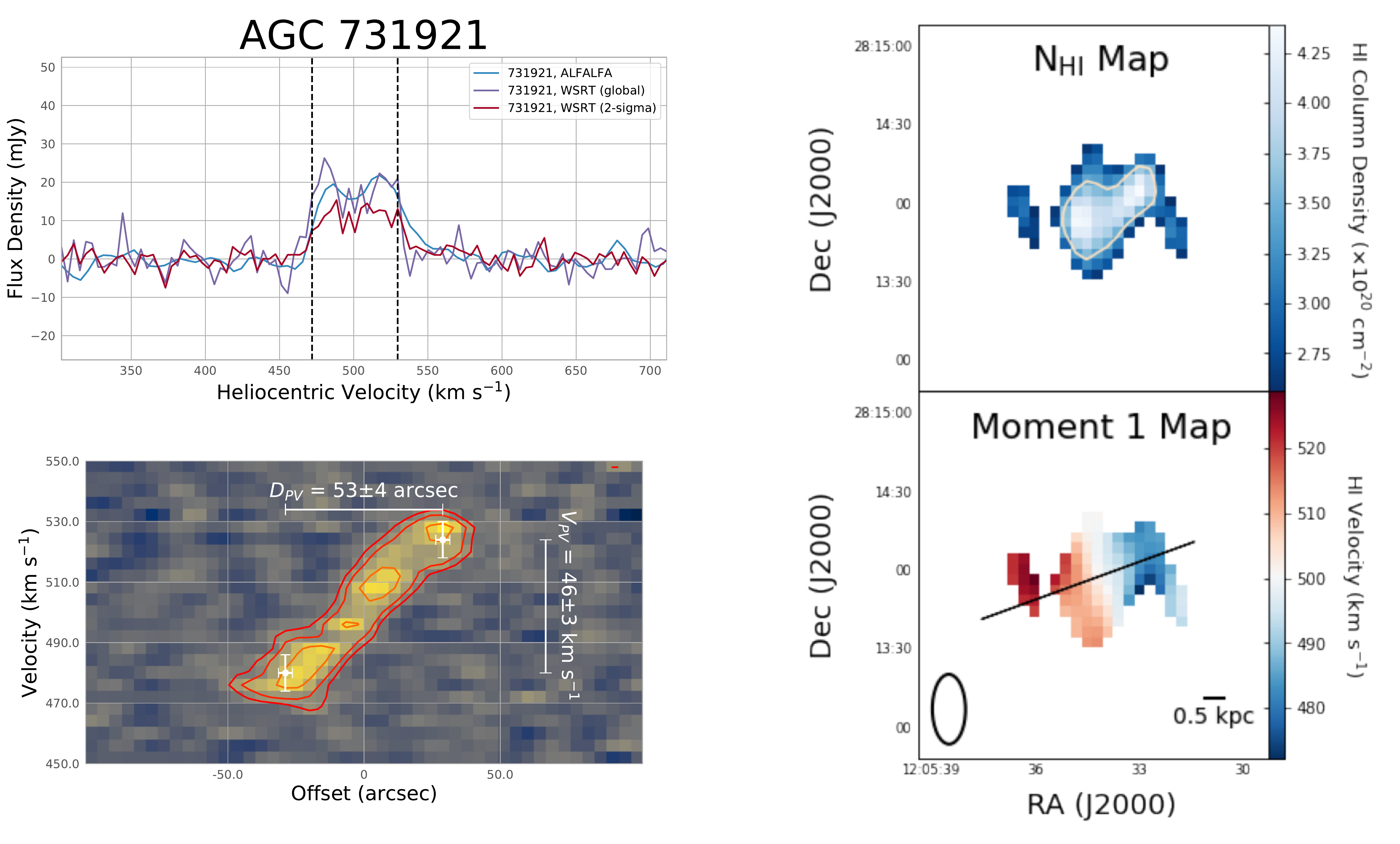} 
\end{centering}
\caption{Clockwise from the top left: Comparison of the WSRT global spectrum and the WSRT spectrum clipped at the 2-$\sigma$ level with the ALFALFA spectrum for AGC~731921. The velocity range used in cleaning the spectrum is marked with vertical dashed lines. Top right: Example of the \hi\ column density map for AGC~731921 with \hi\ column density contours overlaid at the 4, 8, 12-$\sigma$ detection levels based on the rms listed in Table~\ref{tab:obs}. Bottom right: the \hi\ velocity field map.  The identified major axis of rotation passing through the kinematic center used for making a position$-$velocity (PV) diagram is marked as a black line. Physical scale marker and \hi\ beam size are also shown. The field of view in each panel is $3\times$ the optical diameter. Bottom left: Example of spatially resolved PV diagram across the major axes of AGC~731921 taken along the black line shown in the velocity field with contours overlaid at 2, 3, 5, 7-$\sigma$ detection levels based on the rms measured off-source in the PV slice. The spatial extent of the \hi\ (D$_{\rm PV}$) and the range in velocity measured from the PV slice (V$_{\rm PV}$), marked in the figure, as well as the position angle (PA) measured east of north of the PV slice, are listed in Table~\ref{tab:gas}. An atlas of the \hi\ data for the full SHIELD~II sample is presented in Appendix~\ref{app:HI_atlas}.}
\label{fig:hi}
\end{figure*}

\subsection{Measurements of the HI Kinematics}\label{sec:kinematics}
While the moment one maps provide information on the gas kinematics, the small spatial extent of the \hi\ and the less clearly defined rotation in the gas make it difficult to extract the kinematic information using methods traditionally applied to more massive galaxies \citep[i.e., fitting tilted ring models using codes such as $^{\rm 3D}${\sc Barolo} to derive rotation curves;][]{DiTeodoro2015}. Instead, we measure the rotational velocities and spatial extent of the \hi\ by analyzing spatially-resolved PV slices through the cubes, at the angle of greatest velocity spread. 

An example PV diagram is shown in the bottom left of Figure~\ref{fig:hi} for AGC~731921; Appendix~\ref{app:HI_atlas} includes PV diagrams for the rest of the sample. Contours are overlaid at the 2, 3, 5, 7-$\sigma$ level of the rms values measured from the PV slice in an off-source region, after spatially smoothing with a 1-pixel boxcar. All PV slices have a width of 50\arcsec, which is larger than the major axis beam size in all cases. This ensures that the slices are representative of the bulk projected motion and are less sensitive to the absolute slice position or PA than a narrower (e.g., a few pixels wide) slice would be. 

Here, we introduce a new methodology that enables the robust determination of the maximum velocity detected in the \hi\ data for each system and its spatial extent, including associated uncertainty estimates, from PV diagrams. This new method is an improvement over the technique originally used for the SHIELD I galaxies in \citet{McNichols2016} where the maximum velocity of the \hi\ gas was empirically estimated based on the maximum identified extent of the emission in the PV diagram.  

Briefly, the PV diagram is sliced in orthogonal bins and the maximum range in velocity is determined by taking bins along the spatial position offset and fitting the velocity profile with a Gaussian. The difference between the center velocity of the Gaussians in the two furthest spatial bins is the maximum range in velocity, which we label V$_{\rm PV}$. The maximum spatial extent is obtained in an analogous way by taking bins along the velocity axis and fitting a Gaussian to the spatial dimension. The difference between the minimum and maximum values fit with a Gaussian is taken as the total extent, or diameter, which we label D$_{\rm PV}$. The returned velocity and spatial extent values are only considered meaningful if the spatial extent is larger than the effective beam of the PV diagram. 

Figure~\ref{fig:hi} shows an example PV slice with the maximum velocity value and spatial extent found by this methodology for a galaxy that is well-resolved with high S/N data. There is a projected velocity gradient from north to south with a magnitude of 46$\pm3$ km\,s$^{-1}$ over an extent of 53$\pm4$\arcsec. The source is resolved by the \hi\ beam so this gradient can be interpreted as projected rotation. The velocities appears to reach a plateau in the outer edges of the detected \hi, suggesting the data are reaching the flat part of the rotation curve. 
 
This new methodology and its application to both the SHIELD~I and SHIELD~II galaxies is described in detail in Appendix \ref{app:pvvel}. We also present PV slices and their derived maximum velocities and extents in Appendices \ref{app:HI_atlas} and \ref{app:pvvel} for SHIELD~II and SHIELD~I, respectively. Note that the velocity and spatial extent could not be measured for all galaxies; Appendix \ref{app:pvvel} also describes the criteria we used to apply our new methodology and which galaxies did not meet our criteria.

It is important to note that these velocity measurements represent estimates of the bulk motion of the gas, and while they are good indicators of the rotational velocities, they are not as robust as values determined from kinematic modeling. J.\ Fuson et al.\ (in preparation) will present a detailed comparison of this approach to rotation curve modeling based on higher angular resolution observations.

Final rotational velocity values were calculated using half the difference between the maximum and minimum velocity of gas we could attribute to the source, corrected for inclination (i.e., V$_{\rm rot} = \frac{1}{2} $V$_{\rm PV}~/~sin~i$) The inclinations were derived from the HST optical images. Specifically, the F814W image was edited by hand to remove obvious foreground and background contaminants, smoothed, and then fit with an ellipse to determine the axial ratio and position angle. This follows the approach used for the SHIELD~I sample in \citet{Teich2016}. The values of $i$, D$_{\rm PV}$, V$_{\rm PV}$, and V$_{\rm rot}$, as well as the radius R$_{\rm PV} = \frac{1}{2}$ D$_{\rm PV}$ converted to a physical scale in kpc, are listed in Table~\ref{tab:gas}. We also include our updated values for the SHIELD~I galaxies in Table~\ref{tab:gas}, and, for completeness, we list the \hi\ heliocentric velocities, line widths, and fluxes as reported in the ALFALFA catalog \citep{Haynes2018}. The \hi\ masses for the galaxies were calculated based on the measured fluxes from the ALFALFA survey and adopting our TRGB distances (see \S\ref{sec:trgb}). 

For nearly all the galaxies, the rotational velocities measured are still on the rising part of the rotation curve in the inner radii of the galaxies. Thus, our values of V$_{\rm rot}$ are lower limits for the full rotational velocity of the galaxies which can only be determined at larger radii on the flat part of a rotation curve. In addition, we have not accounted for the velocity dispersion in the gas nor have we made any asymmetric drift corrections; given the low rotational velocities measured in the galaxies, the inclusion of dispersion and asymmetric drift may impact the rotational velocities significantly. These corrections, as well as analysis of the rotational velocities, dynamical masses, and the stellar and gas content of the sample will be the focus of a future publication on the BTFR (K. McQuinn et al.\ in preparation). 

\section{Tip of the Red Giant Branch Distances}\label{sec:trgb}
\subsection{TRGB Measurements}
The RGB sequence of stars is well-defined in the CMDs of SHIELD~II galaxies, enabling robust distances to be measured using the TRGB method\footnote{The exception, as noted above, is AGC~198691, which has a TRGB distance measurement from deeper HST data presented in \citet{McQuinn2020}.}. The TRGB is a primary, Population~II distance indicator that uses the predictable luminosity peak of low-mass stars just before the helium flash as a standard candle \citep[e.g.,][]{Lee1993, Sakai1996}. The luminosity of stars as they approach the TRGB has some dependency on the metal content of the stars, and, to a lesser extent, the stellar age (or mass), due to bolometric corrections. This dependency is modest in the I-band (or I-band equivalent filters such as the F814W), compared with other wavelengths \citep[e.g.,][]{Salaris2005, McQuinn2019a}, and can be accounted for in calibrations as discussed below. High-resolution imaging from HST has revolutionized our ability to efficiently measure high-quality distances using the TRGB method out to $\sim$15 Mpc \citep[e.g.,][]{Tully2013}; since the TRGB is brighter in the infrared, JWST has the potential to reach larger distances than HST, more efficiently \citep{Beaton2018, McQuinn2019a}.

TRGB distances are determined by identifying the discontinuity in the extinction-corrected F814W luminosity function from stars pre-selected from a CMD to be in the RGB region and then translating that luminosity to a distance. The dashed boxes in the CMDs of Figure~\ref{fig:image1} and of Appendix~\ref{app:HST_atlas} highlight the stars used in our distance determinations; all boxes encompass the same color range (0.4$<$ F606W$-$F814W$<$1.5) with luminosity limits individually chosen for each galaxy to include stars that are just above the limit of detection to $\sim2$ mag brighter than the TRGB. 

We measure the F814W luminosity function discontinuity corresponding to the TRGB using a maximum likelihood (ML) technique, following the same approach employed for the SHIELD~I galaxies in \citet{McQuinn2014}. Briefly, the ML approach fits a parametric RGB luminosity function to the observed F814W luminosity function. The probability estimation takes into account the photometric error distribution and completeness from the artificial star tests \citep{Makarov2006}, which is particularly important in data with limited photometric depth. The assumed theoretical form of the luminosity function is: 

\begin{subequations}
\begin{empheq}[left={P = }\empheqlbrace]{alignat=2}
        & 10^{(A*(m - m_{TRGB}) + B)}, & \quad \text{if m - m$_{TRGB} \geq 0$}\\
        & 10^{(C*(m - m_{TRGB}))}, & \quad \text{if m - m$_{TRGB} < 0$}
\end{empheq}
\label{eq:ml_form}
\end{subequations}

\noindent where A, B, and C are treated as free parameters in the majority of fits, and A and C have normal priors of 0.30 ($\sigma=0.07$) and 0.30 ($\sigma=0.2$), respectively. In six cases (AGC~205590, AGC~229053, AGC~229379, AGC~238890, AGC~731448, AGC~747826), the data did not suitably constrain the three parameters and A and C were held fixed to their priors. The range in solutions returning log(P) within 0.5 of the maximum gives the uncertainty. We also independently checked that the ML results agreed within the uncertainties with the simpler approach of using a Sobel filter. The benefit of the ML technique is the more robustly quantified uncertainties \citep[see][for details]{McQuinn2014}. The best-fits to the data are shown as a solid black line in the CMDs of Figure~\ref{fig:image1} and of Appendix~\ref{app:HST_atlas} with the measurement uncertainties shaded in red. 

The measured TRGB luminosities in the F814W filter were calibrated to an absolute distance scale by applying the modest, color-based correction for metallicity and zero-point for ACS-specific filters from \citet{Rizzi2007a}:

\begin{align}
& M_{F814W}^{ACS} = -4.06(\pm0.02) \nonumber \\
& + 0.20(\pm0.01) \cdot [(F606W-F814W) - 1.23] \label{eq:trgb}
\end{align}

\noindent where F606W$-$F814W is the average color of the TRGB stars. The distances using the \citet{Rizzi2007a} calibration are also consistent with the newer calibration from \citet{Jang2017}.\footnote{The \citet{Jang2017} calibration includes a higher-order correction that has a greater impact on more metal-rich stellar populations typical of spiral galaxies.} 

\begin{figure}
\includegraphics[width=0.48\textwidth]{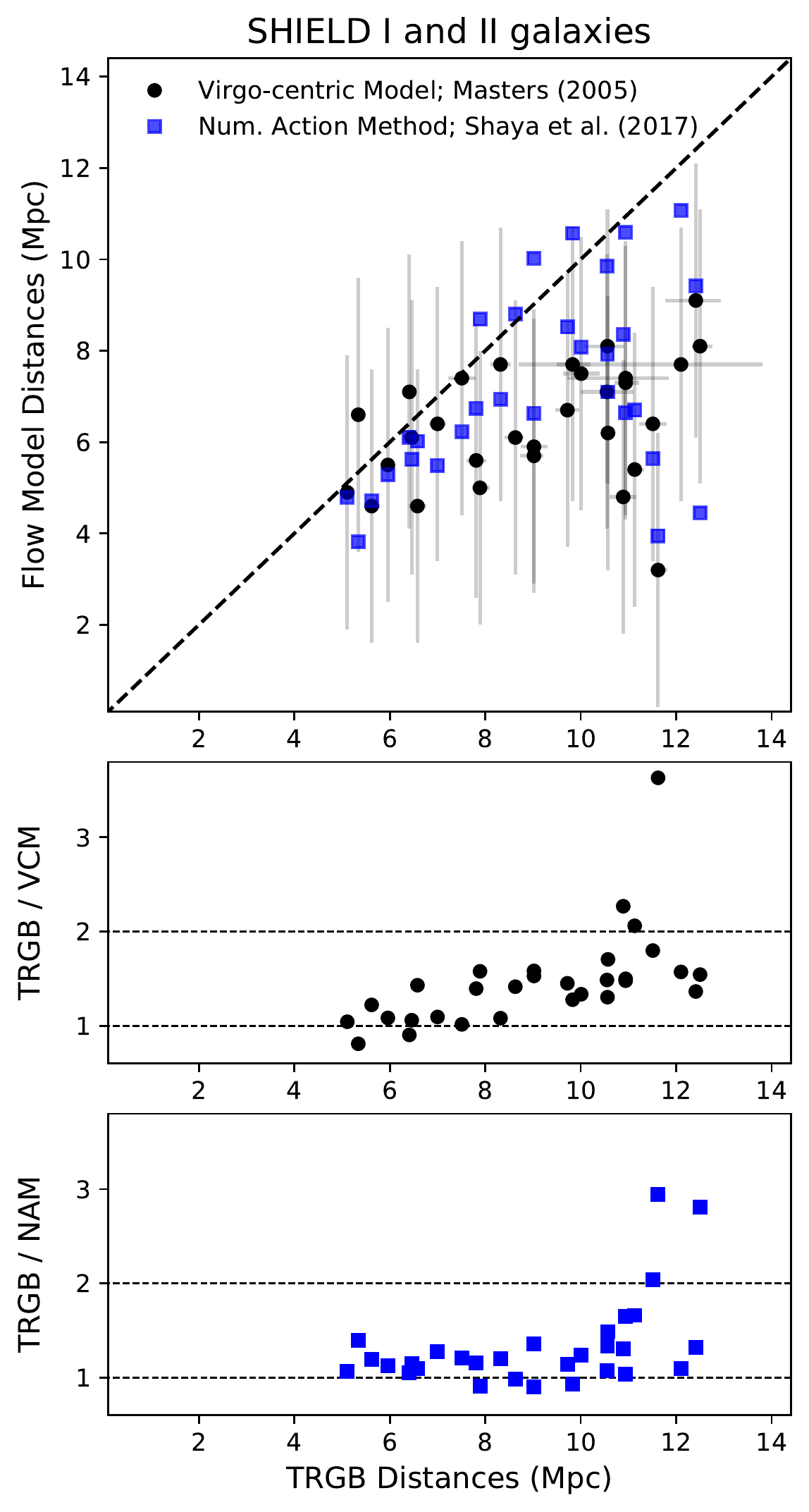} 
\caption{Top panel: Distances determined from a Virgo-centric flow model using the \hi\ velocities of the galaxies \citep{Masters2005} versus the distances determined using the TRGB method (black circles). The dashed line represents one-to-one agreement between the two methods. While, overall, the flow model distances are within 1-$\sigma$ of our measured TRGB distances, the discrepancy for any individual galaxy be as much as a factor of four. Also shown are distance estimates based on the numerical action method from \citet[][blue squares]{Shaya2017}. Lower two panels: The ratio of the TRGB distances to the Virgo-centric flow model (TRGB / VCM) and the numerical action method (TRGB / NAM) as a function of TRGB distance. These comparisons highlight the challenges of using velocity-based distances in the nearby universe for even inter-sample comparisons of galaxy properties.}
\label{fig:flow_trgb}
\end{figure}

The measured distances for the SHIELD~II galaxies range from $\sim$5 to $\sim$12 Mpc. Table~\ref{tab:stars} lists the apparent magnitude of the identified discontinuity in the F814W luminosity function before correcting for extinction, the calculated distance moduli, and the corresponding distance for each galaxy. The uncertainties are based on adding in quadrature the uncertainties from the ML technique, photometry, artificial star tests, and the TRGB calibration. For AGC~198691, we list the distance determined from \citet{McQuinn2020}. For completeness, Table~\ref{tab:stars} also lists the values for the SHIELD~I sample from \citet{McQuinn2014}.

\begin{deluxetable}{cLCC}
\tablecaption{Comparison of TRGB and Flow-Model Distances \label{tab:flow_models}}
\tablehead{
\CH{Galaxy} & \CH{D$_{\rm TRGB}$} & \CH{D$_{\rm M05}$} & \CH{D$_{\rm S17}$}
}
\setlength{\tabcolsep}{15pt}
\colnumbers
\startdata
\hline \\
\multicolumn{4}{c}{SHIELD I Galaxies}\\
\hline \\
AGC~ 110482  &  7.82 \pm 0.21	 		&  5.6  &  6.7  \\
AGC~ 111164  &  5.11 \pm 0.07 	 		&  4.9  &  4.8  \\
AGC~ 111946  &  9.02 ^{+ 0.2 }_{- 0.29 } 		&  5.7  &  6.6  \\
AGC~ 111977  &  5.96 ^{+ 0.11 }_{- 0.09 } 	&  5.5  &  5.3  \\
AGC~ 112521  &  6.58 \pm 0.18 			&  4.6  &  6.0 \\
AGC~ 174585  &  7.89 ^{+ 0.21 }_{- 0.17 } 	&  5.0  &  8.7 \\
AGC~ 174605  &  10.89 \pm 0.28 			&  4.8  &  8.4 \\
AGC~ 182595  &  9.02 \pm 0.28 			&  5.9  &  10.0  \\
AGC~ 731457  &  11.13 ^{+ 0.2 }_{- 0.16 } 	&  5.4  &  6.7 \\
AGC~ 748778  &  6.46 ^{+ 0.14 }_{- 0.17 } 	&  6.1  &  5.6  \\
AGC~ 749237  &  11.62 ^{+ 0.2 }_{- 0.16 } 	&  3.2  &  4.0 \\
AGC~ 749241  &  5.62 ^{+ 0.17 }_{- 0.14 } 	&  4.6  &  4.7 \\
\hline \\
\multicolumn{4}{c}{SHIELD II Galaxies}\\
\hline \\
AGC~ 102728  &  12.41 ^{+ 0.64 }_{- 0.53 } 	&  9.1  &  9.4 \\
AGC~ 123352  &  9.72 \pm 0.25 			&  6.7  &  8.5 \\
AGC~ 198507  &  10.94 ^{+ 0.91 }_{- 1.22 } 	&  7.4  &  10.6 \\
AGC~ 198508  &  9.83 ^{+ 0.38 }_{- 0.33 } 	&  7.7  &  10.6 \\
AGC~ 198691  &  12.1 ^{+ 1.7 }_{- 3.4 } 		&  7.7  &  11.1 \\
AGC~ 200232  &  10.57 ^{+ 0.12 }_{- 0.15 } 	&  6.2  &  7.1 \\
AGC~ 205590  &  10.55 \pm 0.55 			&  7.1  &  9.9 \\
AGC~ 223231  &  8.32 \pm 0.21 			&  7.7  &  6.9 \\
AGC~ 223254  &  6.41 ^{+ 0.09 }_{- 0.07 } 	&  7.1  &  6.1 \\
AGC~ 229053  &  12.50 ^{+ 0.26 }_{- 0.17 } 	&  8.1  &  4.5 \\
AGC~ 229379  &  7.51 \pm 0.29 			&  7.4  &  6.2 \\
AGC~ 238890  &  5.34 \pm 0.07 			&  6.6  &  3.8 \\
AGC~ 731448  &  10.01 \pm 0.38 			&  7.5  &  8.1 \\
AGC~ 731921  &  11.51 \pm 0.29 			&  6.4  &  5.6 \\
AGC~ 739005  &  8.63 ^{+ 0.18 }_{- 0.22 } 	&  6.1  &  8.8 \\
AGC~ 740112  &  10.56 ^{+ 0.36 }_{- 0.5 } 	&  8.1  &  7.9 \\
AGC~ 742601  &  7.00 \pm 0.18 			&  6.4  &  5.5 \\
AGC~ 747826  &  10.94 ^{+ 0.28 }_{- 0.23 } 	&  7.3  &  6.6 \\
\enddata 
\tablecomments{Distances in Mpc measured using the TRGB method in this work compared with distance estimates from the Virgo-centric flow model of \citet[][D$_{\rm M05}$]{Masters2005} and the Numerical Action Method of \citet[][D$_{\rm S17}$]{Shaya2017} for the SHIELD I and II samples. Uncertainties on the flow-model distances of \citet{Masters2005} are estimated to be $\pm3$ Mpc. A graphical comparison of the values is shown in Figure~\ref{fig:flow_trgb} and discussed in the text in \S\ref{sec:trgb}.}
\end{deluxetable}

\subsection{Comparison with flow model Distances}\label{sec:flow}
Figure~\ref{fig:flow_trgb} compares our measured TRGB distances with the Virgo-centric flow model distances adopting the reported minimum uncertainties of $\pm3$ Mpc \citep{Masters2005}. With two exceptions, the TRGB distances to the SHIELD galaxies are farther than the original ALFALFA estimates based on the flow model. The larger number of under-estimated distances is due to our sample selection criteria and the Malmquist bias. Because we imposed an \hi\ upper mass cutoff, galaxies with flow model distances larger than their true distance would have higher \hi\ mass estimates and, thus, are excluded from our study. Also shown in Figure~\ref{fig:flow_trgb} is a comparison of our measured TRGB distances to estimates from the numerical action method (NAM) model that is based on galaxy orbit reconstruction \citep{Shaya2017} with the NAM tool \citep{Kourkchi2020} and is used in the CosmicFlows-3 program \citep{Tully2013}. The NAM distance estimates are a better match, although there are still a number of points that disagree by a factor of 2. 

The lower two panels in Figure~\ref{fig:flow_trgb} show the ratio of the TRGB distances to the Virgo-centric flow models (TRGB / VCM) and NAM method (TRGB / NAM) as a function of TRGB distance. These comparisons highlight the importance of primary distance measures in the local universe since the NAM distances are still discrepant. While parametric multi-attractor flow models work to minimize the impact, the complicated peculiar motions in the local universe make individual distances inferred from velocity based models highly uncertain. For even inter-sample comparisons, measuring distances via reliable methods such as the TRGB method are critical for accurate interpretation of galaxy properties as discrepancies at the factor of 2 level translate to differences of a factor of 4 for many galaxy properties that rely on the square of the distance (e.g., luminosity, galaxy masses, etc.).

\section{CMD-Based Star Formation Rates}\label{sec:sf}
In addition to providing TRGB distances, the CMDs in Figure~\ref{fig:image1} and in Appendix~\ref{app:HST_atlas} also contain information on the star formation histories of the galaxies. Qualitatively, differences in the overall stellar mass assembly of the galaxies can be discerned in the CMDs. The total number of stars detected in the CMDs gives an indication of the total stellar mass and constrains the lifetime SFRs. The structure of the brightest stellar evolution sequences gives an indication of recent star formation activity and points to obvious differences between galaxies in the sample. For example, as mentioned in \S\ref{sec:photometry}, there is evidence of recent star formation activity in AGC~223231, based on the strong upper main sequence and helium burning sequences. In contrast, there has been comparatively less recent star formation in AGC~747826.  

Quantitatively, the recent and lifetime star formation properties of the galaxies can be measured by fitting the CMD with a series of synthetic simple stellar populations of different ages and metallicities. The best-fitting modeled CMD represents the most likely star formation history (i.e., SFR(t)) of the system. Measuring detailed star formation histories with high time resolution requires both sufficient photometric depth and sufficiently populated CMDs. As the photometric depth of the SHIELD observations is limited to $\sim1$ mag below the expected TRGB by design, we limit the temporal resolution of the star formation histories to two time bins. Specifically, we measure the average SFR over the last 200 Myr (i.e., $\langle$SFR$\rangle_{\rm 200~Myr}$) and the average SFR over the lifetime of the galaxy (i.e., $\langle$SFR$\rangle_{\rm life}$). 

The SFRs were measured using the CMD-fitting software \textsc{match} \citep{Dolphin2002a}. We followed the same methodology employed in the analysis on the SHIELD~I galaxies, as described in detail in \citep{McQuinn2015a}, with some modifications of the inputs. We assumed a Kroupa initial mass function (IMF) from \citet{Kroupa2001} and a binary fraction of 35\% with a flat secondary mass function distribution. The distances were fixed to the TRGB distances listed in Table~\ref{tab:stars} and the mean metallicities were constrained to be a continuous, non-decreasing function with time. We assumed the Galactic foreground extinction from \citet{Schlafly2011}, listed in Table~\ref{tab:stars}, and assumed internal extinction values determined by iteratively fitting the CMDs with an extinction parameter in increments of 0.05 mag. The majority of galaxies were best fit without internal extinction; final values are listed in Table~\ref{tab:stars}. Random uncertainties were estimated by applying a hybrid Markov Chain Monte Carlo simulation \citep{Dolphin2013}. Systematic uncertainties from the stellar evolution models were estimated by applying shifts in luminosity and temperature to the observed stellar populations through 50 Monte Carlo simulations per solution \citep{Dolphin2012}. 

\begin{deluxetable*}{llcccc}
\tablecaption{Galaxies within 500 kpc of the SHIELD I and II samples \label{tab:neighbors}}
\tablehead{
\CH{SHIELD} 	& \CH{Neighbor}	& \CH{Neighbor}	& \CH{Neighbor}	& \CH{Dist.} 	& \CH{3-D Separation}\\
 \CH{Galaxy}	& \CH{Galaxy }		& \CH{RA (J2000)}	& \CH{Dec (J2000)} 	& \CH{(Mpc)}	& \CH{(kpc)}
 }
\startdata
\hline \\
AGC~110482	& IC~1727 	& 1.791638	& 27.333021	& 7.45	& 420 \\
AGC~111164 	& NGC0784 	& 2.021333	& 28.843611	& 4.97 	& 140 \\
AGC~111164 	& UGC01281 	& 1.825639 	& 32.592500	& 4.94 	& 420 \\
AGC~112521 	& KK14 		& 1.745206	& 27.288457	& 7.01 	& 440 \\
AGC~223254	& UGC07603 	& 12.47891	& 22.820347	& 6.79 	& 390 \\
AGC~739005 	& D564-08 	& 9.048332 	& 20.074716	& 8.83 	& 440 \\
\hline
\enddata
\tablecomments{Galaxies within 500 kpc of the SHIELD I and II samples identified in the Cosmicflows-3 \citep{Tully2013} and Updated Nearby Galaxy Catalog \citep{Karachentsev2013}. AGC~111164 appears to be paired with a slightly more massive neighbor and resides in a void \citep{Pustilnik2016}.}
\end{deluxetable*}

We experimented with deriving the CMD fits with different stellar libraries including PARSEC \citep{Bressan2012} and MIST \citep{Choi2016}. Both the lifetime and recent SFRs from the two libraries agreed within the uncertainties. For the remainder of the analysis, we adopt the SFRs derived using the PARSEC library.

Table~\ref{tab:stars} lists the recent and lifetime SFRs as well as the present-day stellar masses for the sample. The present-day stellar masses were calculated based on the total stellar mass formed in each galaxy derived from the CMD fits with a Kroupa IMF \citep[adjusted to the mass limits of $0.1 - 100$ \msun;][]{Telford2020} and assuming a correction for the amount of mass returned from stars via stellar winds and supernovae.  We adopt a return mass fraction of 55\% based on a Kroupa IMF at low metallicity.

As noted above, the star formation properties for the SHIELD~I sample were previously derived using the same technique, but the CMD-fits were based on the Padua stellar evolutionary models from \citet{Marigo2008} with AGB tracks from \citet{Girardi2010}, assumed a Salpeter IMF \citep{Salpeter1955}, and adopting a lower return mass fraction of 30\% \citep{Kennicutt1994}. To ensure the SHIELD galaxies can be inter-compared without a systematic bias due to these different assumptions, we have re-derived the star formation properties of SHIELD~I using the PARSEC library, assuming a Kroupa IMF, and the higher return fraction of 55\%. The change from a Salpeter to a Kroupa IMF decreases the total stellar mass by a factor of $\sim$0.66, while the difference in return fraction has a smaller impact. After adjusting for these changes, we found the originally derived stellar mass using the Padua library agreed within the uncertainties with those found using PARSEC in 10 of the 12 SHIELD~I galaxies. For the remainder of our analysis for the SHIELD~I galaxies, we adopt the star formation properties derived with the PARSEC library and the updated assumptions; those values can be found in Table~\ref{tab:stars}.

As a consistency check, we compared our $\langle$SFR$_{200~\rm Myr}\rangle$ to values determined using GALEX far ultraviolet imaging available for the SHIELD~I galaxies \citep{Teich2016} and adopting an empirically determined UV-SFR scaling relation \citep{McQuinn2015c}. All values agreed within the uncertainties. We also compared our measured stellar masses to values previously determined using Spitzer 3.6 and 4.5$\micron$ imaging \citep{Haurberg2015}. Here, we found that nearly all values also agreed within the uncertainties of the CMD-based stellar mass values. There were two exceptions that differed by 0.10 and 0.30 dex, which are still in reasonable agreement as the stellar masses reported from the Spitzer imaging do not have reported uncertainties.

Note that, while all galaxies have measurable star formation activity over the past 200 Myr, 6 of the 18 SHIELD~II galaxies are not detected in H$\alpha$ (M.~Shepley et al.\ in preparation). The lack of a massive star population needed to ionize hydrogen implies either that the SFR is fluctuating on short timescales, the upper IMF is not fully populated at these low SFRs, or both. 

\begin{figure*}
\includegraphics[width=0.99\textwidth]{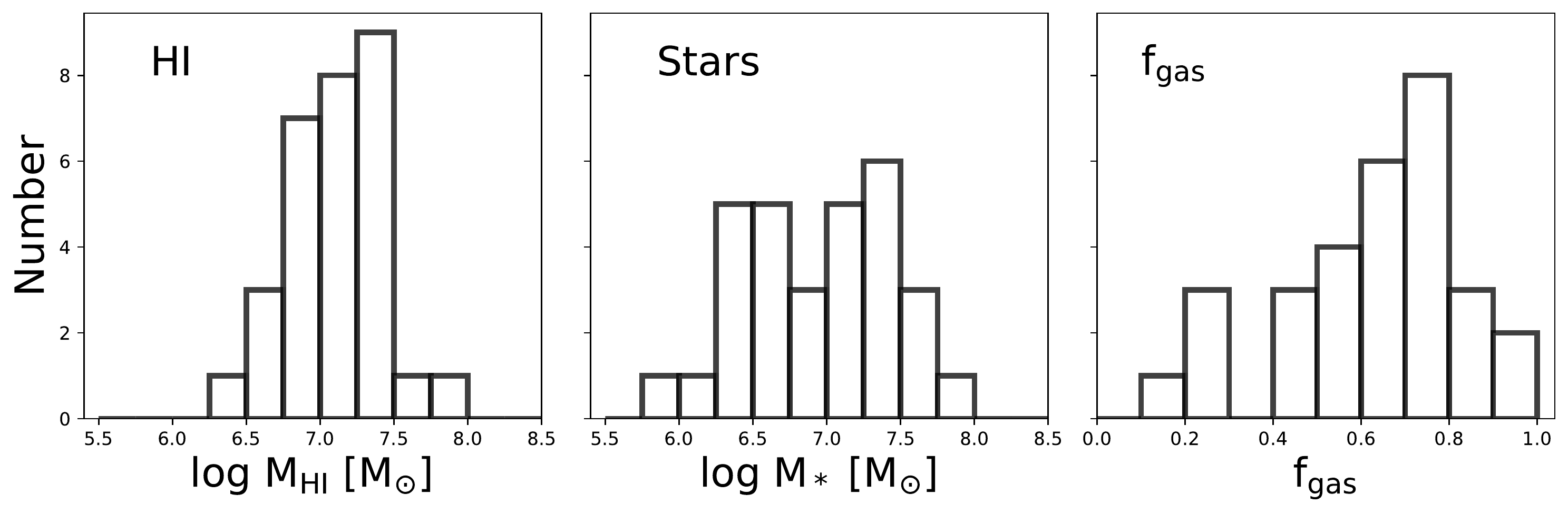}
\caption{The number of galaxies as a function of \hi\ mass, stellar mass, and gas fractions (f$_{\rm gas}$ $=$ M$_{\rm gas}$/M$_{\rm bary}$) which includes scaling the \hi\ by 1.33 to account for the mass of helium. The mean \hi\ mass of the SHIELD sample is $1.6\times10^7$ \msun\ with a standard deviation of $1.1\times10^7$. The stellar mass has a flatter distribution with a mean value of $1.6\times10^7$ \msun, and a standard deviation of $1.6\times10^7$. The gas fractions similarly span a wide range.}
\label{fig:histo_stars_gas}
\end{figure*}

\section{The SHIELD Neighborhoods}\label{sec:neighbors}
In this section, we describe our investigation of the environments around the SHIELD galaxies. We searched for the nearest neighbors of our sample using the CosmicFlows-3 database \citep{Tully2013} and the Updated Nearby Galaxy Catalog \citep{Karachentsev2013}. Wherever possible, we used robust distances determined, for example, from the TRGB or Cepheids. Some of the catalogued distances, however, are from less precise indicators. Regardless, the search for the nearest neighbors provides an overall picture of whether the SHIELD galaxies reside in highly dense versus sparse environments. 

No galaxies were identified within 1 Mpc of 5 SHIELD galaxies (AGC~102728, AGC~123352, AGC~174605, AGC~198507, and AGC~198691). Of these, AGC~102728, AGC~176405, and AGC~198691 are located in Void Numbers 25, 8, and 12 respectively, identified by \citet{Pustilnik2019}. Note, however, the in-depth study of AGC~198691 suggests that a galaxy nearby-on-the-sky and also located in Void Number 12, UGC~5186, may be much closer than 1 Mpc and interacting \citep{McQuinn2020}. We also note that AGC~748778, which is identified as having two neighbors within 1 Mpc \citep[i.e., UGC~0075 separated by 0.7 Mpc, and the newly discovered Pisces~A separated by 0.9 Mpc;][]{Tollerud2016} is also located within Void Number 25 \citep{Pustilnik2019}. Of the remaining 24 SHIELD galaxies, 6 galaxies have neighbors within 500 kpc as identified in the Cosmicflows-3 catalog, listed in Table~\ref{tab:neighbors}. 

From the study of the SHIELD~I galaxies, \citet[][see their Figure 6]{McQuinn2014} reported 6 systems that appear to be part of 2 galaxy groups that are aligned in a single structure extending $\sim4$ Mpc. The main galaxies in these two loose groups are also dwarf galaxies, namely NGC~784 and NGC~672. The NGC~784 group is located in Void Number 8 by \citet{Pustilnik2019}. Despite being in low-density environments, loose associations of dwarf galaxies are not uncommon \citep[e.g.,][]{Tully2006}.

One SHIELD galaxy is within 200 kpc of its nearest neighbors and warrants additional discussion. AGC~111164 is located 140 kpc from NGC~784, the main galaxy in one of the aforementioned groups. NGC~784 is a starburst dwarf galaxy with a stellar mass of $6\times10^8$ \msun\ \citep{McQuinn2010b}. At this 3D physical separation, the evolution of each of the systems in this dwarf galaxy pair may be impacted by the gravitational presence of the other. 

In summary, the SHIELD galaxies are located in sparsely populated environments. Seven (23\%) SHIELD galaxies reside in voids catalogued by \citet{Pustilnik2019}, five (17\%) galaxies have no clearly identified neighbors within 1 Mpc, six (20\%) reside in two loosely associated dwarf galaxy groups \citep{McQuinn2014}, and one system appears to be loosely paired with another dwarf galaxy. 

\begin{deluxetable}{cCCCC}
\tablecaption{Comparison of HI and Stellar Content \label{tab:gas_stars}}
\tablehead{
\CH{Galaxy} & \CH{M$_{\rm HI}$/M$_*$} & \CH{f$_{\rm gas}$} & \CH{b} & \CH{log(M$_{\rm bary}$/\msun)}
}
\colnumbers
\startdata
\hline \\
\multicolumn{5}{c}{SHIELD I Galaxies}\\
\hline \\
AGC~110482	& 0.76^{+0.44}_{-0.37}	& 0.50^{+0.15}_{-0.12} & 1.1 $^{+ 0.8 }_{-0.9 }$	& 7.70^{+0.13}_{-0.10} \\
AGC~111164	& 1.14^{+0.33}_{-0.38}	& 0.60^{+0.08}_{-0.09} & 0.5 $^{+ 0.3 }_{-0.3 }$	& 6.94^{+0.05}_{-0.06} \\
AGC~111946	& 2.23^{+1.03}_{-1.23}	& 0.75^{+0.10}_{-0.12} & 1.6 $^{+ 0.9 }_{-1.3 }$	& 7.41^{+0.05}_{-0.06} \\
AGC~111977	& 0.53^{+0.20}_{-0.22}	& 0.42^{+0.09}_{-0.10} & 0.7 $^{+ 0.4 }_{-0.4 }$	& 7.36^{+0.09}_{-0.10} \\
AGC~112521	& 2.40^{+1.02}_{-1.08}	& 0.76^{+0.11}_{-0.11} & 0.9 $^{+ 0.7 }_{-0.7 }$	& 7.09^{+0.05}_{-0.05} \\
AGC~174585	& 2.19^{+0.49}_{-0.82}	& 0.74^{+0.09}_{-0.11} & 2.1 $^{+ 0.8 }_{-1.3 }$	& 7.15^{+0.04}_{-0.05} \\
AGC~174605	& 1.49^{+2.15}_{-0.70}	& 0.66^{+0.33}_{-0.12} & 1.2 $^{+ 1.9 }_{-0.8 }$	& 7.57^{+0.21}_{-0.07} \\
AGC~182595	& 0.32^{+0.12}_{-0.18}	& 0.30^{+0.08}_{-0.12} & 0.7 $^{+ 0.4 }_{-0.5 }$	& 7.56^{+0.11}_{-0.17} \\
AGC~731457 & 0.28^{+0.03}_{-0.21}	& 0.27^{+0.02}_{-0.15} & 1.0 $^{+ 0.1 }_{-0.9 }$	& 7.95^{+0.02}_{-0.24} \\
AGC~748778	& 3.36^{+0.94}_{-1.68}	& 0.82^{+0.11}_{-0.13} & 2.2 $^{+ 1.0 }_{-1.8 }$	& 6.87^{+0.04}_{-0.05} \\
AGC~749237	& 2.27^{+2.05}_{-1.63}	& 0.75^{+0.17}_{-0.14} & 1.6 $^{+ 1.6 }_{-1.3 }$	& 8.01^{+0.10}_{-0.08} \\
AGC~749241 & 3.11^{+1.11}_{-0.95}	& 0.81^{+0.09}_{-0.08} & 1.3 $^{+ 0.8 }_{-0.8 }$	& 6.97^{+0.04}_{-0.03} \\
\hline \\
\multicolumn{5}{c}{SHIELD II Galaxies}\\
\hline \\
AGC~102728	& 2.95^{+1.12}_{-2.22}	& 0.80^{+0.16}_{-0.18} & 1.4 $^{+ 0.7 }_{-1.2 }$	& 7.27^{+0.06}_{-0.08} \\
AGC~123352	& 2.95^{+0.75}_{-1.57}	& 0.80^{+0.08}_{-0.11} & 1.5 $^{+ 0.5 }_{-1.1 }$	& 7.40^{+0.03}_{-0.05} \\
AGC~198507	& 10.35^{+2.89}_{-6.42}	& 0.93^{+0.23}_{-0.30} & 5.1 $^{+ 1.8 }_{-3.5 }$	& 7.43^{+0.07}_{-0.10} \\
AGC~198508	& 1.54^{+0.54}_{-0.79}	& 0.67^{+0.13}_{-0.15} & 2.1 $^{+ 0.8 }_{-1.5 }$	& 7.23^{+0.06}_{-0.08} \\
AGC~198691	& 24.93^{+10.47}_{-20.44}	& 0.97^{+0.40}_{-0.77} & 	\nodata		& 7.40^{+0.12}_{-0.24} \\
AGC~200232	& 0.51^{+0.22}_{-0.31}	& 0.40^{+0.11}_{-0.15} & 0.5 $^{+ 0.3 }_{-0.4 }$	& 7.88^{+0.11}_{-0.16} \\
AGC~205590	& 0.79^{+0.14}_{-0.66}	& 0.51^{+0.09}_{-0.22} & 1.3 $^{+ 0.1 }_{-1.4 }$	& 7.39^{+0.04}_{-0.18} \\
AGC~223231	& 2.47^{+0.32}_{-1.40}	& 0.77^{+0.07}_{-0.12} & 2.1 $^{+ 0.8 }_{-1.5 }$	& 7.44^{+0.02}_{-0.06} \\
AGC~223254 & 1.63^{+0.33}_{-0.57}	& 0.68^{+0.06}_{-0.08} & 1.7 $^{+ 0.5 }_{-1.0 }$	& 7.34^{+0.03}_{-0.05} \\
AGC~229053	& 1.14^{+0.55}_{-0.76}	& 0.60^{+0.12}_{-0.16} & 0.4 $^{+ 0.3 }_{-0.3 }$	& 7.79^{+0.09}_{-0.12} \\
AGC~229379	& 1.91^{+0.65}_{-0.90}	& 0.72^{+0.13}_{-0.15} & 0.8 $^{+ 0.4 }_{-0.7 }$	& 6.97^{+0.06}_{-0.07} \\
AGC~238890 & 0.18^{+0.04}_{-0.08}	& 0.19^{+0.04}_{-0.07} & 0.2 $^{+ 0.1 }_{-0.1 }$	& 7.21^{+0.06}_{-0.16} \\
AGC~731448	& 1.04^{+0.38}_{-0.64}	& 0.58^{+0.10}_{-0.16} & 0.9 $^{+ 0.4 }_{-0.7 }$	& 7.71^{+0.07}_{-0.11} \\
AGC~731921	& 0.87^{+0.13}_{-0.65}	& 0.54^{+0.05}_{-0.19} & 0.5 $^{+ 0.1 }_{-0.5 }$	& 7.99^{+0.03}_{-0.15} \\
AGC~739005 & 1.60^{+0.25}_{-0.98}	& 0.68^{+0.06}_{-0.14} & 1.2 $^{+ 0.5 }_{-1.0 }$ 	& 7.60^{+0.03}_{-0.09} \\
AGC~740112	& 0.28^{+0.07}_{-0.22}	& 0.27^{+0.06}_{-0.16} & 0.1 $^{+ 0.0 }_{-0.1 }$	& 7.73^{+0.07}_{-0.25} \\
AGC~742601	& 3.53^{+0.54}_{-1.79}	& 0.82^{+0.09}_{-0.12} & 1.2 $^{+ 0.4 }_{-1.0 }$	& 7.22^{+0.03}_{-0.05} \\
AGC~747826 & 0.60^{+0.16}_{-0.49}	& 0.44^{+0.07}_{-0.20} & 0.4 $^{+ 0.1 }_{-0.4 }$	& 7.70^{+0.06}_{-0.20} \\
\enddata 
\tablecomments{\hi\ to stellar mass ratios, gas fractions, birthrate parameter (i.e., b$=\langle$SFR$\rangle_{\rm 200~Myr}/\langle$SFR$\rangle_{life}$) and baryonic masses for the SHIELD I and II samples. Note that f$_{\rm gas}$ $=$ M$_{\rm gas}$/M$_{\rm bary}$, M$_{\rm bary} = $ M$_{\rm gas} +$ M$_*$, and M$_{\rm gas}$ = 1.33 $\times$ M$_{\rm HI}$.}
\end{deluxetable}

\section{Stellar and Gas Content}\label{sec:properties}
\subsection{Characterizing the SHIELD Sample}
The distributions of \hi\ masses, stellar masses, and gas fractions are  presented in Figure~\ref{fig:histo_stars_gas}. The SHIELD~I and II samples span $\sim1.5$ decades in \hi\ mass (6.25 $\ltsimeq$ log(M$_{\rm HI}$/\msun) $\ltsimeq$7.75) and more than $\sim2$ decades in stellar mass (5.5 $<$ log(M$_*$/\msun) $<$ 8). While the galaxies are extremely low-mass (log$\langle$ M$_{\rm HI}$/\msun $\rangle = 7.2$), the upper mass range extends to higher values than the original limit of 10$^{7.2}$ \msun\ used to define the SHIELD galaxies. The sample was selected from the ALFALFA catalog based on the \hi\ line flux and distance estimates. Since the measured TRGB distances are nearly all farther than the original estimates (see \S\ref{sec:trgb} and Figure~\ref{fig:flow_trgb}), the \hi\ masses of the galaxies are also larger. Despite this problem, all but two of current sample of 30 SHIELD galaxies have revised \hi\ masses less than $10^{7.5}$ \msun.

The resulting gas fractions, defined as f$_{\rm gas}$ $=$ M$_{\rm gas}$/M$_{\rm bary}$, where M$_{\rm bary} = $ M$_{\rm gas} +$ M$_*$ and M$_{\rm gas}$ = 1.33 $\times$ M$_{\rm HI}$ to account for the helium content in the interstellar medium, range from f$_{\rm gas} = 19$ to 97\%. Note that, despite some of the galaxies having lower gas-fractions, all of the galaxies are considered gas-rich. The gas fractions, as well as the \hi-to-stellar mass ratios (M$_{\rm HI}$/M$_*$), are listed in Table~\ref{tab:gas_stars}.

The stellar masses span a larger range than the \hi\ masses and have a flatter distribution, seen in the middle panel of Figure~\ref{fig:histo_stars_gas}, which may also be a result of our imposed \hi\ mass cutoff in our selection criteria. Unsurprisingly, given that SHIELD is an \hi\ selected survey and includes galaxies previously overlooked in optical surveys, the gas fractions are predominantly high.

The declining number of sources below log(M$_{\rm HI}$/\msun) $\sim$ 7 reflects, in part, the growing incompleteness of the ALFALFA catalog at these very low masses \citep{Haynes2011, Jones2018}. While the \hi\ mass function at lower masses and, in particular, the turnover, are not well quantified, larger numbers of low-mass galaxies than detected are expected. See \S\ref{sec:survey} for a calculation of the number density of SHIELD galaxies and implications for finding lower mass galaxies in future surveys.

\subsection{Star Formation and Gas Properties}
\begin{figure*}
\includegraphics[width=0.33\textwidth]{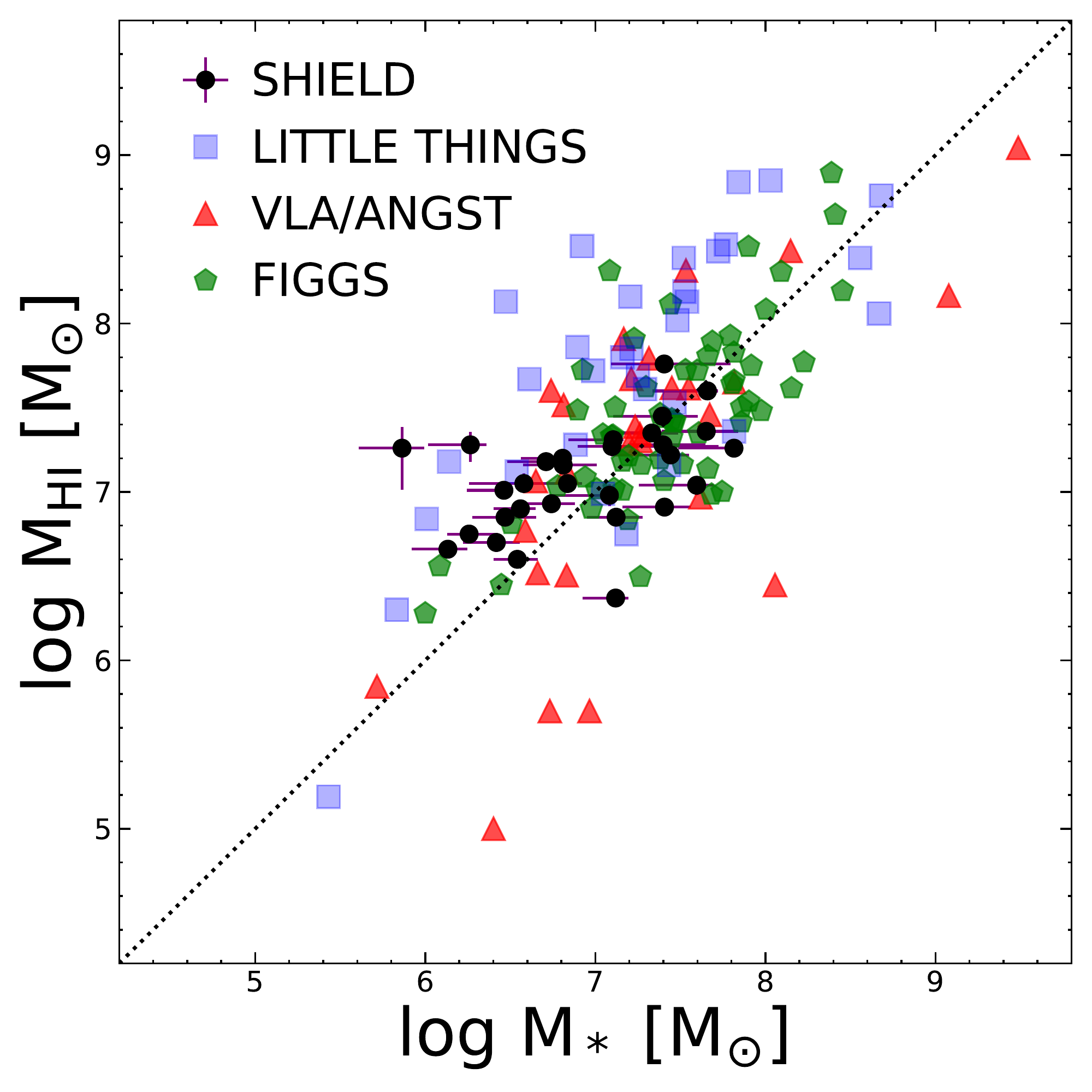} 
\includegraphics[width=0.33\textwidth]{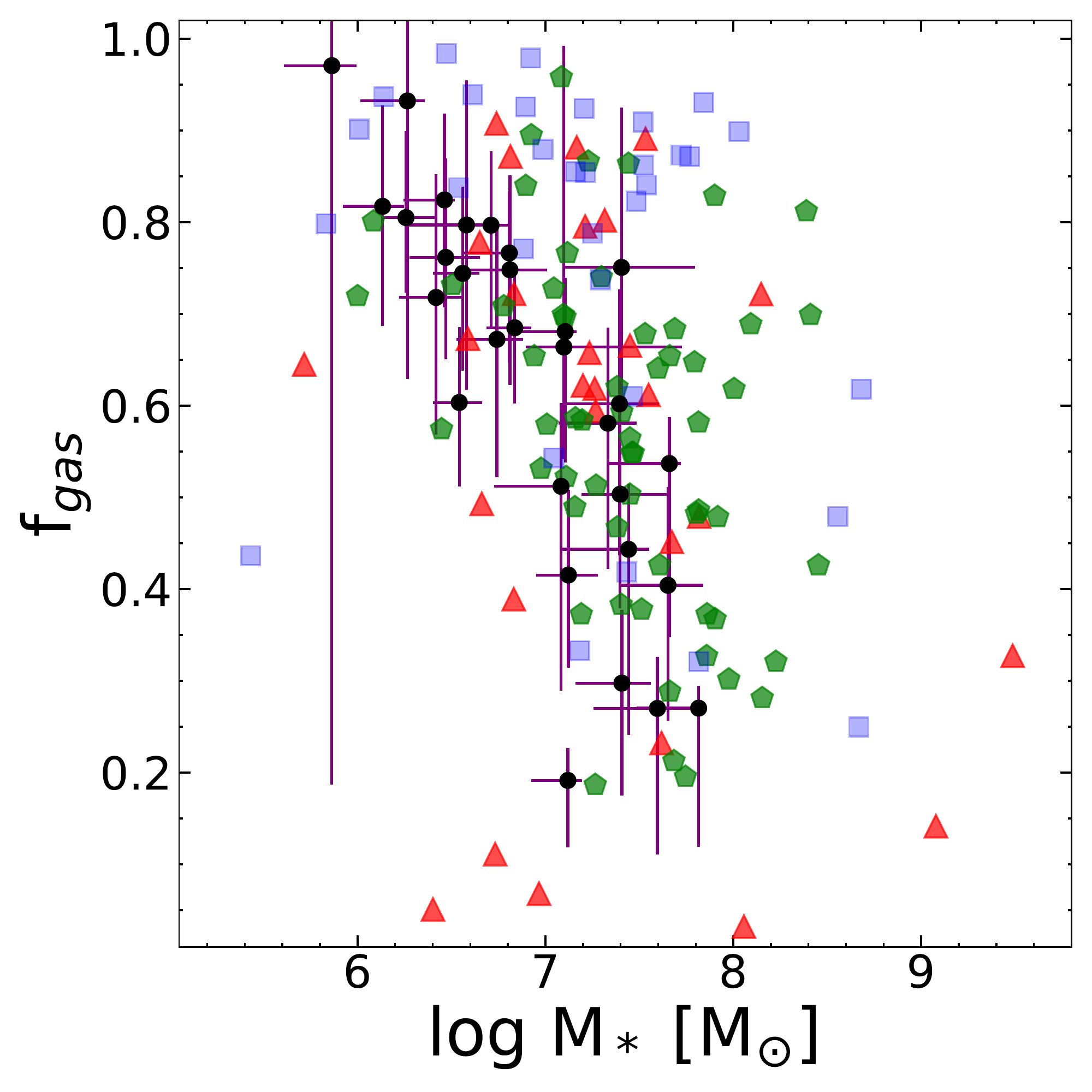} 
\includegraphics[width=0.33\textwidth]{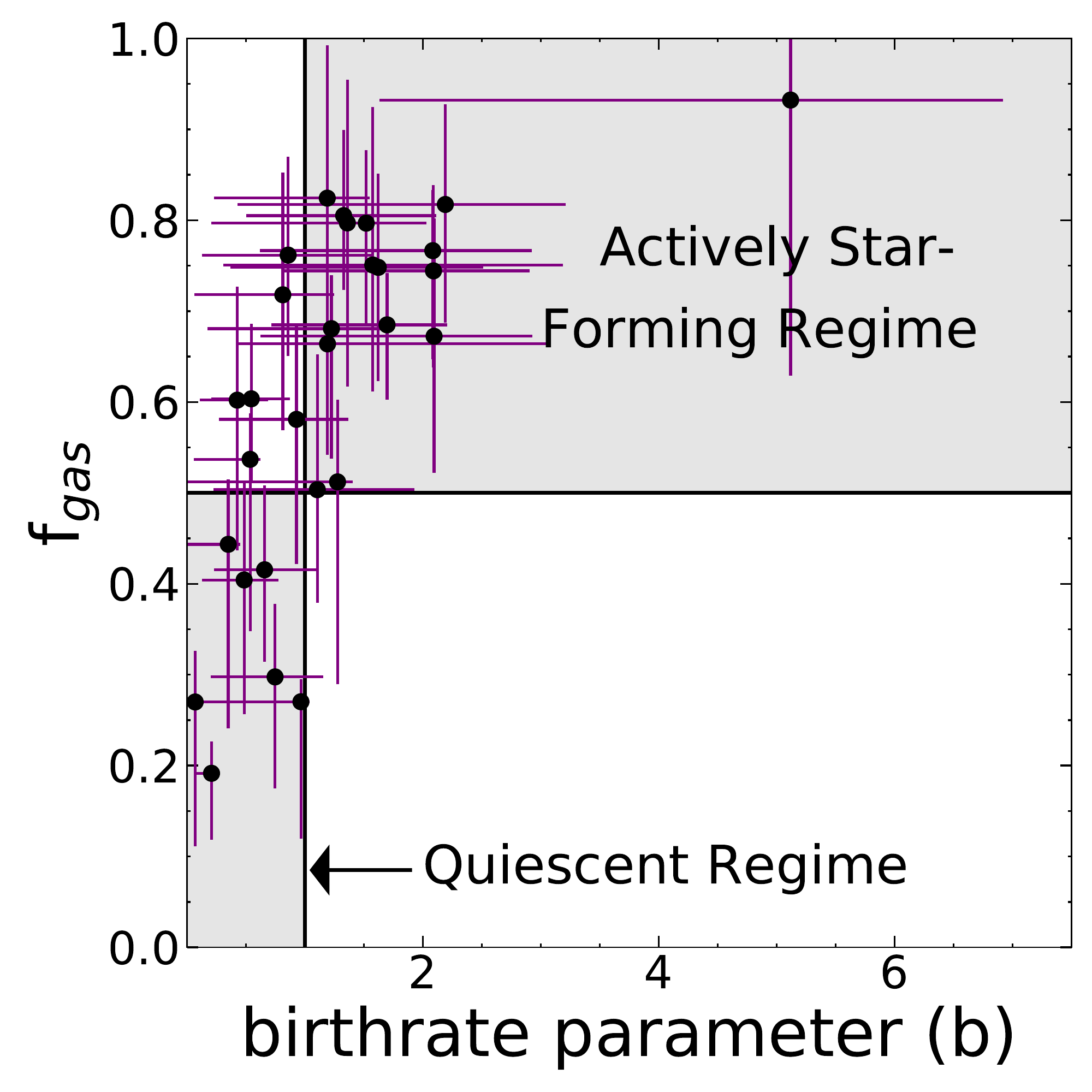} 
\caption{Left: The \hi\ masses as a function of stellar mass for the SHIELD~I and II galaxies (black points). Samples from the literature are shown including from LITTLE THINGS \citep[blue squares;][]{Hunter2012, Zhang2012}, VLA/ANGST \citep[red triangles;][]{Ott2012}, and FIGGS \citep[green pentagons;][]{Begum2008}. Middle: Gas fractions as a function of stellar mass. Right: Gas fractions compared with the birthrate parameter for the SHIELD~I and II samples. Galaxies with birthrate parameters b$\geq1$ are labeled actively star-forming in the upper right shaded region whereas systems with b$<1$ are labeled quiescent in the lower left shaded region.}
\label{fig:Mstar}
\end{figure*}

\begin{figure*}
\includegraphics[width=0.33\textwidth]{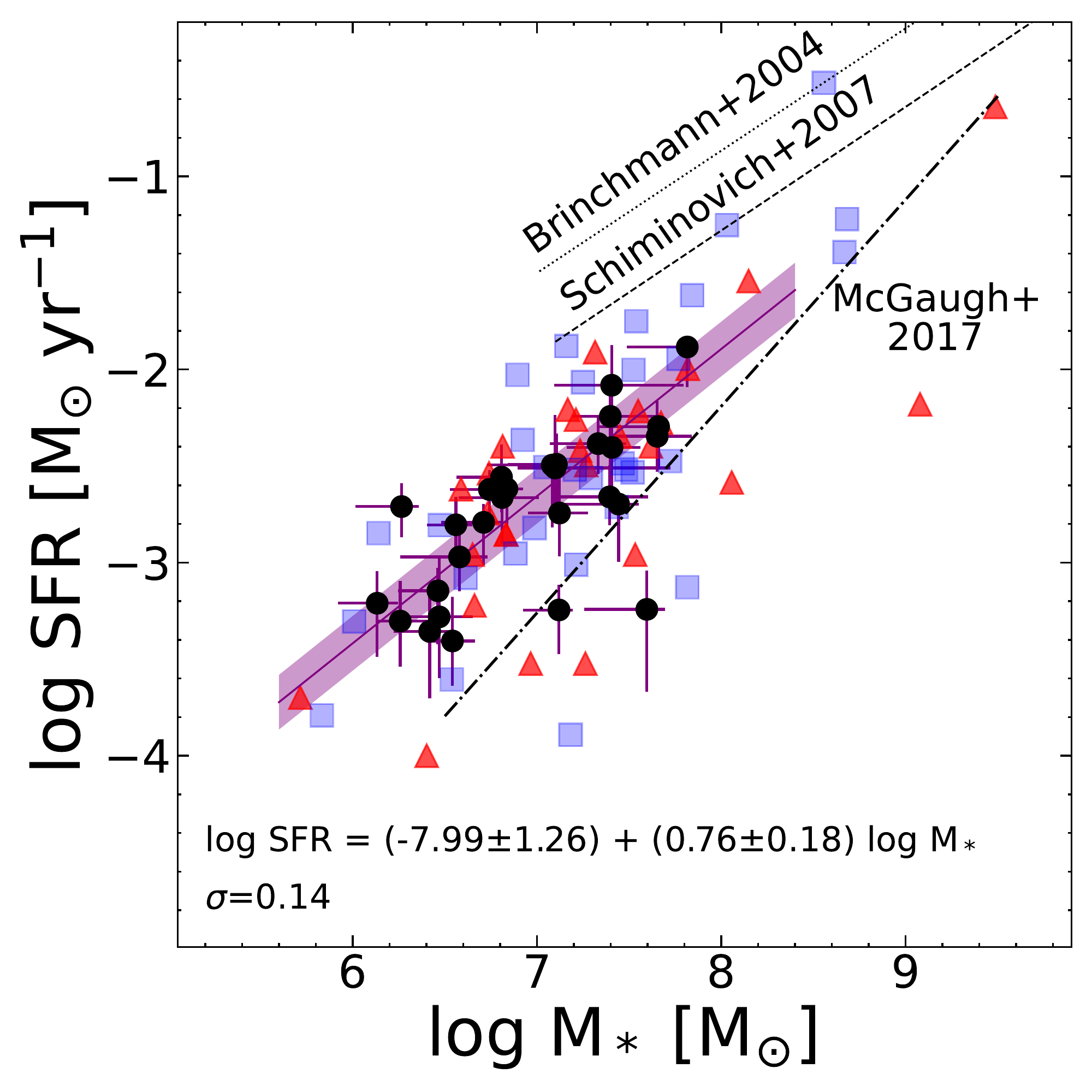} 
\includegraphics[width=0.33\textwidth]{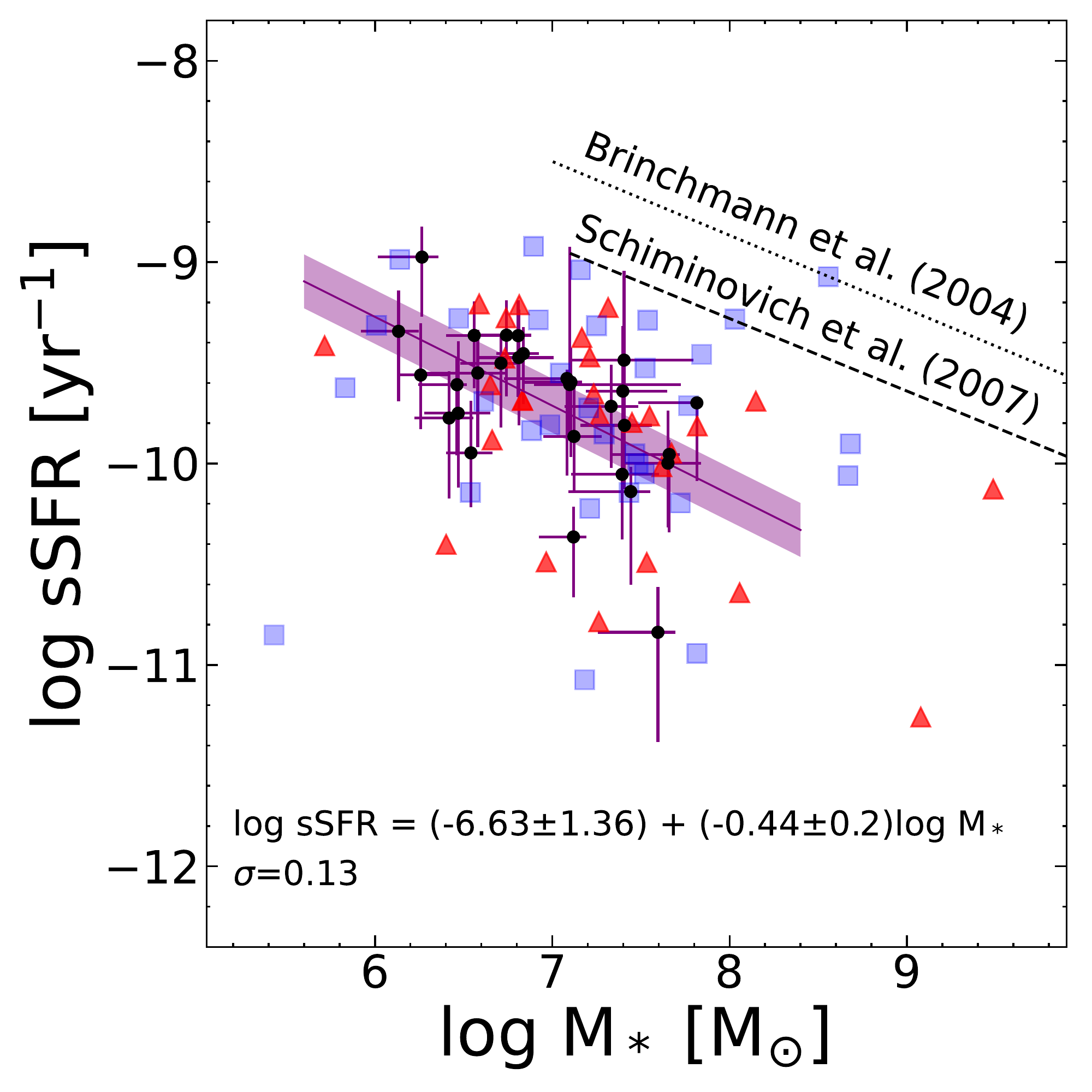} 
\includegraphics[width=0.33\textwidth]{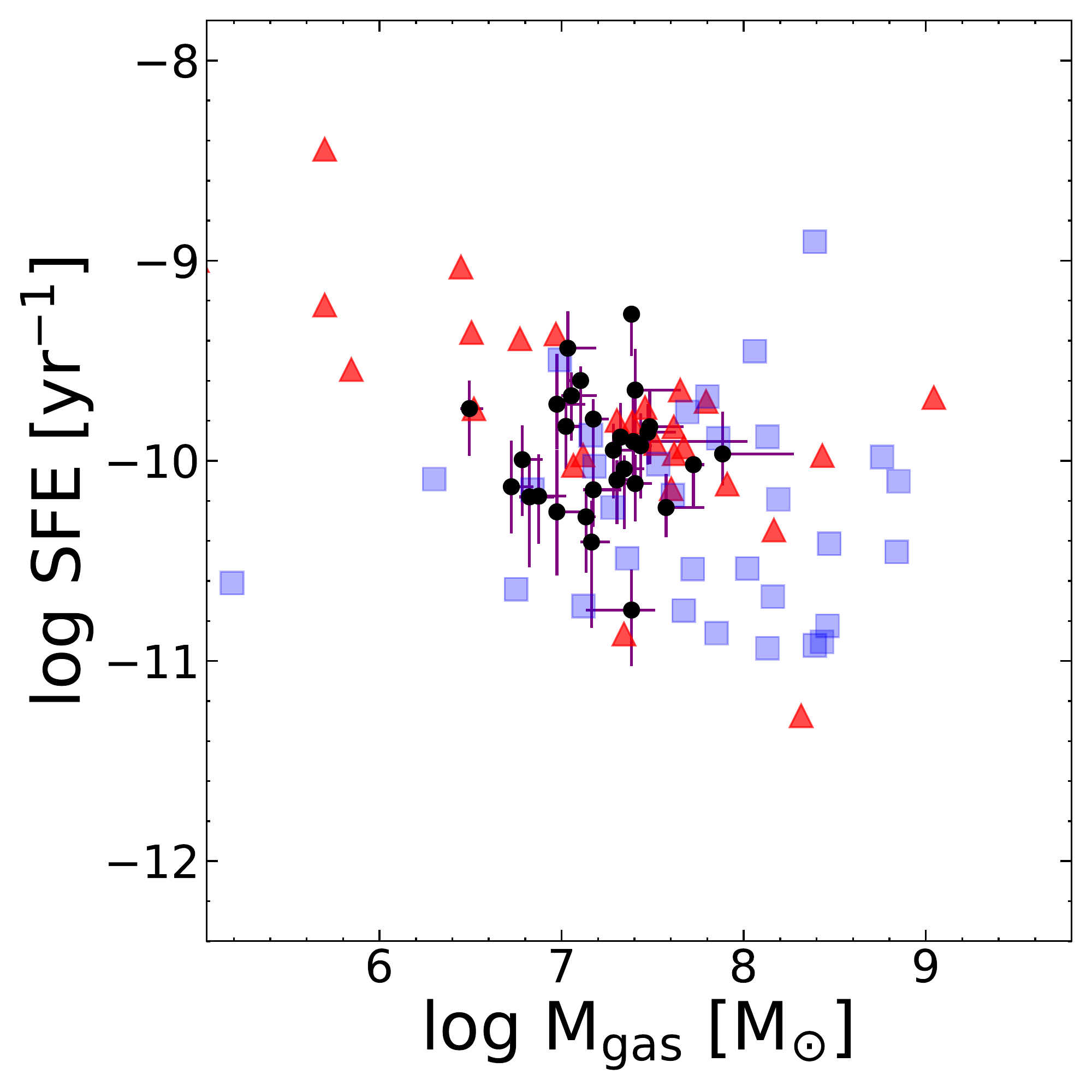} 
\caption{Left: Recent SFRs as a function of stellar mass. Best-fit line to the SHIELD data and the $1-\sigma$ dispersion are shown in the panel. The log SFR$-$log M$_*$ relations extrapolated from more massive galaxies \citep{Brinchmann2004, Schiminovich2007} and from LSB galaxies in the low-mass galaxy regime \citep{McGaugh2017} are also shown. Middle: recent SFRs normalized by stellar mass (sSFR) as a function of stellar mass. The same relations normalized by stellar mass are shown except for the LSB sample which has a flat slope. Right: SFE as a function of gas mass; the SFE axis limits are set to the same range as the sSFRs to allow a direct comparison between normalizations. Symbols are the same as in Figure~\ref{fig:Mstar}.}
\label{fig:sfr}
\end{figure*}

Here, we explore the correlations between the recent star formation properties, stellar content, and gas content for galaxies at the faint end of the \hi\ mass function. We use our measurements of the SHIELD galaxies as well as measurements of other low-mass galaxy samples from surveys introduced in \S\ref{sec:sample}, including the LITTLE THINGS survey \citep{Hunter2012}, the VLA/ANGST survey \citep{Ott2012}, and the FIGGS survey \citep{Begum2008}. Note that, as many of the measurements from the other surveys lack uncertainties, our focus is on the overall qualitative trends between the samples. For ease of comparison, the properties are shown in a series of plots in Figures~\ref{fig:Mstar}$-$\ref{fig:sfr}. 

The first panel in Figure~\ref{fig:Mstar} provides a comparison of the \hi\ and stellar masses probed by the four surveys. The majority (63\%) of the galaxies have M$_{\rm HI}$/M$_*$ ratios approximately equal to or greater than unity, which is expected from \hi\ selected surveys. VLA/ANGST, the only optically selected survey, is the exception and includes galaxies that have notably low \hi\ masses for a given stellar mass. As noted in \S\ref{sec:sample}, the LITTLE THINGS study includes 40 galaxies within the Local Volume (D$\ltsimeq 10$ Mpc), but we limit the comparison to the 32 galaxies for which stellar masses, determined from SED fits, and SFRs, based on the H$\alpha$ emission, are reported \citep{Zhang2012}. The VLA/ANGST sample includes systems within $\sim4$ Mpc. Stellar masses were estimated from CMDs \citep{Weisz2011}, SFRs were estimated using the far-ultraviolet scaling relation \citep{Kennicutt1998} and adopting robust TRGB distances \citep{Dalcanton2009}. The FIGGS sample shown includes 59 low-mass galaxies. Stellar masses were calculated using the B-band luminosity and assuming a mass-to-light ratio of unity; no SFRs were reported. 

The gas fractions as a function of M$_*$ are presented in the middle panel in Figure~\ref{fig:Mstar}. There is a general trend that gas fractions increase at lower stellar masses, but the spread in values is significant with a large range of gas fractions ($\sim0.4-0.9$) at nearly all stellar masses probed. The range in gas fractions at lower stellar mass could be even larger except for our observational bias.

Gas fractions are more directly correlated with higher recent SFRs relative to the lifetime averages. Shown in the final panel in Figure~\ref{fig:Mstar}, the birthrate parameter \citep[b$\equiv$ recent SFR / lifetime SFR;][]{Scalo1986, Kennicutt2005} for the SHIELD systems increases with increasing gas content. Specifically, galaxies with b parameters $\geq 1$ indicating constant or, in cases with b$\gtsimeq2$, a burst of star formation, have f$_{\rm gas} > $50\% \citep[upper right shaded region; see e.g.,][for defining bursting star formation in dwarf galaxies with the birthrate parameter]{McQuinn2009, McQuinn2010a}, whereas systems with lower gas fractions are better characterized by a declining recent SFR relative to their lifetime average (lower left shaded region). As the gas fraction generally increases as stellar mass decreases, the birthrate parameter also increases for lower stellar masses, consistent with galaxy downsizing  down to M$_*\sim10^6$ \msun. The birthrate parameters for the SHIELD galaxies are listed in Table~\ref{tab:gas_stars}.

The recent SFRs, as a function of stellar mass, are shown in the left panel of Figure~\ref{fig:sfr}. The SFRs for the low-mass SHIELD galaxies increase with increasing galaxy stellar masses. The SFR-M$_*$ correlation is similar to the well-established and relatively tight trend found for star-forming galaxies with masses between $\sim10^8\sim10^{11}$ \msun\ \citep{Brinchmann2004, Schiminovich2007} also shown in Figure~\ref{fig:sfr}. The main difference is a lower implied star formation efficiency than extrapolations from the more massive galaxies to the mass regime of the SHIELD galaxies would suggest. The best-fitting line to the SFR$-$M$_*$ relation for the SHIELD galaxies is log SFR $= (-7.99\pm1.26) + (0.76\pm0.18)  \cdot $log M$_*$, with an intrinsic dispersion in SFR of $\sigma=0.14$. The VLA/ANGST sample lies in the same parameter space as the SHIELD galaxies; the LITTLE THINGS sample, which includes starbursting dwarfs \citep{McQuinn2010b}, show a larger range in SFRs for a given M$_*$. Note that, while the lower mass SHIELD galaxies have increasingly lower SFRs, these systems are gas-rich and have generally increasing gas fractions with sufficient material to fuel star formation. None of the galaxies have quenched star formation.

The specific SFRs (sSFR $=$ SFR / M$_*$) as a function of M$_*$ are shown in the middle panel of Figure~\ref{fig:sfr}. Normalizing the SFR by stellar mass quantifies the intensity of star formation and helps to isolate other factors that may help regulate star formation \citep{Brinchmann2004}, including the energy input from supernovae. The best-fitting line to the SHIELD galaxies is log sSFR $= (-6.63\pm1.36) + (-0.44\pm0.20)  \cdot$ log M$_*$, with an intrinsic dispersion in sSFR of $\sigma=0.13$. As with the log SFR$-$ log M$_*$ relation, the log sSFR$-$ log M$_*$ trend found for the SHIELD galaxies has a slope similar to the value of $-0.36$ found for SDSS galaxies in the mass range $10^8 - 10^{11}$ \msun, but with a lower sSFR intercept \citep{Brinchmann2004, Schiminovich2007}. The trend that sSFR increases with decreasing M$_*$ is broader when considering the LITTLE THINGS sample. The larger spread in sSFR, as mentioned above, can be at least partially attributed to differences in sample as LITTLE THINGS includes starbursting dwarfs with intrinsically higher star formation activity for a given stellar mass. In addition, SFRs for LITTLE THINGS were derived using H$\alpha$ emission which has been previously attributed to increasing the scatter in the SFR$-$M$_*$ relation at low galaxy masses \citep{Bothwell2009}. H$\alpha$ emission can under-represent the SFR in systems with star formation activity below $\sim10^{-3}$ \msun\ yr$^{-1}$ where the upper-end of the IMF may not be fully populated \citep{Lee2009a}. 

Note that the SFR$-$M$_*$ relation for the SHIELD galaxies has a shallower slope than what has been reported for a sample of 56 low surface brightness (LSB) galaxies, which has a slope consistent with unity \citep{McGaugh2017}. For the LSB sample, the SFRs were based on H$\alpha$ fluxes and the stellar masses were based on optical mass-to-light ratios yielding the relation log SFR $= (-10.75\pm0.53) + (1.04\pm0.06)  \cdot $log M$_*$, with an intrinsic scatter of $\sigma=0.34$. These authors suggest that the steeper slope for the LSB galaxies relative to more massive galaxies is another manifestation of galaxy downsizing. Interestingly, if we consider only the SHIELD galaxies with a birthrate parameter less than 1, we find log SFR $= (-9.85\pm2.30) + (0.99\pm0.32)  \cdot $log M$_*$, with an intrinsic scatter of $\sigma=0.21$, in excellent agreement with the LSB relation. Thus, while the full SHIELD sample is consistent with a shallower slope more typical of spiral galaxies, a subset with lower recent star formation is consistent with the star-formation properties of LSB galaxies in this mass regime. 

The stellar mass is likely not the dominant factor in regulating star formation in low-mass galaxies. Instead, it is likely that the star formation process has a greater dependency upon both the overall gas mass and gas fraction. To investigate this further, we normalize the SFRs by {\it gas masses}, which is often referred to as the star formation efficiency (SFE) as it calculates the inverse timescale of converting gas to stars. 

The SFE as a function of gas mass is shown in the right panel of Figure~\ref{fig:sfr}. To allow for a direct comparison with the normalization by stellar mass, the axis ranges for the SFE are the same as those used for the sSFR in the middle panel of Figure~\ref{fig:sfr}. The SFEs are increasing for galaxies with lower gas masses, but with a large spread of nearly 2 orders of magnitude at M$_*\sim10^7$ \msun. The gas depletion timescales, calculated by taking the inverse of the SFEs of the galaxies, range from 2$-$56 Gyr, with a mean approximately equal to a Hubble time (12 Gyr). Thus, the SHIELD galaxies have enough gas to continue fueling star formation over long timescales, in contrast to spiral galaxies with short gas depletion timescales that imply gas accretion is required to maintain star formation activity over comparable timescales. 

In summary, the stellar content, gas content, and star forming properties of the gas-rich, low mass galaxies of the SHIELD, VLA/ANGST, LITTLE THINGS, and FIGGS samples, are qualitatively similar to one another. Overall observed trends show higher gas fractions and higher birthrate parameters at lower masses, but with significant scatter. Based on the quantitative fits to the SHIELD galaxies, galaxies in the mass range $\sim10^6- 10^7$ \msun\ are consistent with extrapolations from more massive star-forming galaxies with a lower SFE normalization. Low SFEs are expected at lower surface mass densities by nearly all models of star formation \citep[e.g.,][and reference therein]{Krumholz2012}. SHIELD galaxies with low birthrate parameters follow the SFR $-$ M$_*$ trend identified for LSB galaxies. 

\section{Summary}\label{sec:conclusions}
The SHIELD program includes a complete sample of low-mass, gas-rich galaxies from the cosmological volume in the local universe observed in the ALFALFA survey. The SHIELD galaxies populate the under-explored regime of very low-mass galaxies, offering the opportunity to bridge our knowledge of classical gas-rich dwarfs (i.e., M$_* \sim 10^8$ \msun) to the intrinsically faint and low-surface brightness frontier of galaxies discoverable in, for example, the Rubin Observatory Large Survey of Space and Time (LSST) and SKA eras. 

Using newly obtained HST optical imaging of the resolved stars and WSRT observations of the neutral hydrogen, we measure the TRGB distances, star formation activity, and gas properties for 18 of the 82 SHIELD galaxies. Combined with existing similar measurements of an additional 12 SHIELD systems, we begin to quantify the properties at the faint end of the luminosity function with statistical confidence. 

We introduce a new technique for measuring the rotational velocity and spatial extent of the \hi\ gas in low mass galaxies when the \hi\ has a limited extent and the velocity field is not suitable for dynamical modelling (\S\ref{sec:kinematics} and Appendix~\ref{app:pvvel}). Applying this technique to 30 SHIELD galaxies, we report on their gas kinematics (Table~\ref{tab:gas}); future work will include a comparison of these results with other measures of the \hi\ rotation in low mass galaxies (J. Fuson \etal\ in preparation). 

The TRGB distances place the galaxies in the Local Volume, but at distances farther than estimates from parametric flow models (\S\ref{sec:trgb}). The majority of the SHIELD galaxies are located in under-dense environments (\S\ref{sec:neighbors}), with several residing in voids \citep{Pustilnik2019}. 

From measurements of the resolved stars and \hi\ data, we find the log of the \hi\ and stellar masses, in units of \msun, range from 6.25 $-$ 7.75 and 5.75 $-$ 8.0, respectively (Figure~\ref{fig:histo_stars_gas}). The galaxies are predominantly gas-rich with $\langle$M$_{\rm HI}$/M$_* \rangle = 1.9$, excluding the extreme system AGC~198691, which has a M$_{\rm HI}$/M$_*$ value of 25. Recent SFRs (t$\sim200$ Myr) range from 4$\times10^{-4}$ to $8\times10^{-3}$ \msun\ yr$^{-1}$ (\S\ref{sec:properties}). 

Overall, the properties of the SHIELD~I and II galaxies, as well as the properties of galaxies in the LITTLE THINGS, VLA/ANGST, and FIGGS surveys, appear to be a continuation from higher masses when considering their SFRs, \hi\ and stellar masses, gas fractions, and specific SFRs (Figures~\ref{fig:Mstar}$-$\ref{fig:sfr}), although with lower implied star formation efficiencies. When SHIELD galaxies with low birthrate parameters are considered separately, the SFR $-$ M$_*$ relation is steeper and consistent with the trend measured for LSB galaxies \citep{McGaugh2017}. 

Despite the low baryonic masses, the ongoing star formation in the SHIELD galaxies suggests sufficiently deep potential wells that both enable the galaxies to retain some gas and promote the concentration of gas needed for star formation. Simulations predict that galaxies in this mass regime may have a range of baryon-to-dark matter mass. If the baryon fraction (i.e., M$_{\rm baryon}$ / M$_{\rm halo}$) is lower than the typical fractions estimated in more massive galaxies, this would offer an explanation for the continuation of the scaling relations. In an upcoming paper, we explore the location of the SHIELD galaxies in the Baryonic Tully Fisher Relation and their implied dark matter content (K.~McQuinn et al.\ in preparation). 

The number density of these surveys suggest that current samples are significantly incomplete below M$_* \sim 10^7$ \msun; upcoming facilities such as the Rubin Observatory and the SKA have the potential for discovering larger samples below this mass threshold (\S\ref{sec:survey}). 

\facilities{Hubble Space Telescope, Westerbork Synthesis RadioTelescope}
\software{Astropy \citep{astropy2013, astropy2018}}

\section{Acknowledgments}
KBWM would like to thank Hunter Thu for a helpful exploration of the surface brightness properties of the galaxies. We would also like to thank the referee for helpful comments that improved the manuscript. Support for program HST-GO-13750 was provided by NASA through a grant from the Space Telescope Science Institute, which is operated by the Associations of Universities for Research in Astronomy, Incorporated, under NASA contract NAS~5-26555 and by a COX grant from the University of Texas at Austin. KBWM is supported by related grant HST-GO-15243. MPH is supported by grants NSF/AST-1714828 and from the Brinson Foundation. JMC is supported by NSF/AST-2009894 and by Macalester College. EAKA is supported by the WISE research programme, which is financed by the Netherlands Organisation for Scientific Research (NWO). KLR is supported by NSF grant AST-1615483. The Westerbork Synthesis Radio Telescope is operated by the ASTRON (Netherlands Institute for Radio Astronomy) with support from the Netherlands Foundation for Scientific Research (NWO). This research has made use of NASA Astrophysics Data System Bibliographic Services and the NASA/IPAC Extragalactic Database (NED), which is operated by the Jet Propulsion Laboratory, California Institute of Technology, under contract with the National Aeronautics and Space Administration.\\

\noindent This work is dedicated in loving memory to Anne B. Wingfield. 

\renewcommand\bibname{{References}}
\bibliographystyle{apj}
\bibliography{../../../bibliography.bib}

\appendix
We include three appendices. The first, Appendix~\ref{app:HST_atlas}, is an atlas of the 3-color HST images, CMDs from the resolved stellar populations, and a comparison of the \hi\ column densities with the stellar distributions. The second, Appendix~\ref{app:HI_atlas}, is an atlas of the \hi\ spectra, moment zero maps, moment one maps, and PV diagrams of the galaxies. The third, Appendix~\ref{app:pvvel}, includes a detailed description of our new methodology for measuring the rotational velocity and spatial extent of the \hi\ in data with limited spatial sampling, with applications to the SHIELD I and II galaxies. \\

\section{Atlas of HST Images and CMDs}\label{app:HST_atlas}
We present an atlas of the HST optical images, CMDs, and HST images with WSRT 21-cm contours overlaid for the remaining SHIELD~II galaxies. The figures follow the same format presented in Figure~\ref{fig:image1}; we refer the reader to the description presented in \S\ref{sec:stars} for details. 

\begin{figure*}[b!]
\begin{center}
\includegraphics[width=0.98\textwidth]{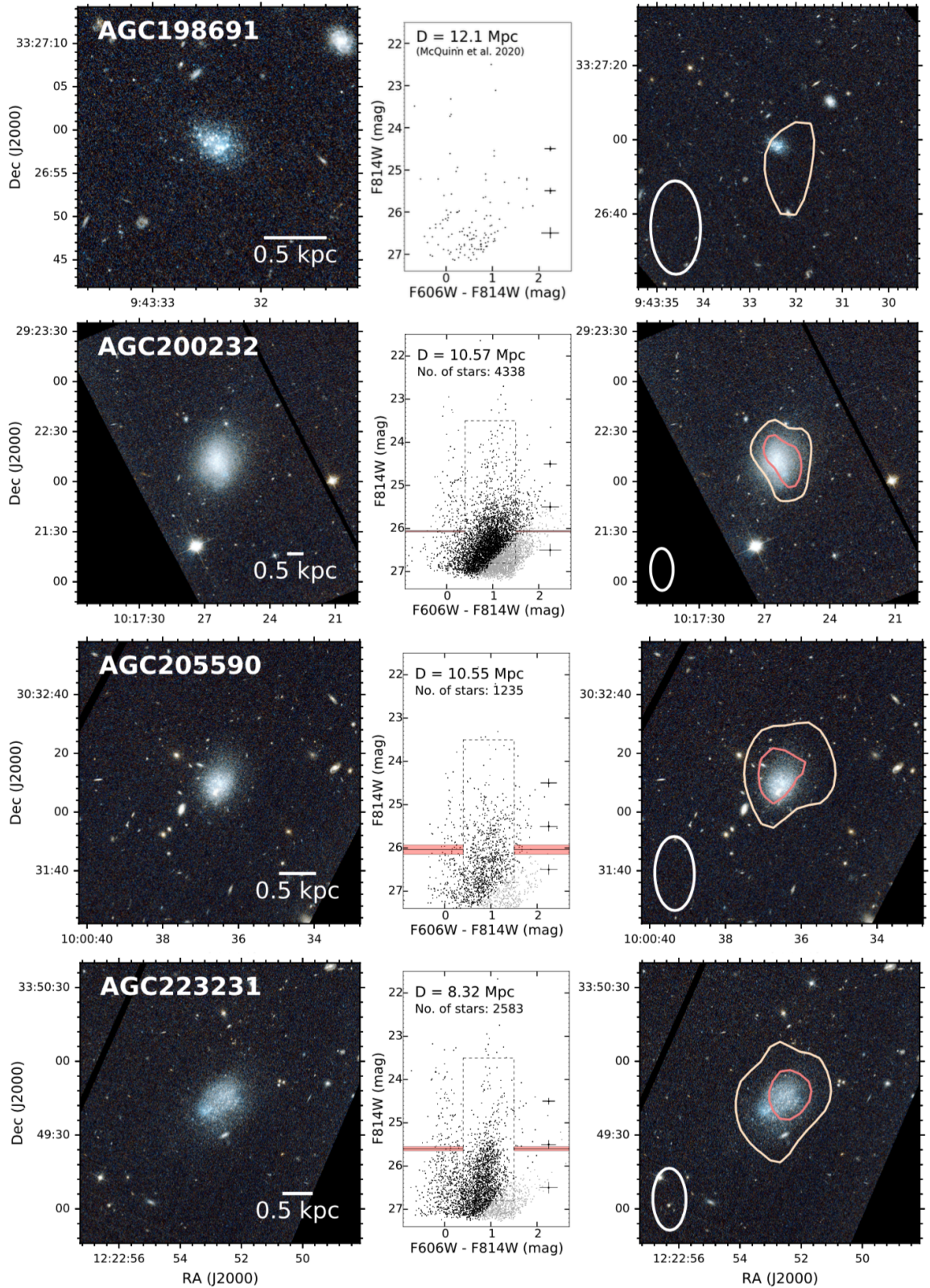}
\end{center}
\vspace{-0.25in}
\caption{HST optical images, CMDs, and WSRT \hi\ data for AGC~198691, AGC~200232, AGC~205590, AGC~223231. See Figure~\ref{fig:image1} caption for details.} 
\label{fig:image2}
\end{figure*}

\begin{figure*}
\begin{center}
\includegraphics[width=0.98\textwidth]{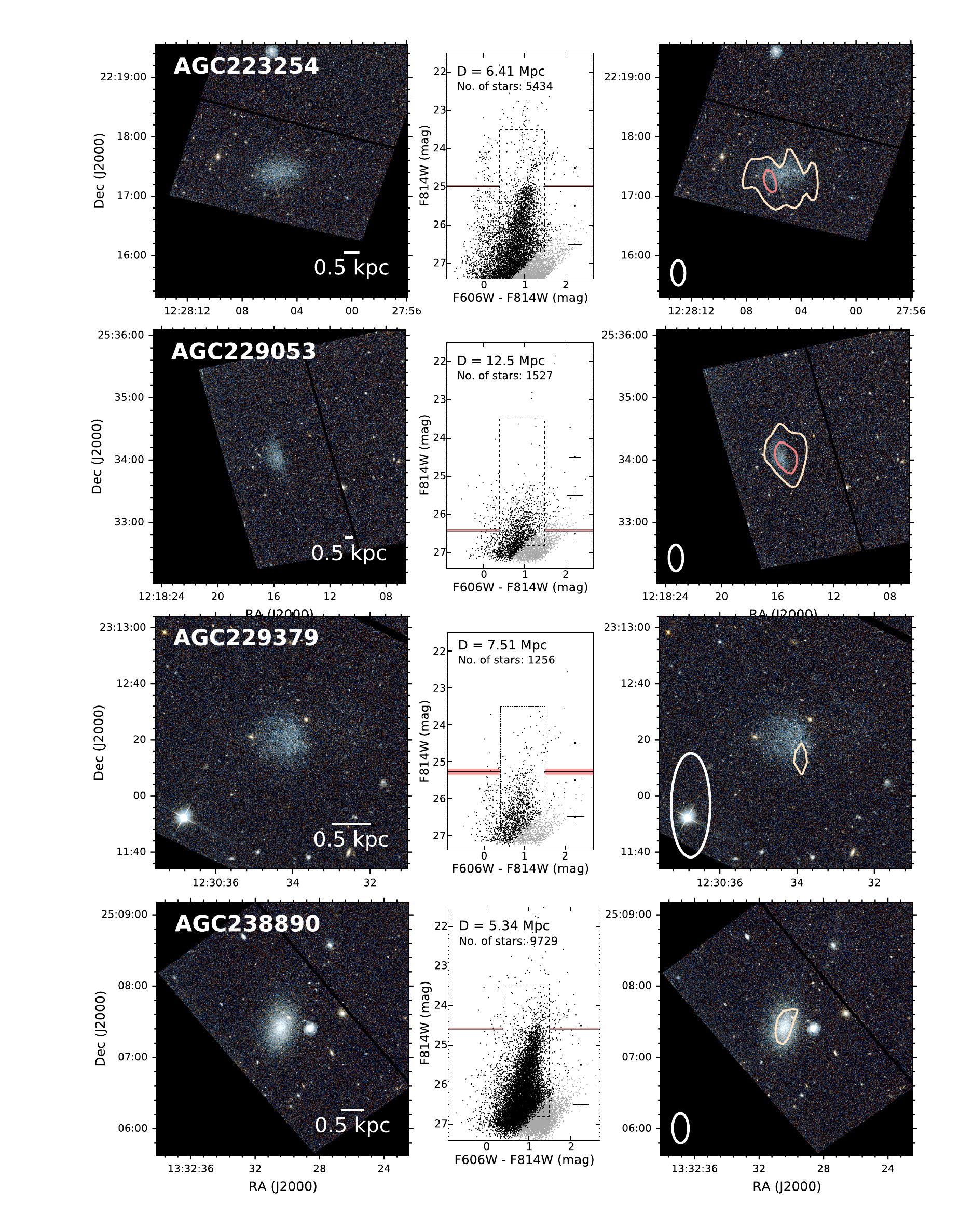}
\end{center}
\caption{HST optical images, CMDs, and WSRT \hi\ data for  AGC~223254, AGC~229053, AGC~229379, AGC~238890. See Figure~\ref{fig:image1} caption for details.}
\label{fig:image3}
\end{figure*}

\begin{figure*}
\begin{center}
\includegraphics[width=0.98\textwidth]{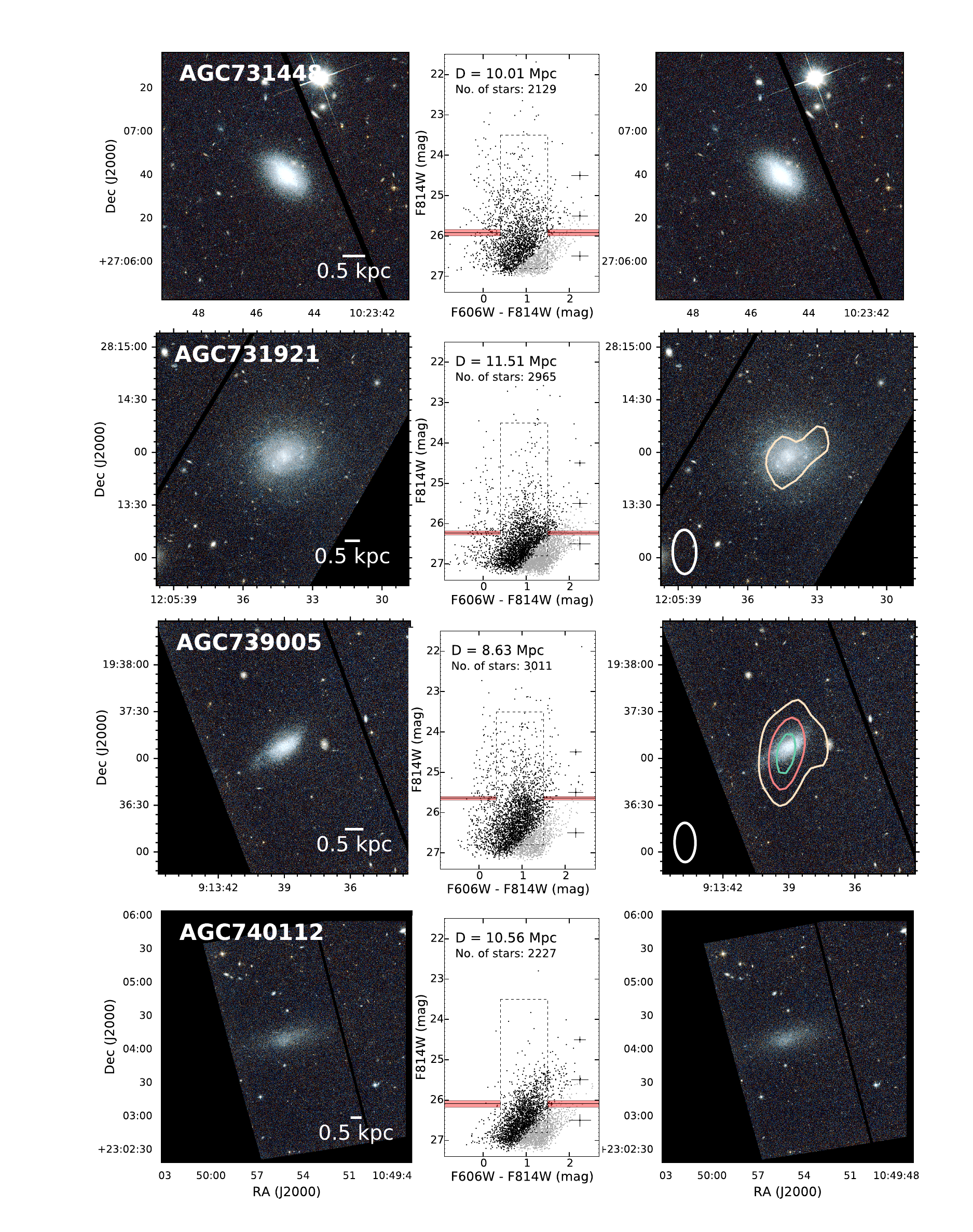}
\end{center}
\vspace{-0.25in}
\caption{HST optical images, CMDs, and WSRT \hi\ data for AGC~731448, AGC~731921, AGC~739005, AGC~740112. See Figure~\ref{fig:image1} caption for details. WSRT \hi\ data are not available for AGC~731448 and AGC~740112; see \S\ref{sec:gas} for details.}
\label{fig:image4}
\end{figure*}

\begin{figure*}
\begin{center}
\includegraphics[width=0.85\textwidth]{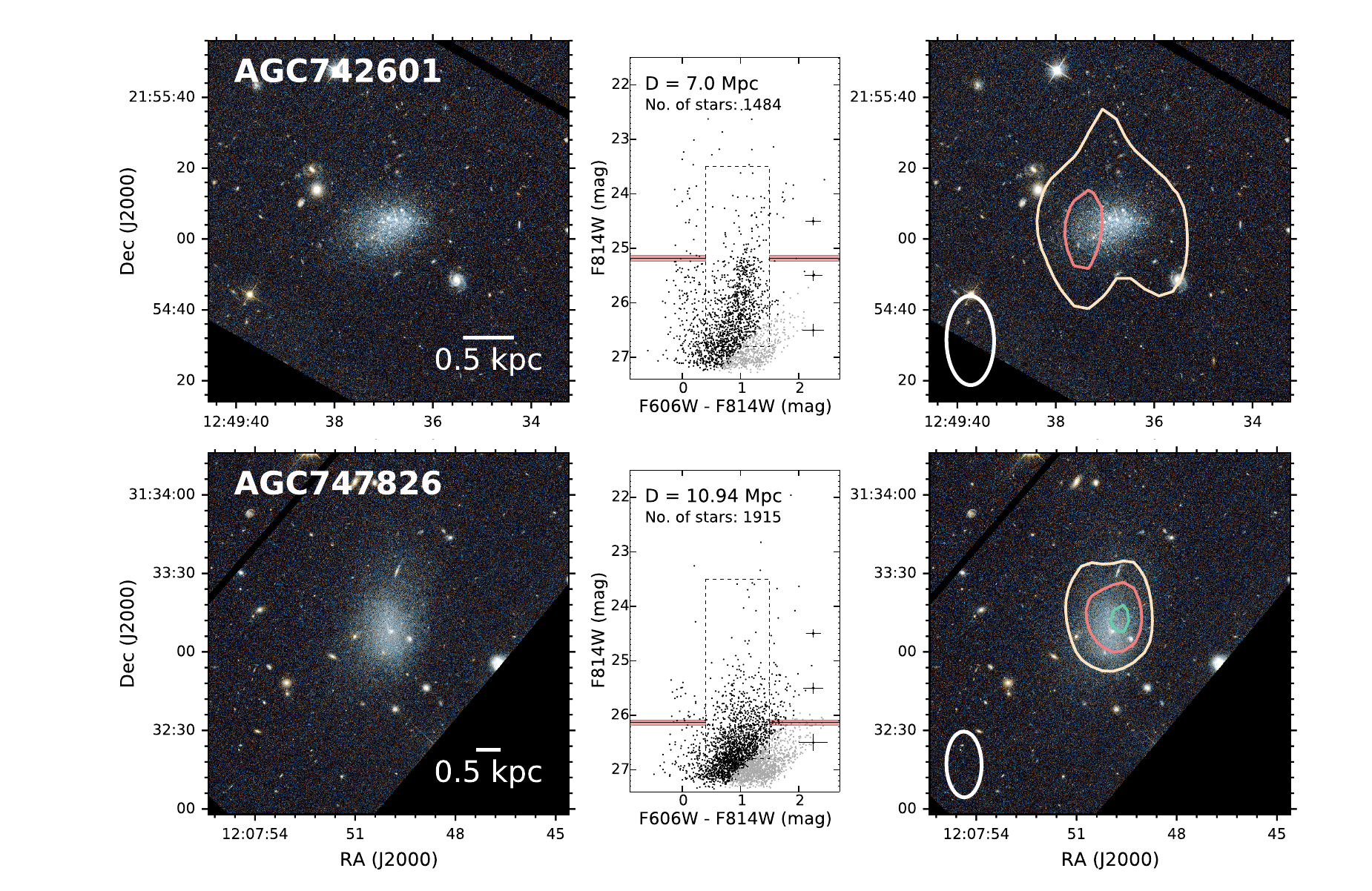}
\end{center}
\vspace{-0.15in}
\caption{HST optical images, CMDs, and WSRT \hi\ data for AGC~742601, AGC~747826. See Figure~\ref{fig:image1} caption for details.}
\label{fig:image5}
\end{figure*}

\clearpage
\section{\hi\ Atlas}\label{app:HI_atlas}
We present an atlas of the WSRT \hi\ observations including the 21-cm spectra, \hi\ column density maps (moment zero maps), \hi\ velocity fields (moment one maps), and PV diagrams for the remaining SHIELD~II galaxies with WSRT detections. The figures follow the same format presented in Figure~\ref{fig:hi}; we refer the reader to the description presented in \S\ref{sec:gas} for details. 

\begin{figure*}[h!]
\begin{centering}
\includegraphics[width=0.98\textwidth]{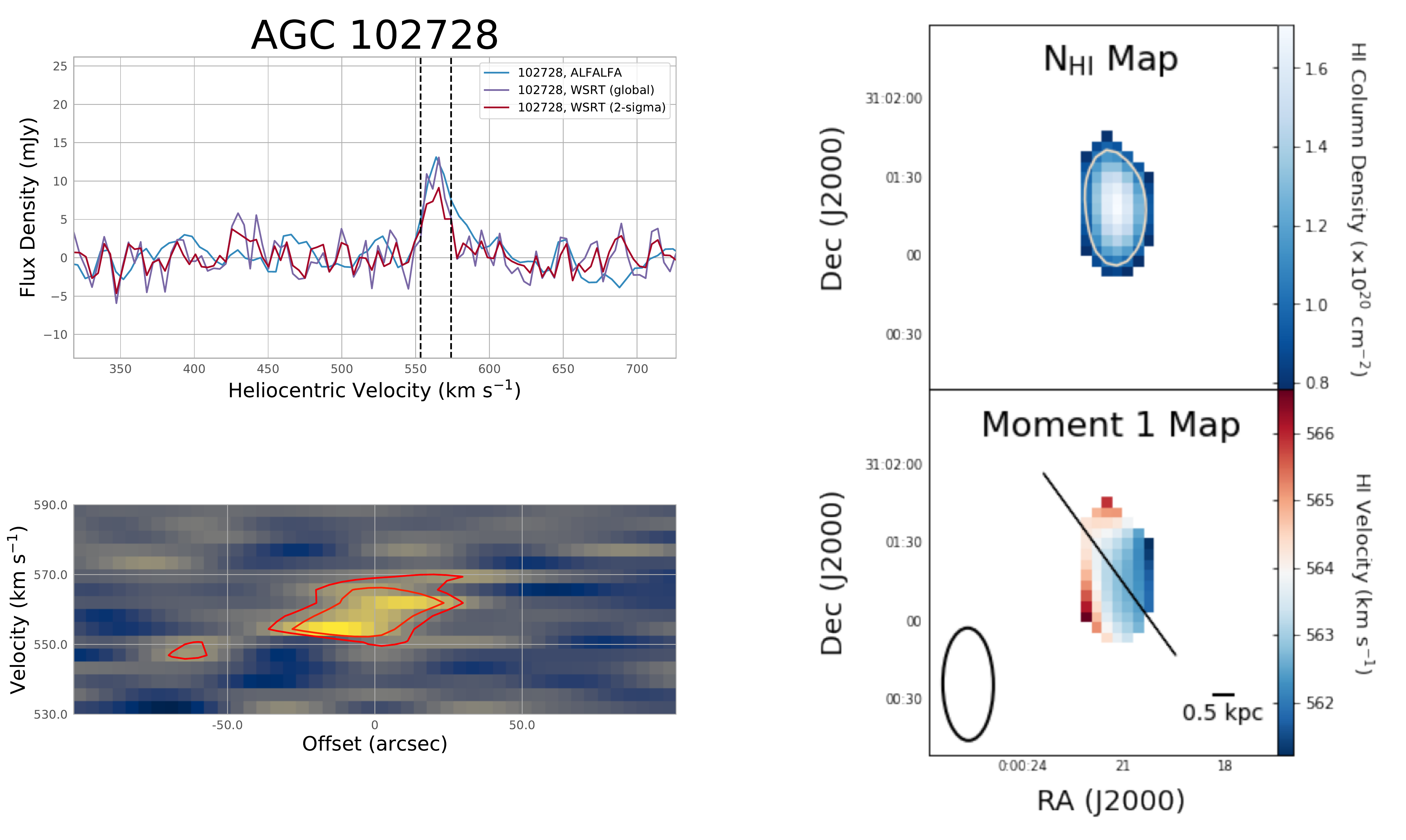}
\end{centering}
\caption{AGC\,102728.  There is a
  modest projected velocity gradient from southwest to northwest with a magnitude of 7 km\,s$^{-1}$.  
  The source is only marginally resolved by the HI beam.}
\label{fig:HI_AGC102728}
\end{figure*}
	
\begin{figure*}
\centering
\includegraphics[width=0.9\textwidth]{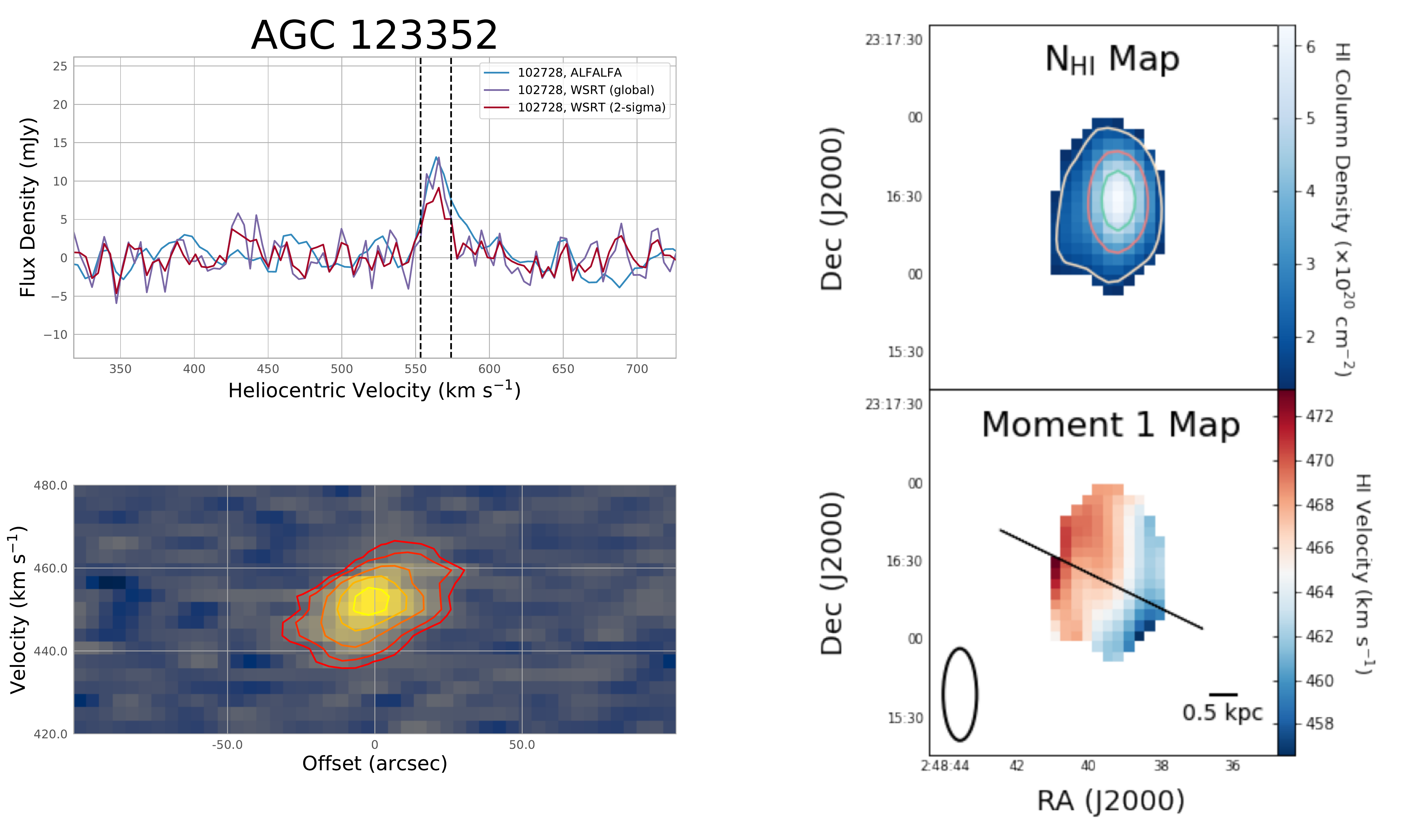}
\caption{AGC\,123352.  There is a
  clear projected velocity gradient from southwest to northwest with a
  magnitude of 13 km\,s$^{-1}$.}
\label{fig:HI_AGC123352}
\end{figure*}

\begin{figure*}
\centering
\includegraphics[width=0.9\textwidth]{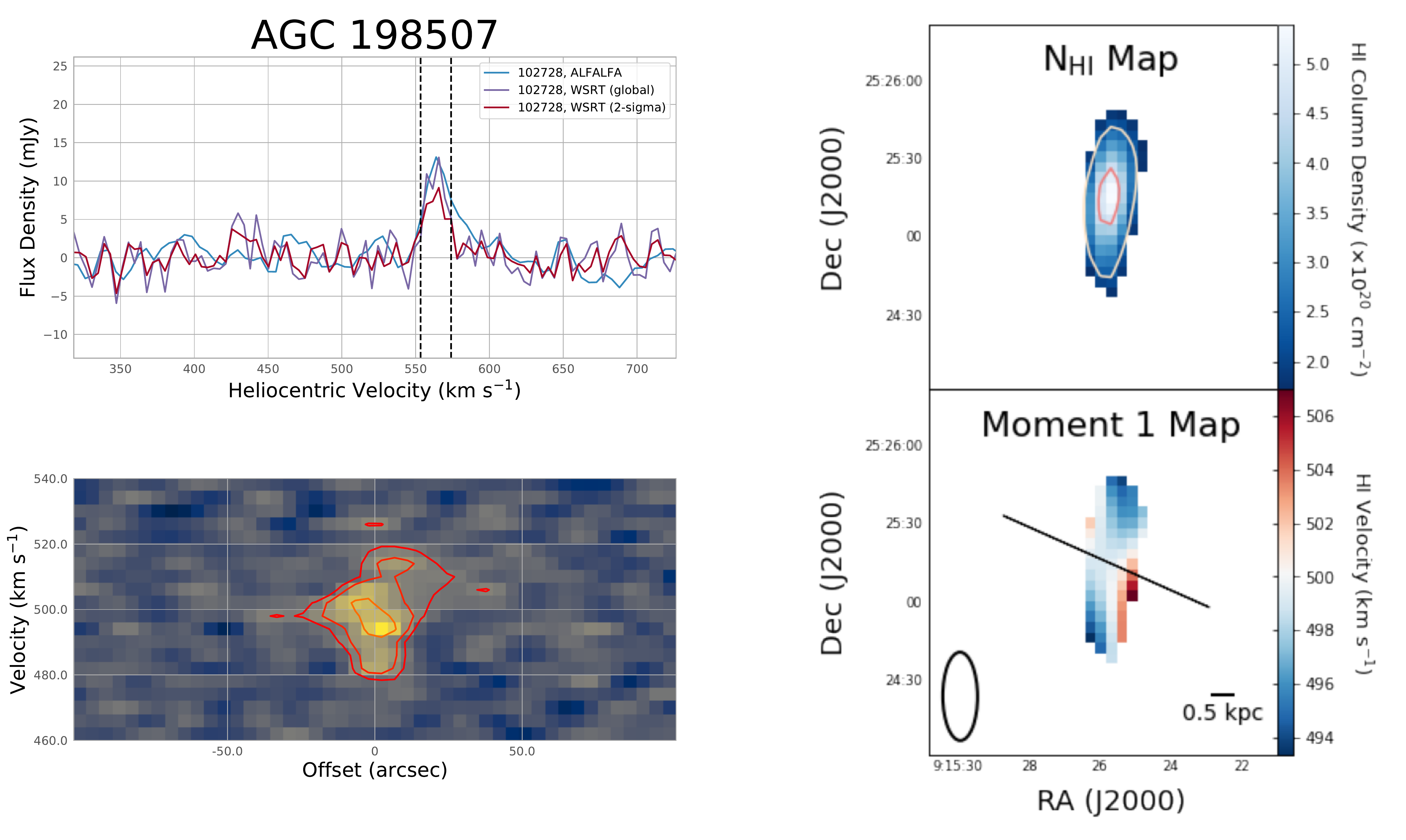}
\caption{AGC\,198507.  The HI
  morphology of this source is curious, with an extension of low
  surface brightness gas to the east and west.  There is weak evidence for
  a projected velocity gradient.}
\label{fig:HI_AGC198507}
\end{figure*}

\begin{figure*}
\centering
\includegraphics[width=0.9\textwidth]{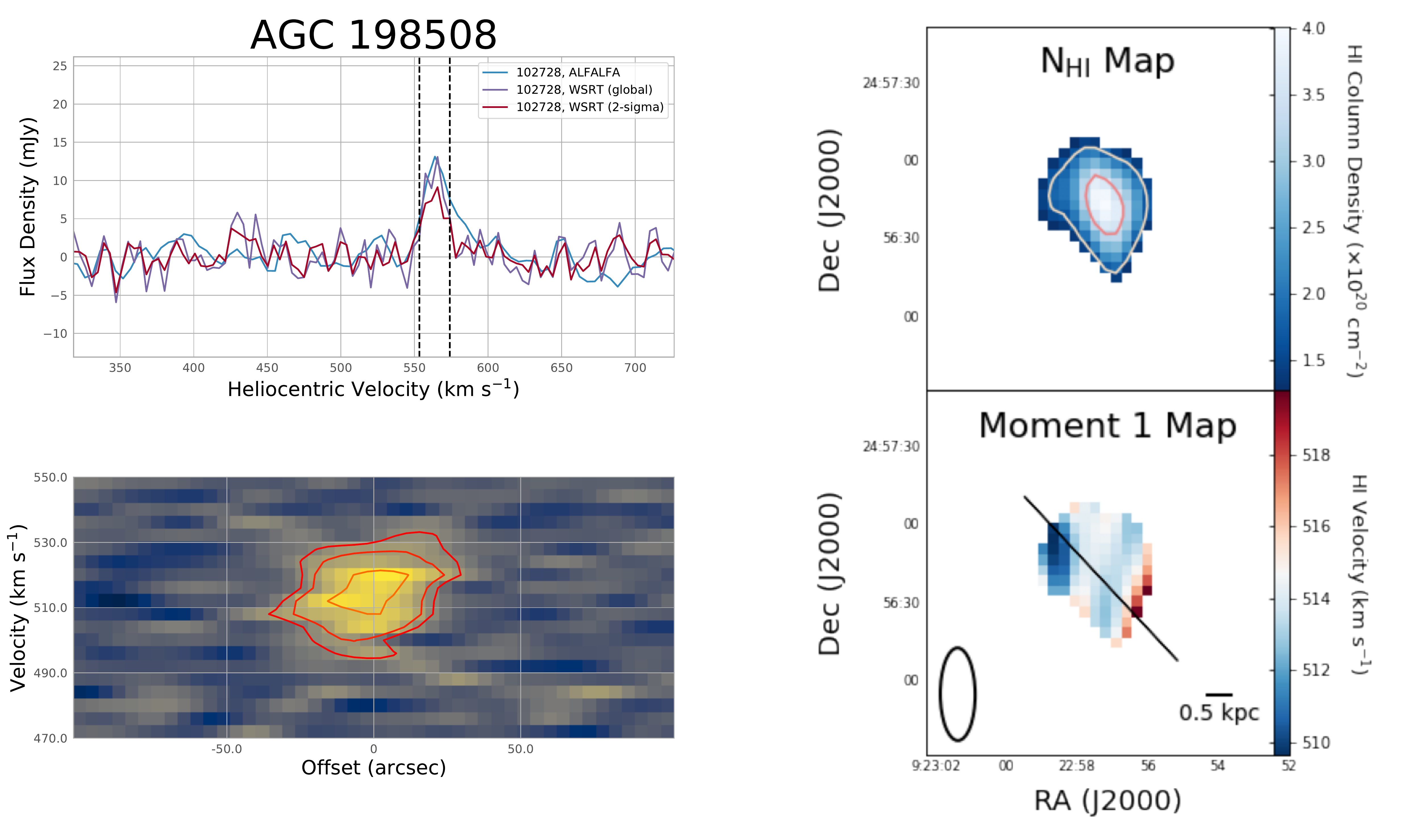}
\vspace{-8pt}
\caption{AGC\,198508.  There is a projected velocity gradient of 10 \kms.}
\label{fig:HI_AGC198508}
\end{figure*}

\begin{figure*}
\centering
\includegraphics[width=0.9\textwidth]{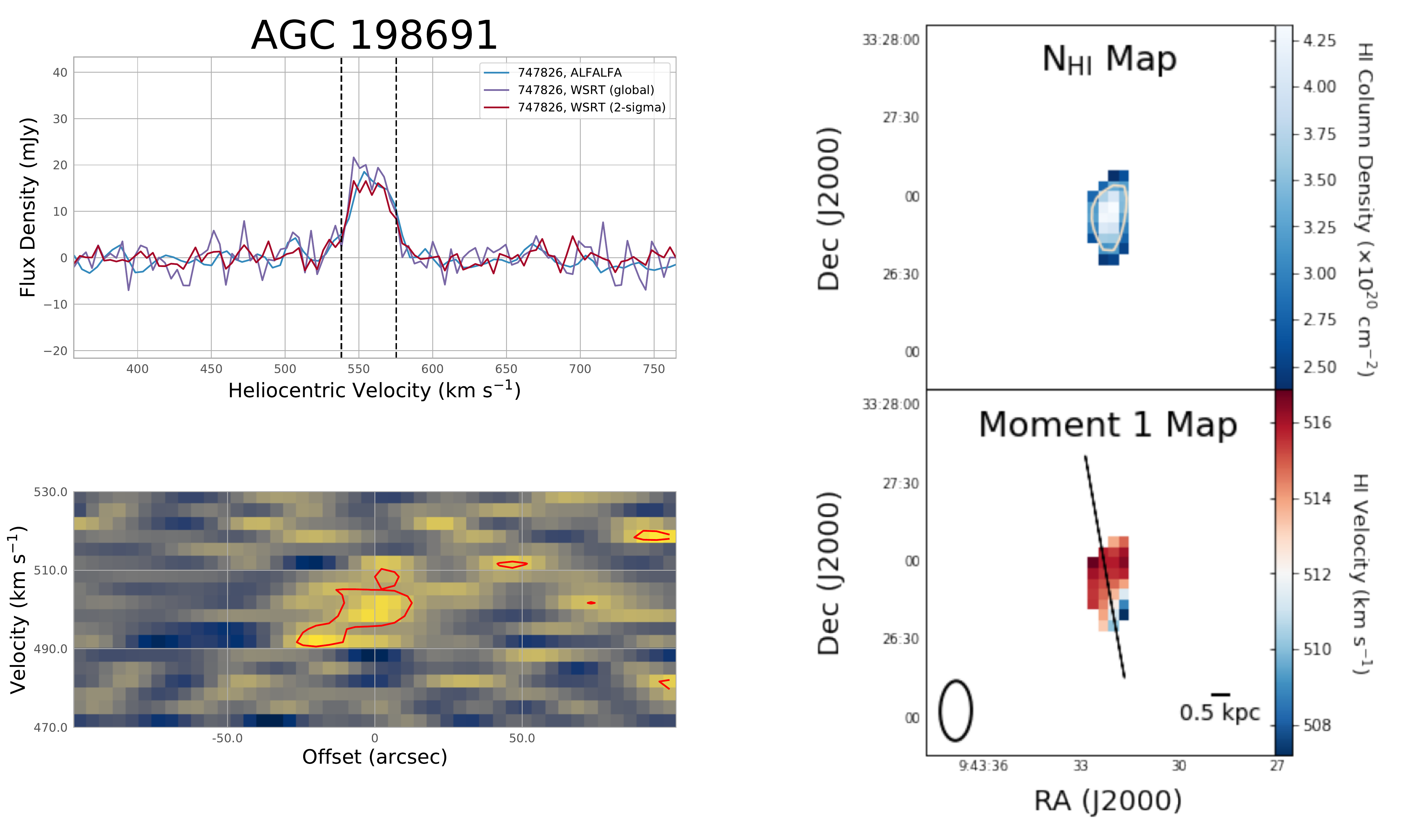}
\vspace{-8pt}
\caption{AGC\,198691.  The HI mass surface density is co-spatial with the stellar population of the source \citep[for details see ][]{McQuinn2020}.  These WSRT images show no clear projected velocity gradient and suggest the presence of low surface density \hi\ gas in the outskirts of the galaxy.  Deep VLA HI images of this galaxy will be presented in Cannon et al. (in preparation).}  
\label{fig:HI_AGC198691}
\end{figure*}

\begin{figure*}
\centering
\includegraphics[width=0.9\textwidth]{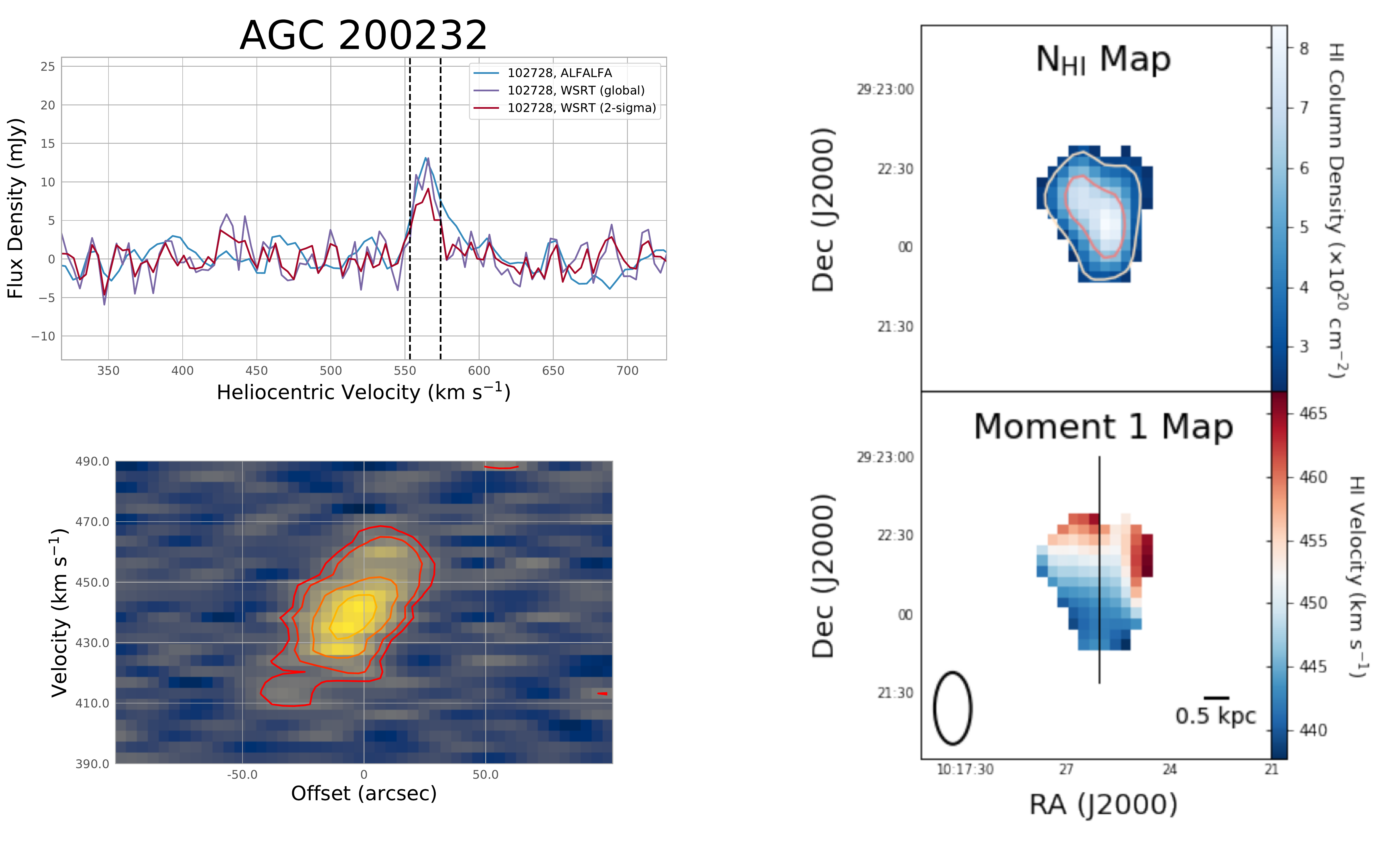}
\vspace{-8pt}
\caption{AGC\,200232.  There is a
  projected velocity gradient from south to north with a
  magnitude of 22 km\,s$^{-1}$.}
\label{fig:HI_AGC200232}
\end{figure*}

\begin{figure*}
\centering
\includegraphics[width=0.9\textwidth]{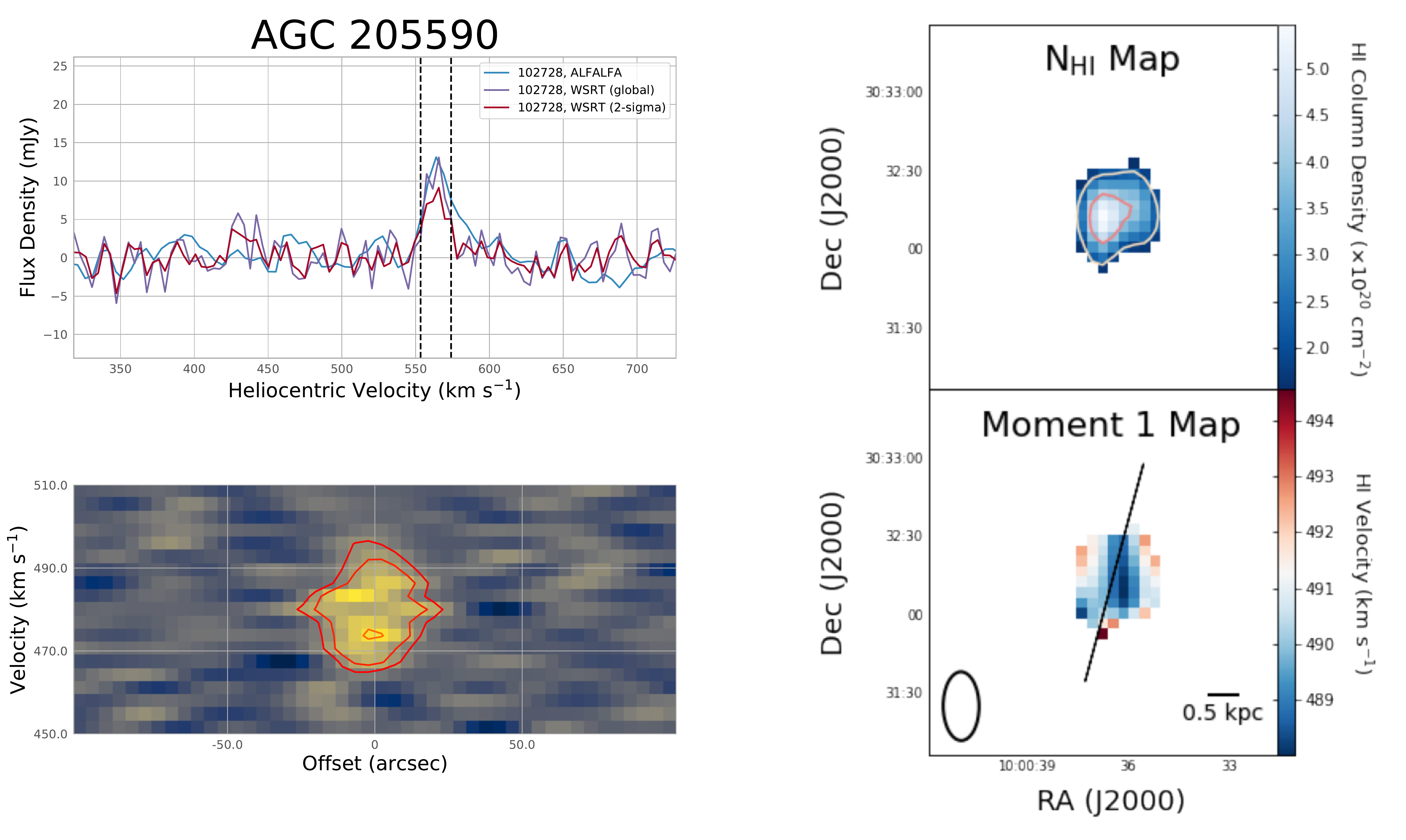}
\vspace{-8pt}
\caption{AGC\,205590.  There is a
  projected velocity gradient from south to north with a
  magnitude of 5 km\,s$^{-1}$.}
\label{fig:HI_AGC205590}
\end{figure*}

\begin{figure*}
\centering
\includegraphics[width=0.9\textwidth]{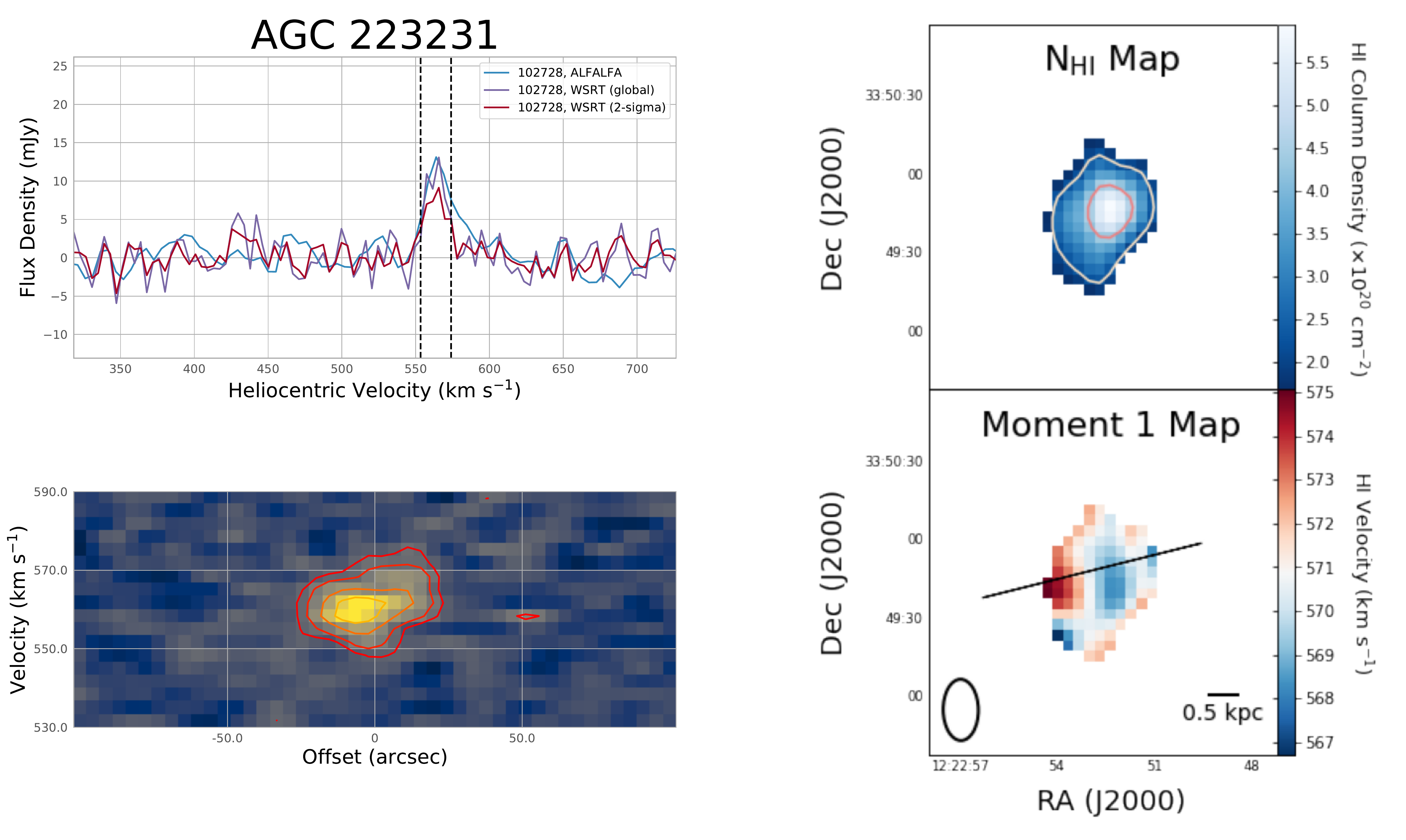}
\vspace{-8pt}
\caption{AGC\,223231.  While the
  source is resolved by the HI beam and the HI mass surface density
  maximum exceeds 5\,$\times$\,10$^{20}$ cm$^{-2}$, there is only weak
  evidence for a coherent projected velocity gradient of 4 \kms.}
\label{fig:HI_AGC223231}
\end{figure*}

\begin{figure*}
\centering
\includegraphics[width=0.9\textwidth]{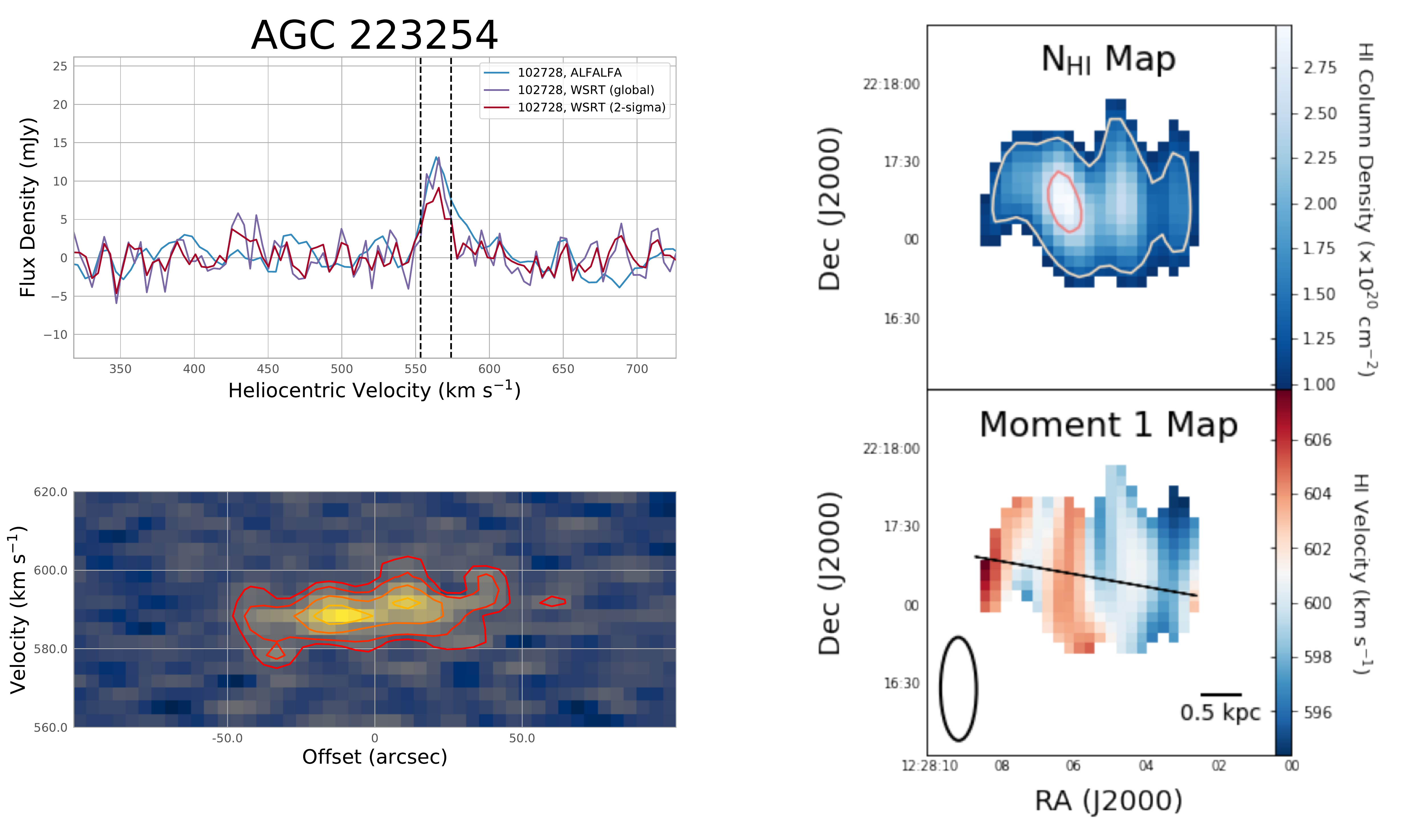}
\caption{AGC\,223254.  There is a
  clear projected velocity gradient from west to east with a
  magnitude of 9 km\,s$^{-1}$.}
\label{fig:HI_AGC223254}
\end{figure*}

\begin{figure*}
\centering
\includegraphics[width=0.9\textwidth]{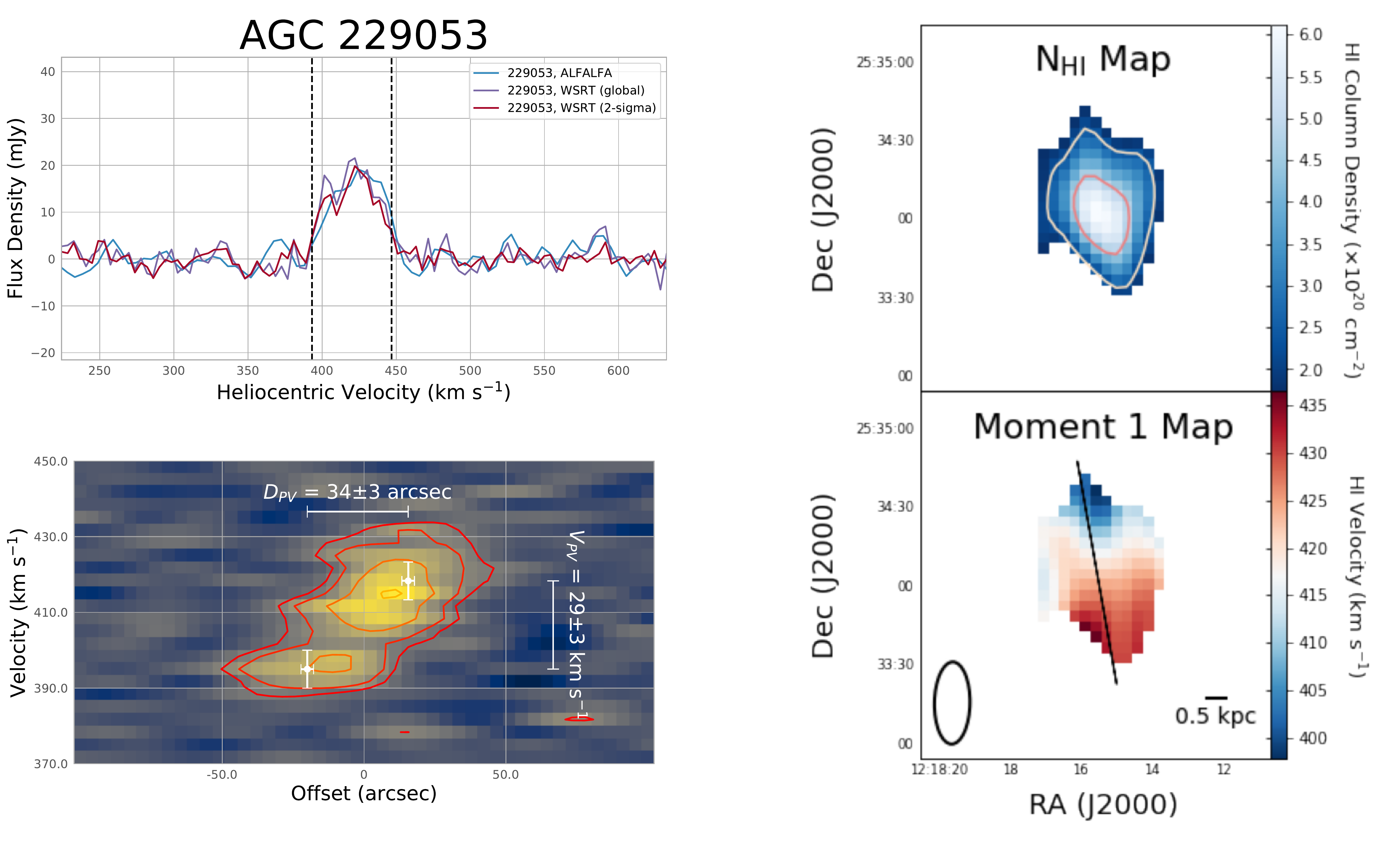}
\caption{AGC\,229053.  There is a
  projected velocity gradient from north to south with a
  magnitude of 29 km\,s$^{-1}$.}
\label{fig:HI_AGC229053}
\end{figure*}

\begin{figure*}
\centering
\includegraphics[width=0.98\textwidth]{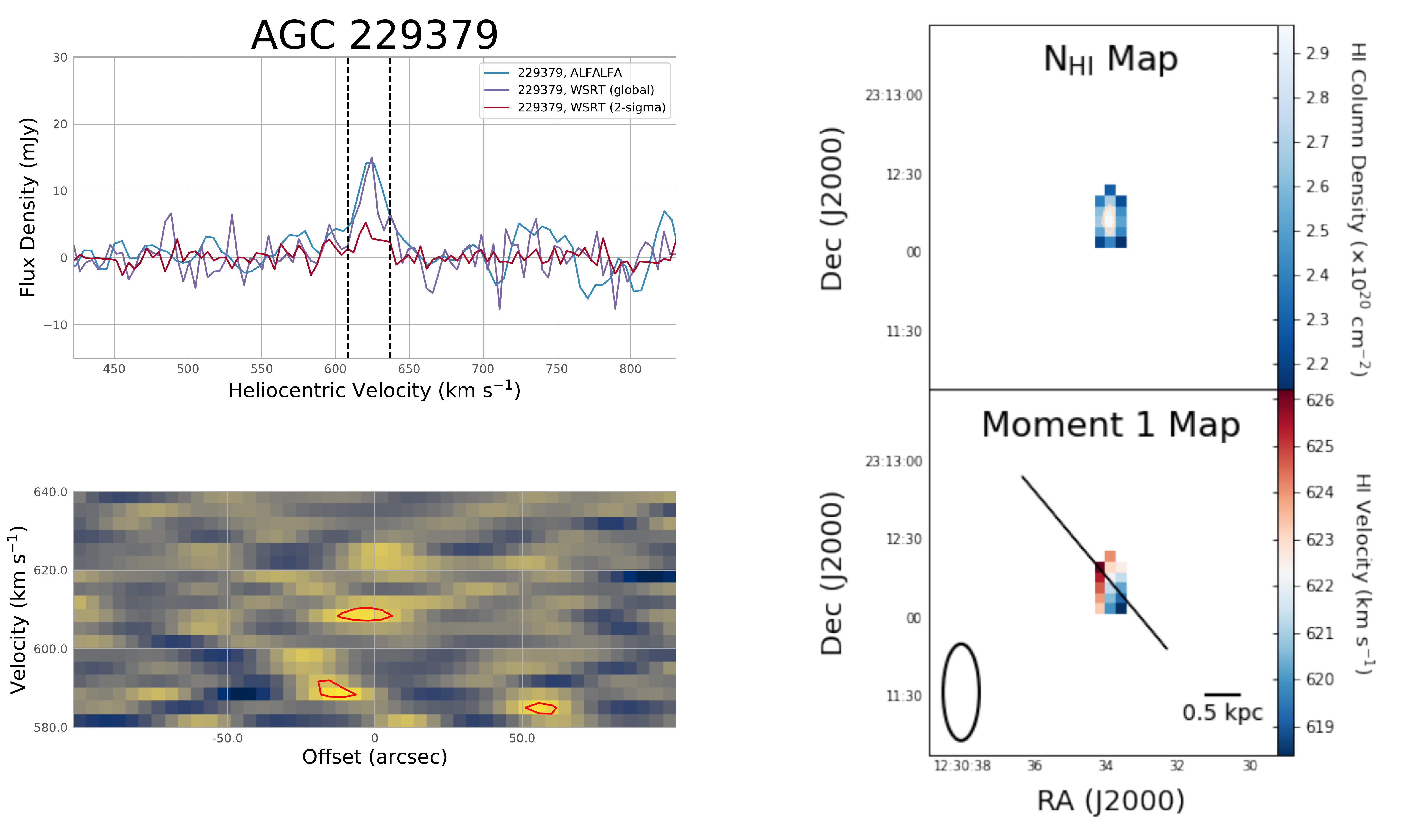}
\vspace{-8pt}
\caption{AGC\,229379.  This source is
  detected at low S/N; there is no evidence for a clear projected
  velocity gradient.}
\label{fig:HI_AGC229379}
\end{figure*}

\begin{figure*}
\centering
\includegraphics[width=0.9\textwidth]{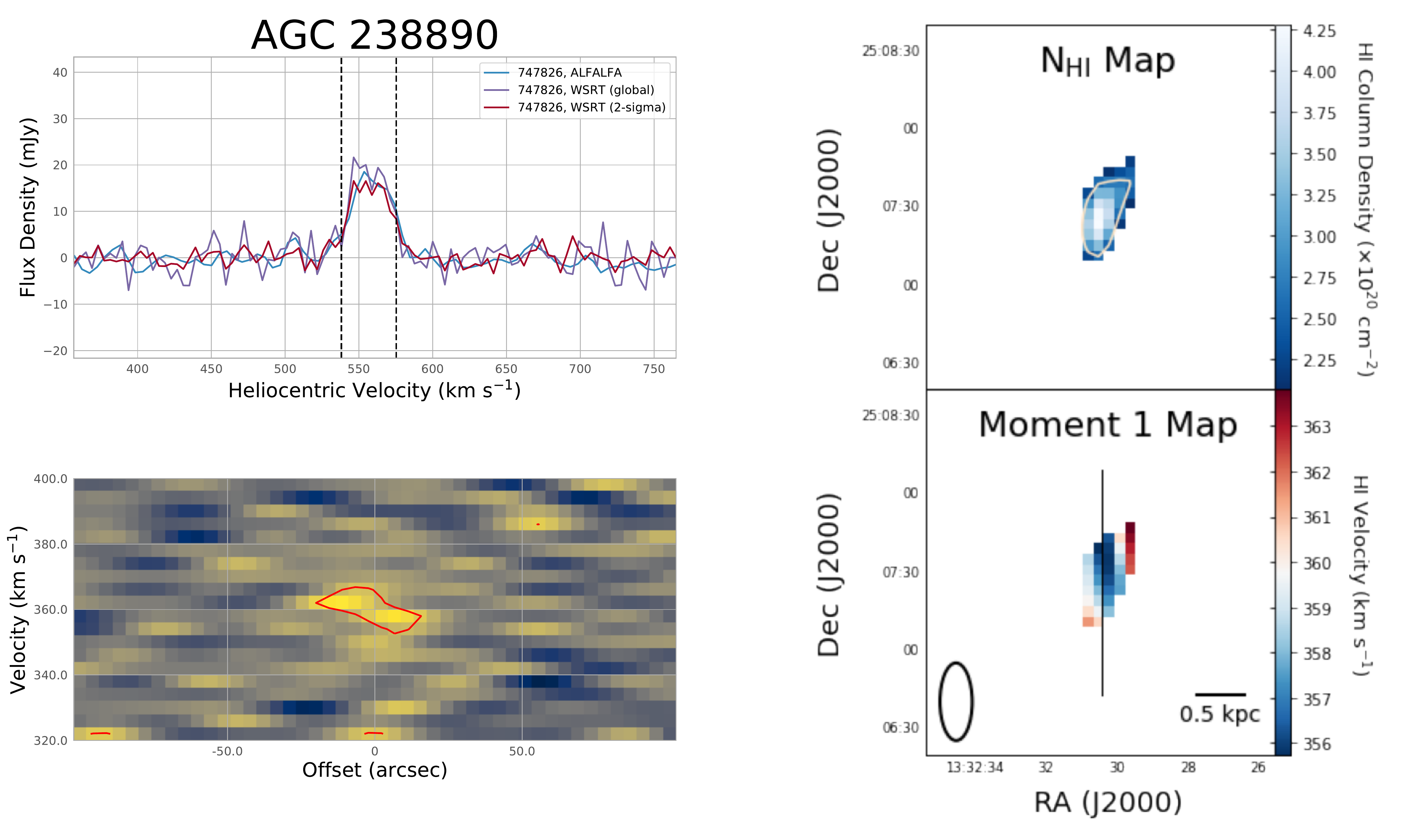}
\vspace{-8pt}
\caption{AGC\,238890.  There is no
  clear projected velocity gradient, and the source is only marginally
  resolved by the HI beam.}
\label{fig:HI_AGC238890}
\end{figure*}

\begin{figure*}
\centering
\includegraphics[width=0.9\textwidth]{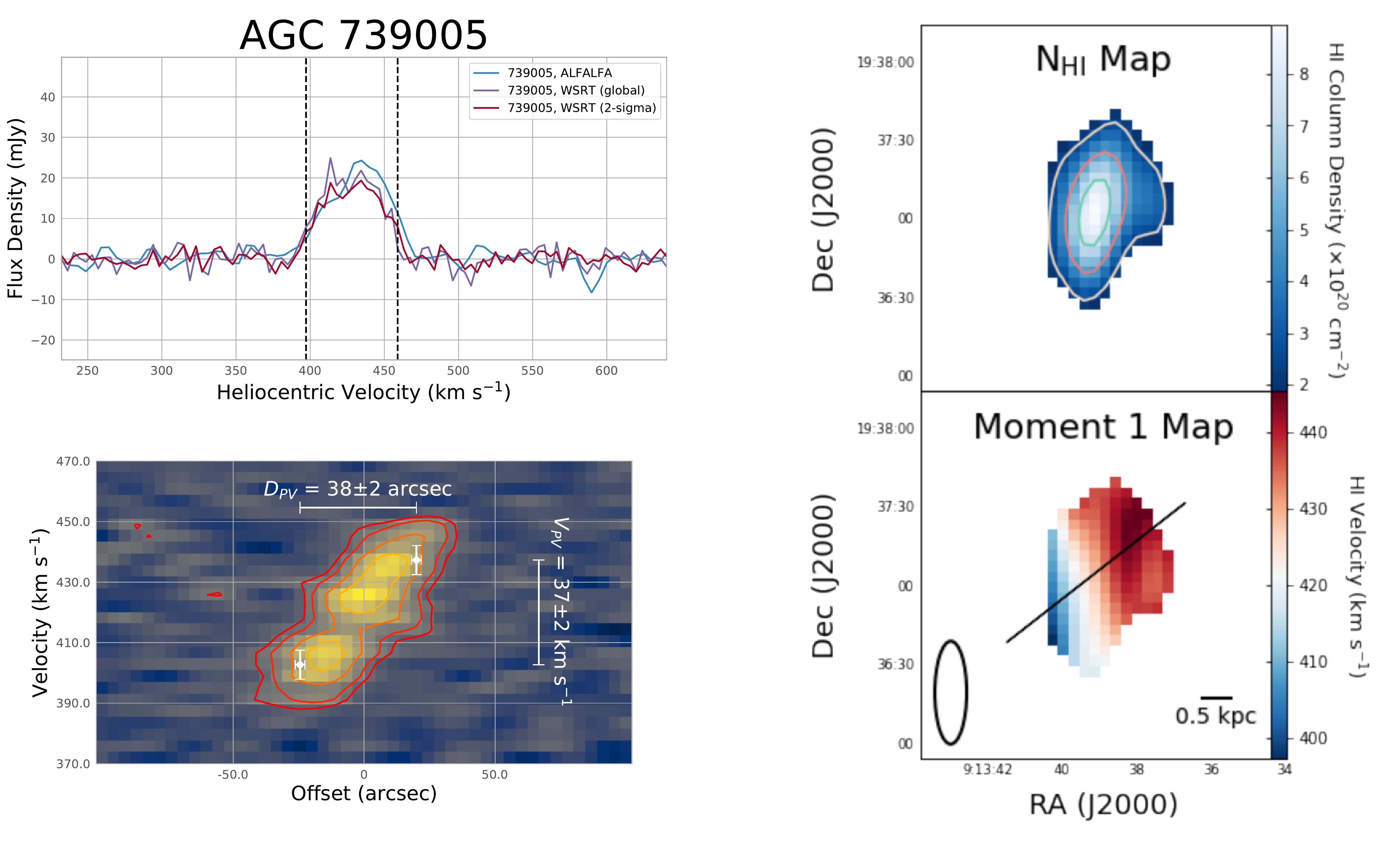}
\vspace{-8pt}
\caption{AGC\,739005.  There is a
  clear projected velocity gradient from east to west with a
  magnitude of 36 km\,s$^{-1}$.}
\label{fig:HI_AGC739005}
\end{figure*}

\begin{figure*}
\centering
\includegraphics[width=0.9\textwidth]{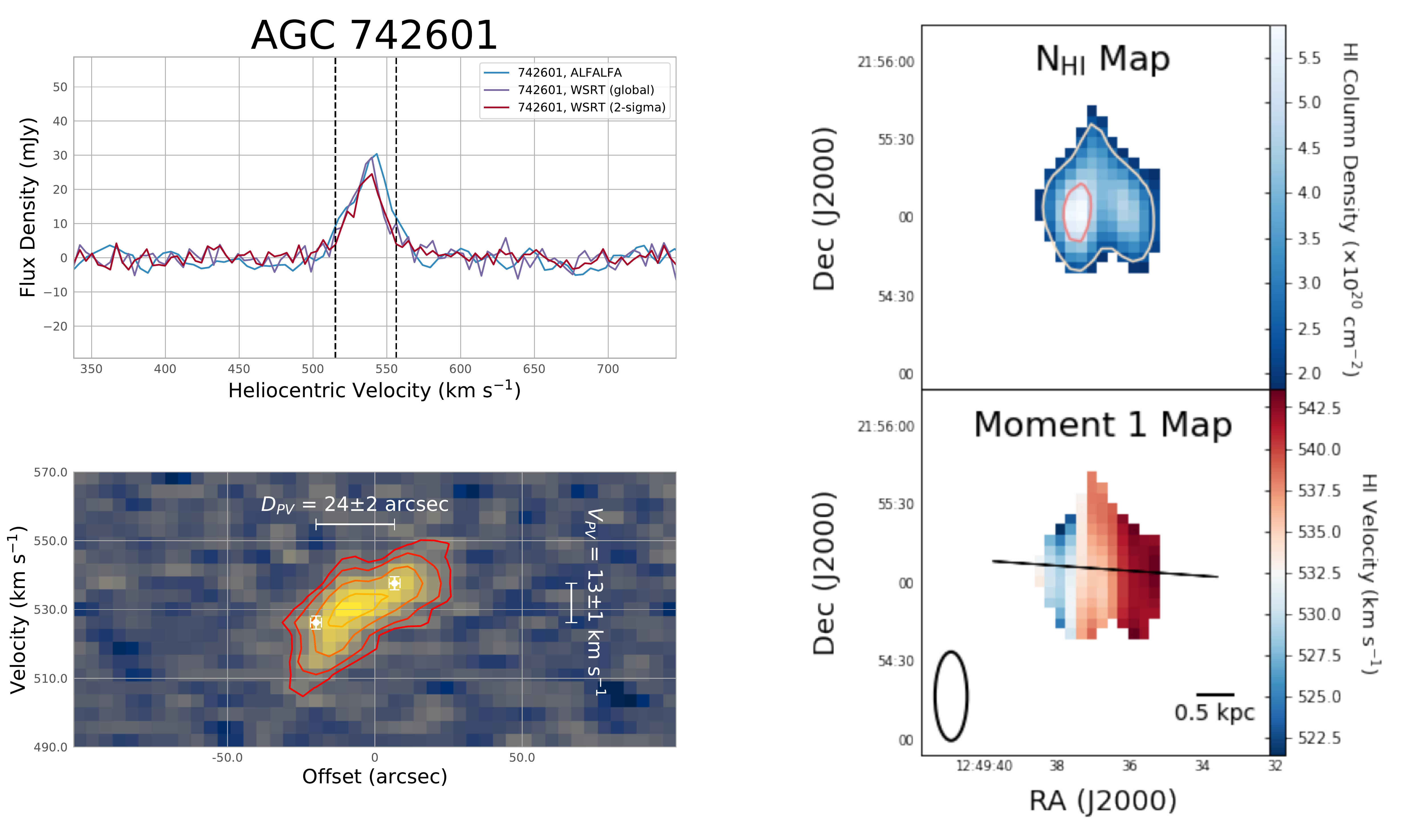}
\caption{AGC\,742601.  There is a
  clear projected velocity gradient from east to west with a
  magnitude of 12 km\,s$^{-1}$.}
\label{fig:HI_AGC742601}
\end{figure*}

\begin{figure*}
\centering
\includegraphics[width=0.9\textwidth]{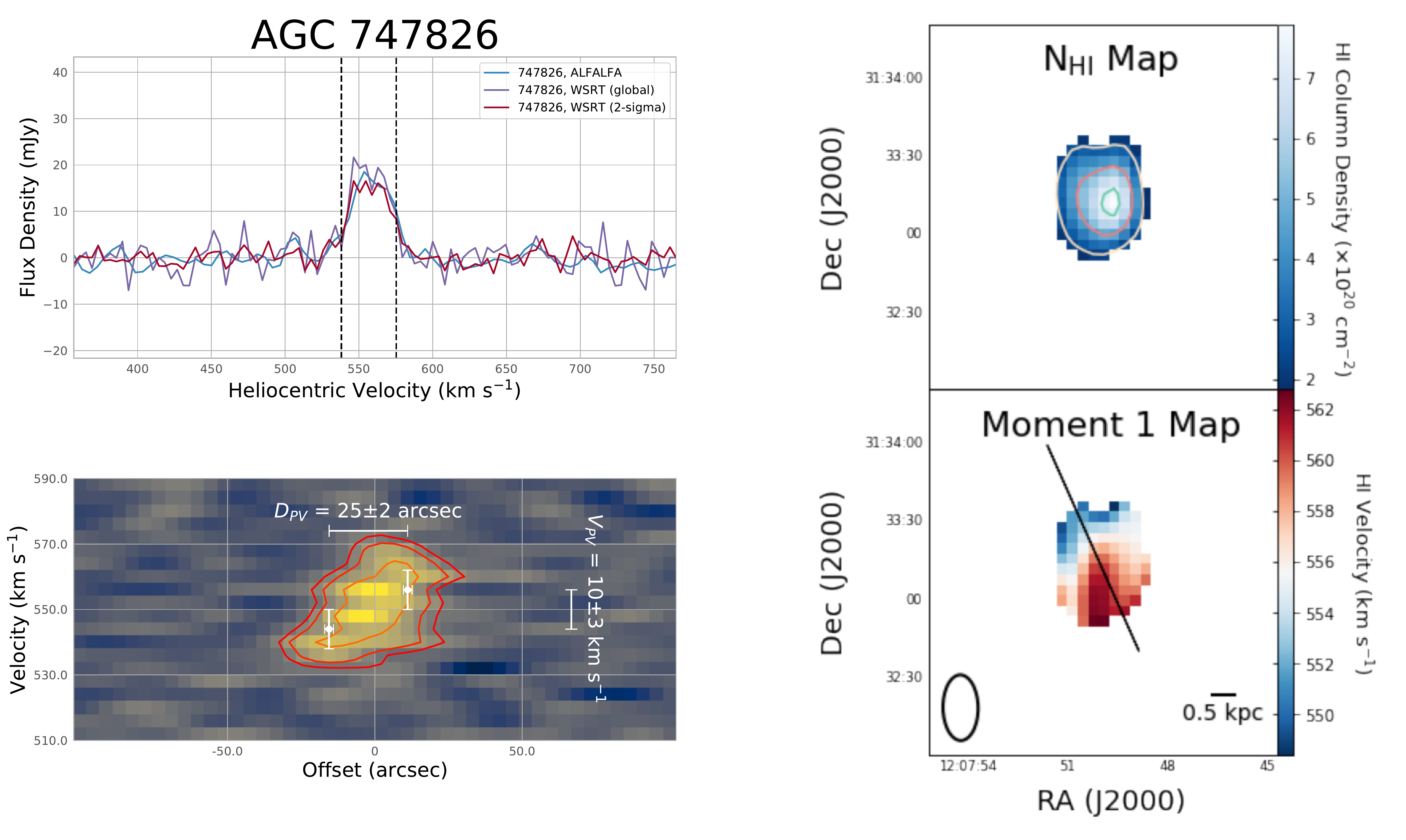}
\caption{AGC\,747826.  There is a clear projected velocity gradient from east to west with a
  magnitude of 10 km\,s$^{-1}$.}
\label{fig:HI_AGC747826}
\end{figure*}

\clearpage

\section{A novel approach for robust maximum velocity and extent determinations from position-velocity slices}\label{app:pvvel}

We have developed a new methodology for robustly constraining the maximum velocity and spatial extent from position-velocity (PV) slices. The idea behind using PV slices is to take advantage of the information available in resolved \hi\ data to provide a better estimate of rotation velocity and the physical extent at which it is measured when modelling a rotation curve from a velocity field is problematic, for example, due to limited spatial sampling and/or in cases where the gas rotational motion is comparable to the dispersion. This new methodology fits Gaussians to orthogonal samples from the PV diagrams to derive maximum velocity and spatial extents, based on the centroid of the Gaussian in different bins. The goal of developing this new methodology was three-fold:
\begin{enumerate}
\item Provide a robust measure of the maximal velocities and positions of the \hi\ gas: The previous methodology applied to SHIELD~I galaxies in \citet{McNichols2016} visually determined the maximum velocity extent that contained emission within the 2-$\sigma$ level in a PV slice; however, this method conflates the velocity dispersion in the gas with the rotational velocity motion. While dispersion support is important in dwarf galaxies, our new method of fitting the centroid of emission returns a velocity measure that is conceptually more similar to a rotation velocity, where an asymmetric drift correction can then be applied. 
\item Provide a method where the results are reproducible: The previously used methodology is responsive to the sensitivity of the data and the subjective determination of the extent. Thus, with different data quality, the results are not necessarily reproducible.
\item Provide a meaningful error on the values of the maximal velocities and positions of the \hi\ gas: Fitting a function to the data allows the opportunity to return an error on the accuracy of the fit, providing meaningful uncertainties.
\end{enumerate}

Below, we briefly describe this new methodology. J. Fuson et al.\ (in preparation) will present a detailed comparison of this methodology to values derived from kinematic modelling and analyze its overall effectiveness as an estimate of rotation velocity and extent. They will also present the code used to undertake this fitting.

\subsection{Deriving the velocity extent}
In order to derive the maximum velocity extent, the PV diagram is binned along the offset axis. The binning is specified by the user in arcseconds and rounded down to a unit number of pixels for numerical ease. A bin is considered eligible for fitting if the maximum value of the velocity spectrum is at least 3 times the rms. This cut-off level was determined through experimentation and comparison with well-determined velocity fields. In cases meeting this criterion, a Gaussian is fit using a Markov Chain Monte Carlo (MCMC) approach implemented via the {\tt emcee} functionality in the {\tt lmfit} python module. The center of the Gaussian is taken as the velocity of the gas in that offset bin. The MCMC approach allows a derivation of uncertainties on the accuracy of the fitting. 

The left panels of Figures~\ref{fig:example_resolved} and \ref{fig:example_unresolved} shows the results of fitting the velocity in this way for two example cases. The velocity extent is determined as the difference between the minimum and maximum velocity values fitted along the offset axis, which we label V$_{\rm PV}$. The uncertainty in V${\rm PV}$ is a combination of the reported uncertainties in the center of the Gaussian fits for the minimum and maximum velocity values. The rotational velocity of the gas can then be determined from half of the value of V$_{\rm PV}$ corrected for inclination (i.e., V$_{\rm rot} = \frac{1}{2}$ V$_{\rm PV}~/~sin~i$). Note that one shortcoming of this approach is that rotational velocities determined from V$_{\rm PV}$ will not account for asymmetries in a velocity field. 

\subsection{Deriving the spatial extent}
The spatial extent is determined in an analogous way to the velocity extent, except that the slicing is done in the orthogonal direction, with bins along the velocity axis of the PV diagram. The binning is specified by the user in \kms\ and rounded down to a unit number of pixels for numerical ease. A bin is considered for fitting if the maximum value of the bin along the offset axis is at least 3 times the rms. In those cases, a Gaussian is fit as above, where the center of the Gaussian is taken as the offset extent for that velocity bin.  
 
\begin{figure}
\centering
\includegraphics[width=0.45\linewidth]{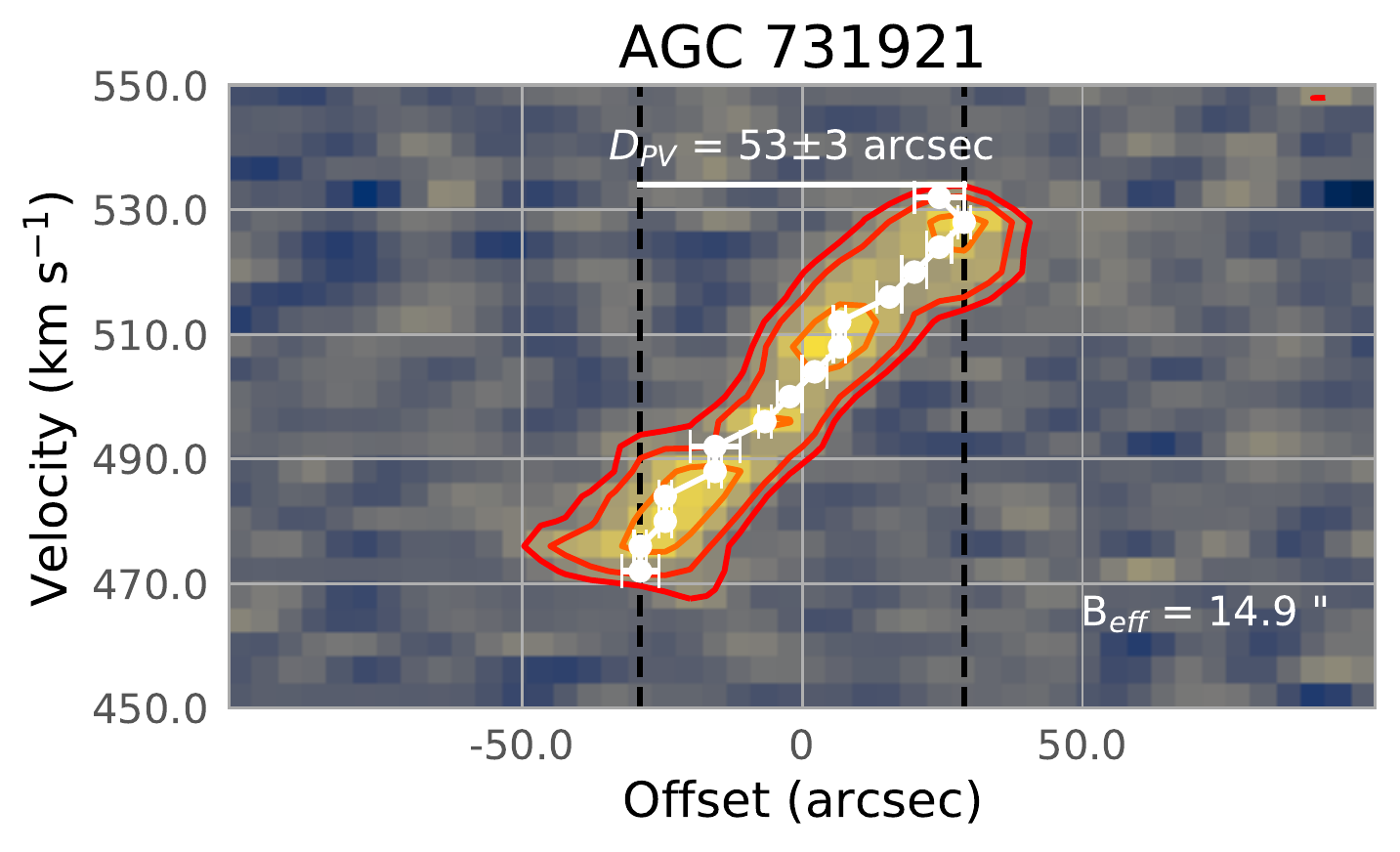}
\quad
\includegraphics[width=0.45\linewidth]{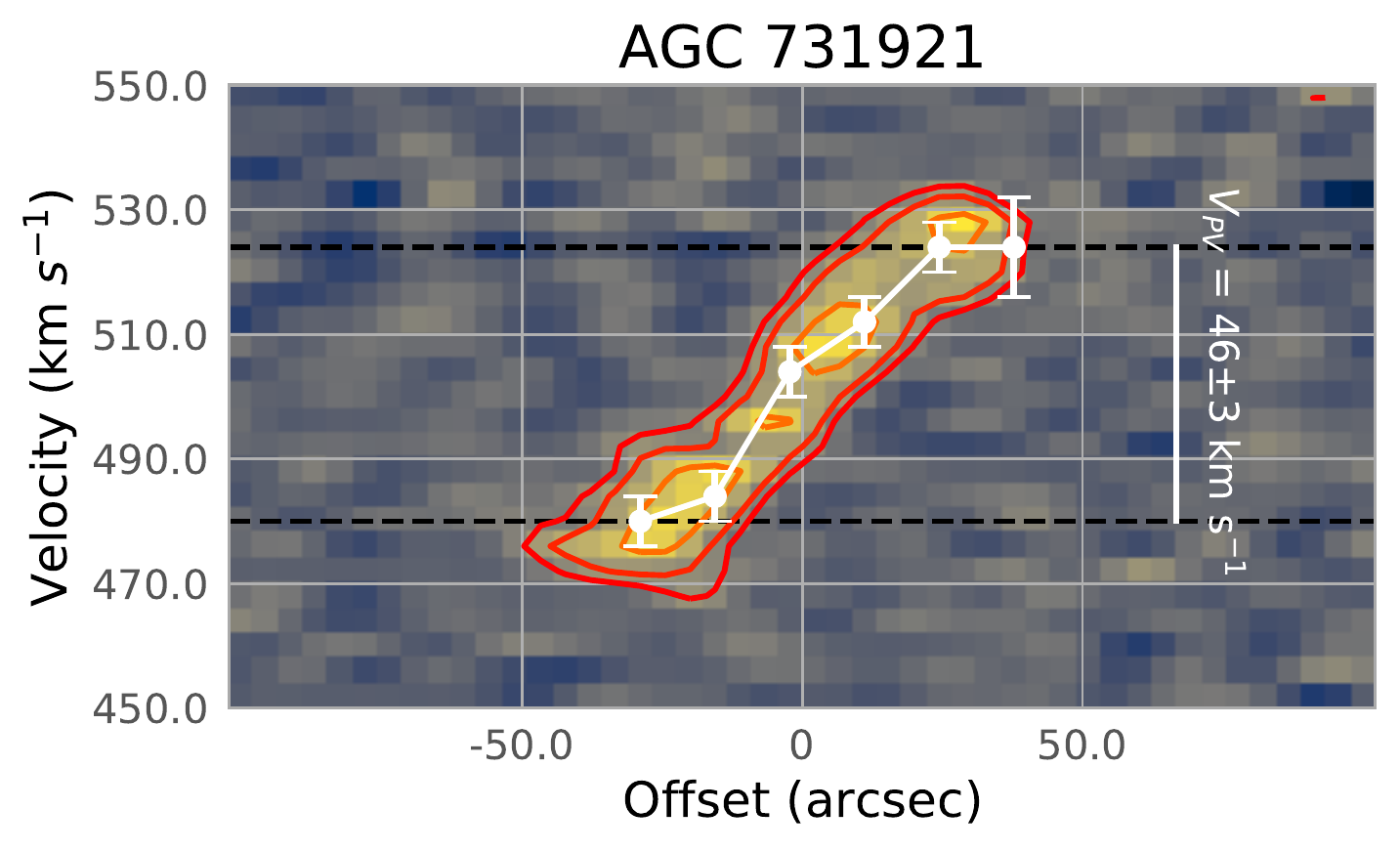}    
\includegraphics[width=0.45\linewidth]{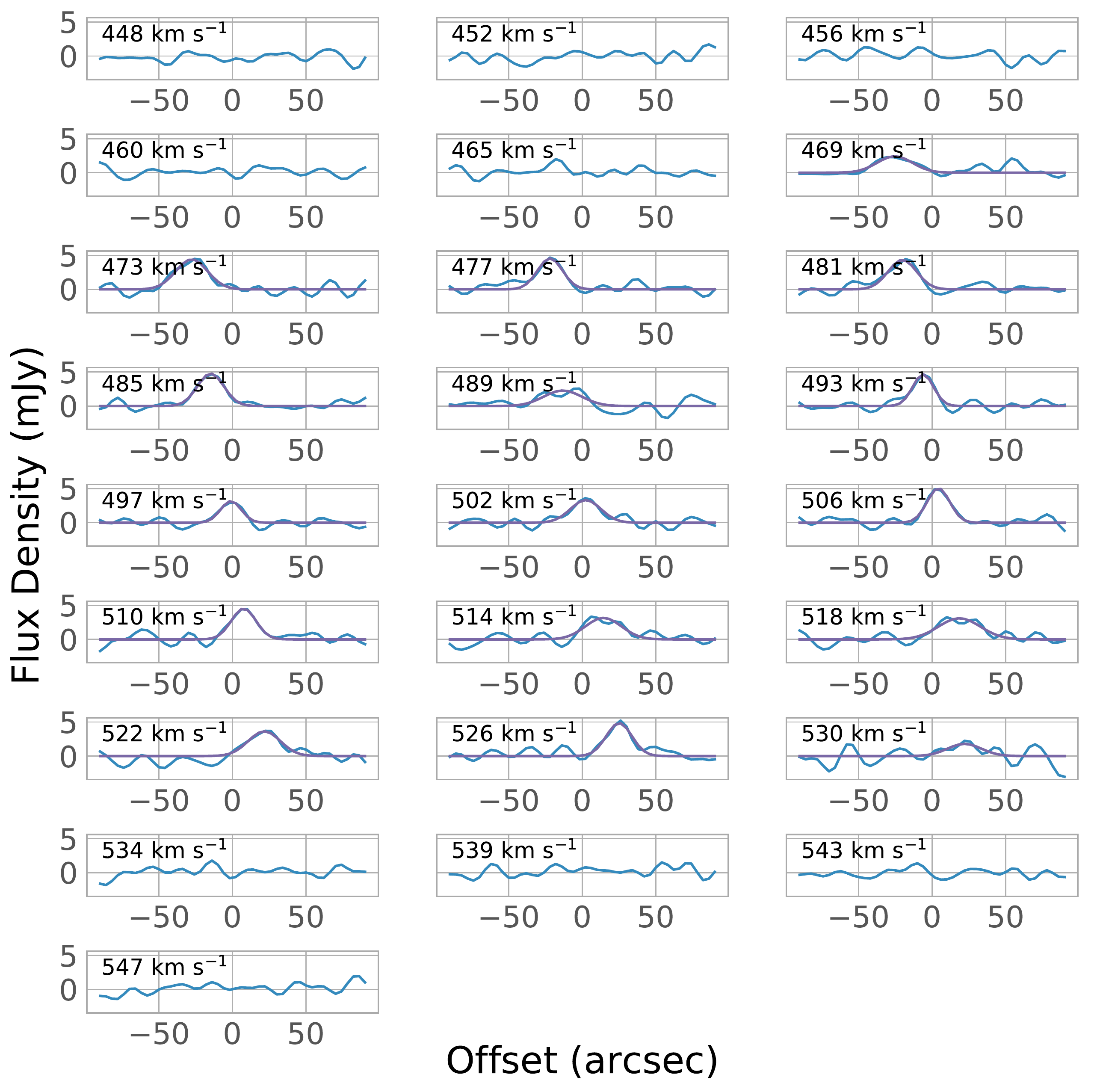}
\quad
\includegraphics[width=0.45\linewidth]{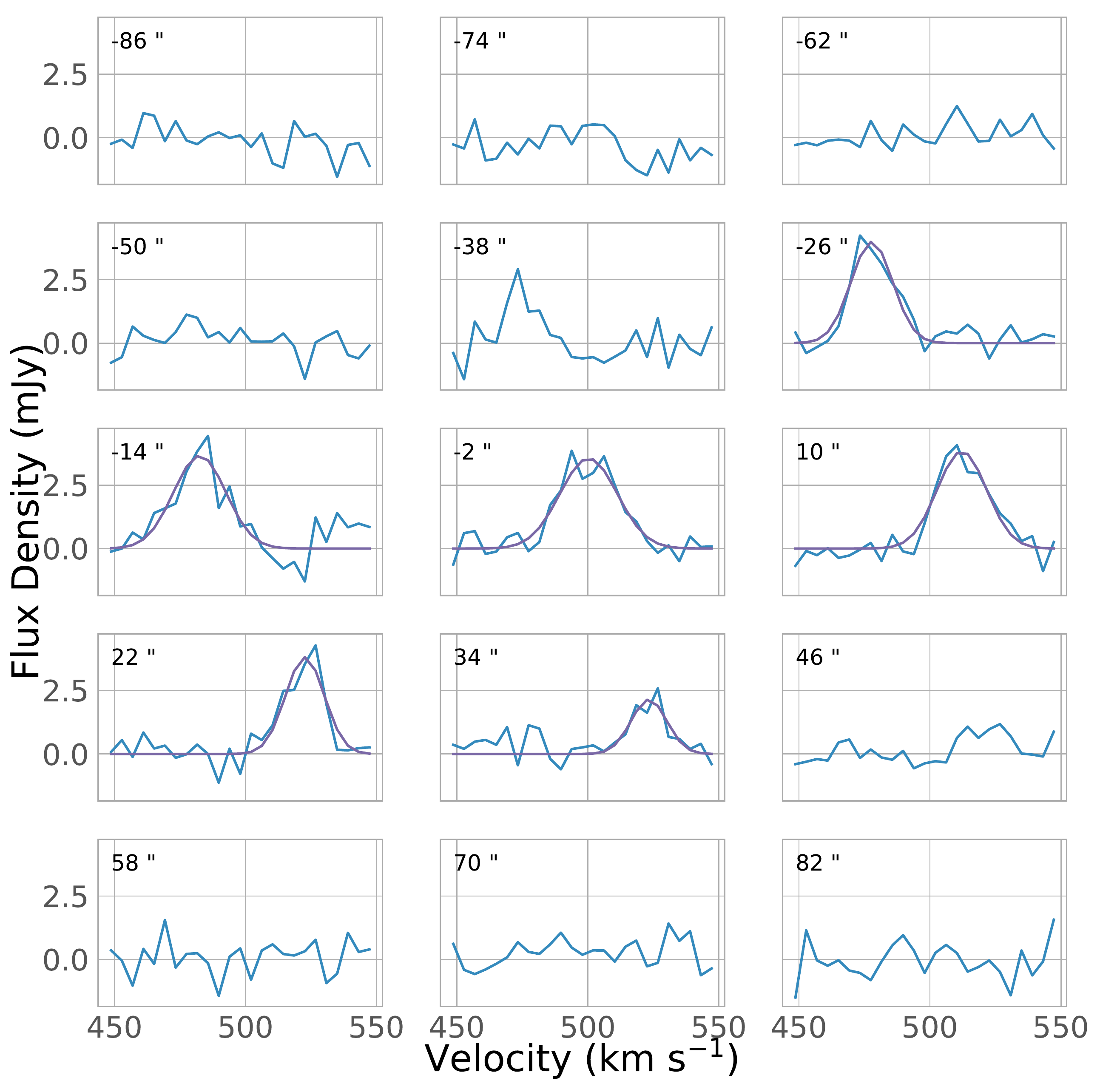}
\caption{Examples of deriving the spatial (left) and velocity (right) extent for a well-resolved case (AGC\,731921). Top panels: the PV diagrams overlaid with \hi\ column density contours at the 2, 3, 5-$\sigma$ detection levels. The measured spatial and velocity extents are indicated on each panel with black dashed lines; final values of the velocity (V$_{\rm PV}$) and diameter (D$_{\rm PV}$), with uncertainties, as well as the calculated effective beam size ($B_{eff}$) are also listed. Bottom panels: The ``spectra" for each bin (blue), with Gaussian fits when the peak value is $>3 \times$ rms (purple). The location of each bin in velocity and spatial position offsets are marked with filled white circles in the top panels; uncertainties are based on the reported uncertainty in the fits to the center of the Gaussian.}
\label{fig:example_resolved}
\end{figure}

\begin{figure}
\centering
\includegraphics[width=0.45\linewidth]{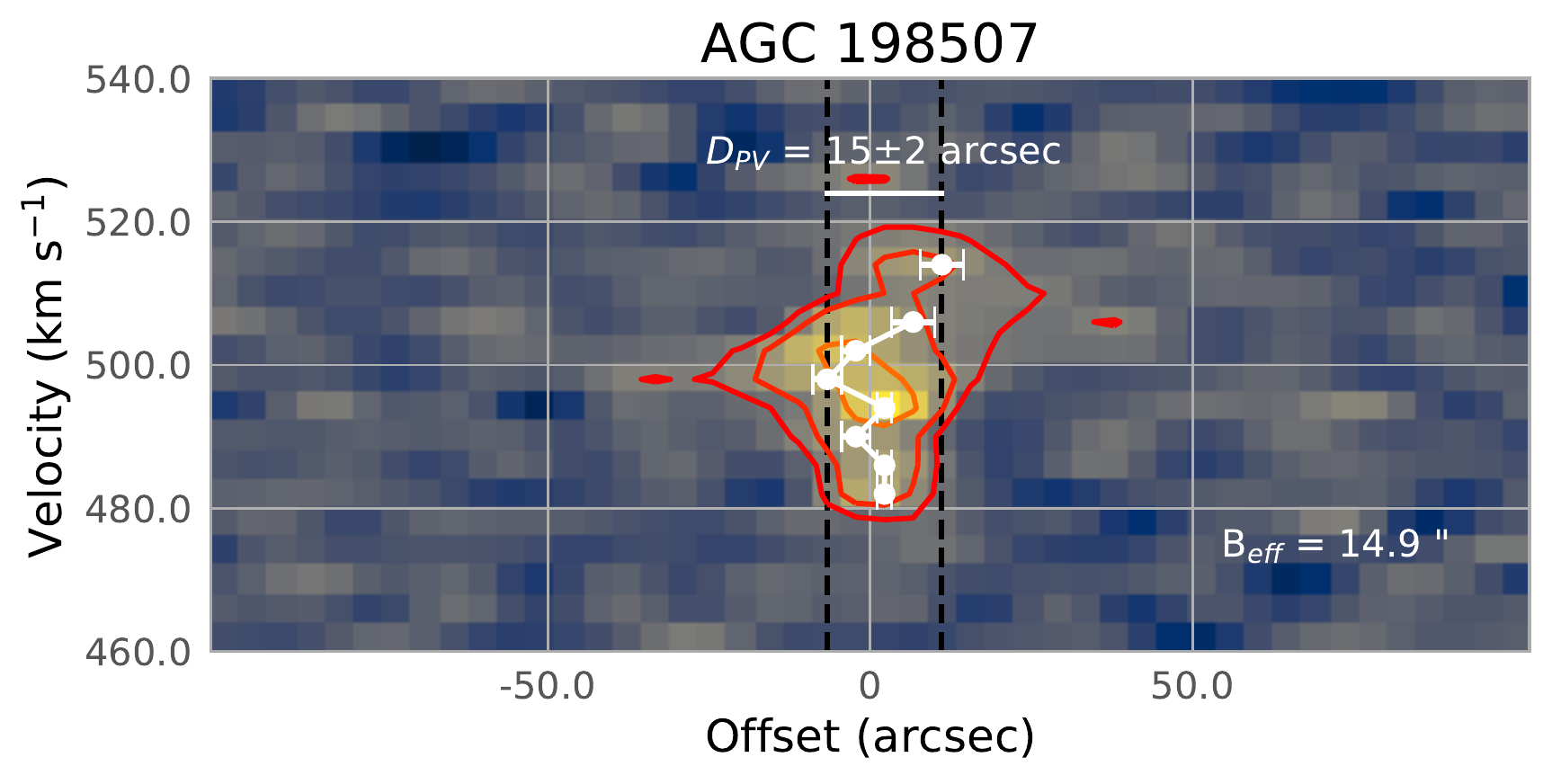}
\quad
\includegraphics[width=0.45\linewidth]{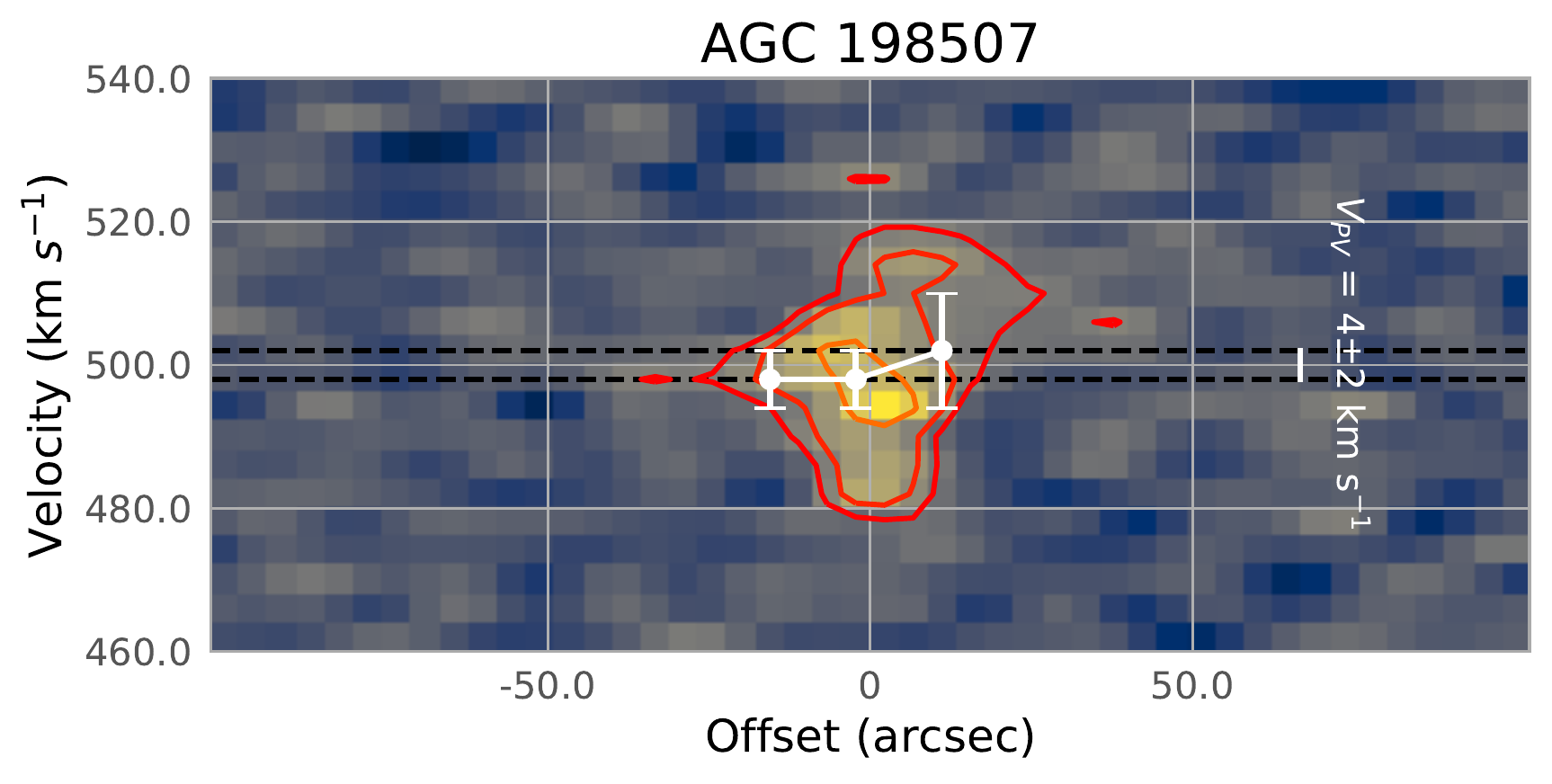}
\includegraphics[width=0.45\linewidth]{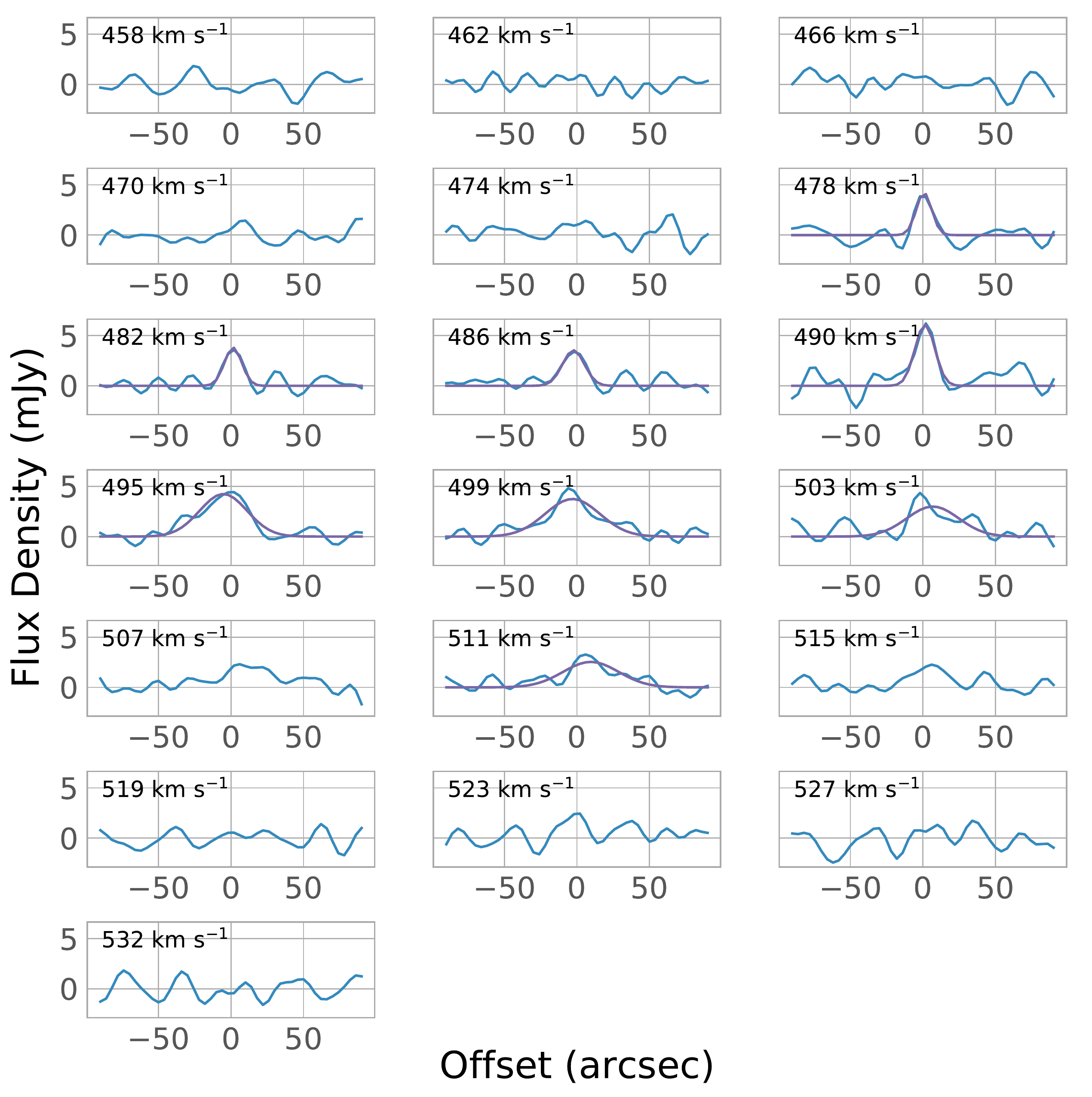}
\quad
\includegraphics[width=0.45\linewidth]{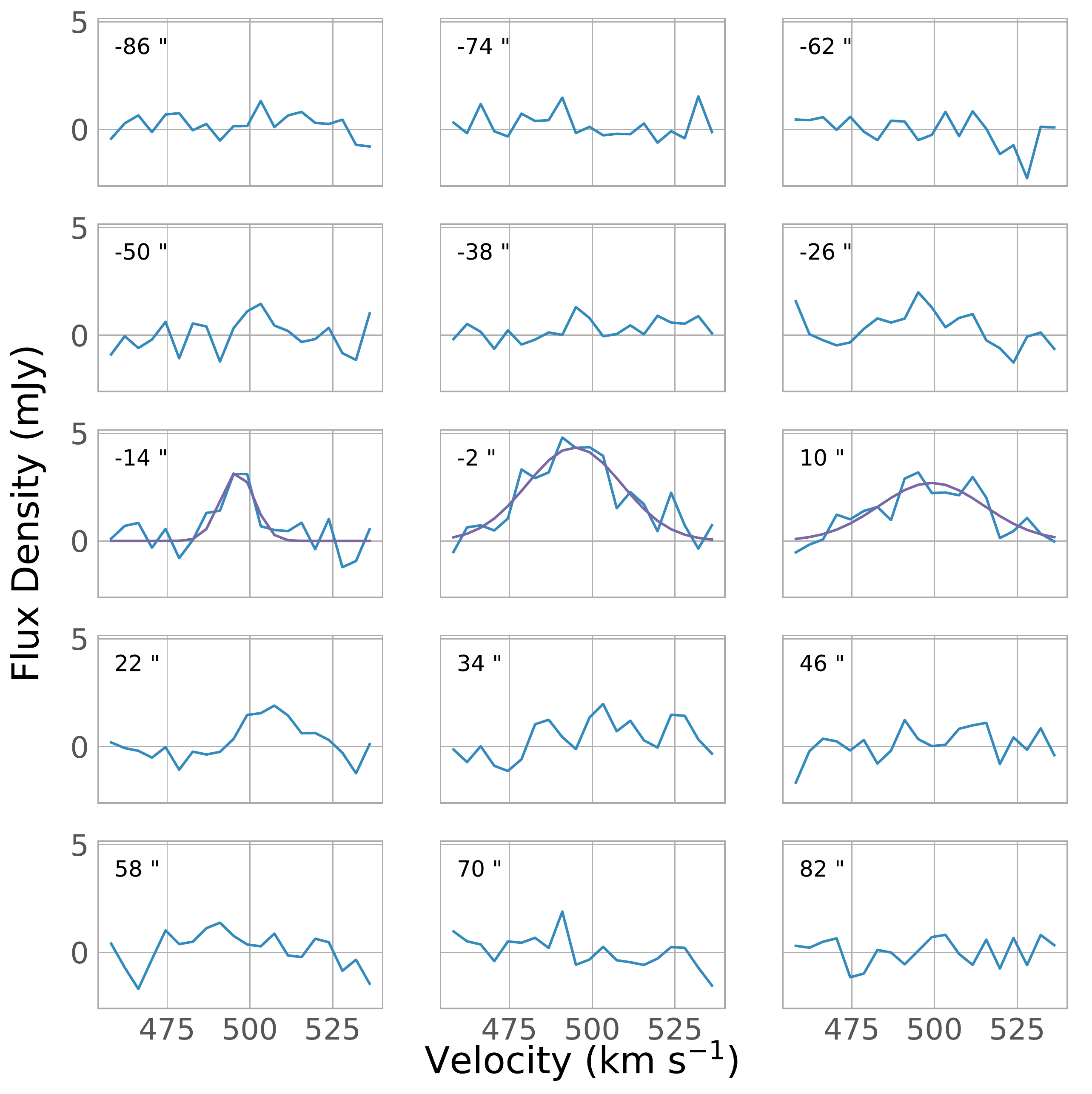}
\caption{Examples of deriving the spatial (left) and velocity (right) extent for an unresolved case (AGC\,198507). The format and labels are the same as in Figure~\ref{fig:example_resolved}. As the spatial position offset, D$_{\rm PV}$, is less than the effective beam, $B_{eff}$, the measured velocity and spatial extents are not meaningful.}
\label{fig:example_unresolved}
\end{figure}

The right panels of Figures \ref{fig:example_resolved} and \ref{fig:example_unresolved} shows the results of fitting the spatial position offset in this way for two example cases. The spatial extent corresponding to the measurement of $V_{\rm PV}$ is determined as the difference between the minimum and maximum offset values that are returned from the Gaussian fitting, which we label D$_{\rm PV}$. The uncertainty in D$_{\rm PV}$ is a combination of the reported uncertainties in the center of the Gaussian fits for the minimum and maximum offset values. 

\subsection{Accepting a fit}
A fit is only accepted if the derived spatial extent is at least as large as the effective beam across the PV diagram. Otherwise, the data is not well resolved and the fitting is considered not meaningful. In order to determine the effective resolution across the PV diagram, we first defined an effective position angle of the slice relative to the beam:

\begin{equation}
\phi = PA - B_{PA}
\end{equation}
where PA is the angle of the PV slice (listed in Table~\ref{tab:gas}) and $B_{PA}$ is the position angle of the restoring beam (listed in Table~\ref{tab:obs}). Then, the effective diameter of the beam across the PV slice is:

\begin{equation}
B_{eff} = \sqrt{(B_{maj} \cdot \cos{\phi})^{2} + (B_{min} \cdot \sin{\phi})^{2} }
\end{equation}

It is worth noting that in many cases when a galaxy is not resolved by this criterion, the returned velocity extent is also suspiciously small, i.e., 5 \kms\ or less. This is consistent with the idea that the data are not resolved enough to provide meaningful measurements using this methodology. It could also be an indication that the smallest dwarf galaxies will always pose a challenge for meaningful measures of rotational velocity.

Figure \ref{fig:example_resolved} demonstrates a well-resolved case, where the derived values clearly track well with the PV emission. Figure \ref{fig:example_unresolved} shows an unresolved case with the measured spatial extent less than the effective beam size; the results of the fitting are clearly not robust. In the latter case for AGC\,198507, while we show the measured velocity and spatial extents, these are not well-measured values of the bulk motion of the gas or its extent and should not be used as representative kinematic information. New VLA observations in the B configuration have recently been obtained on a subset of the full SHIELD sample; we expect these higher resolution data will enable rotational velocities and spatial extents of the \hi\ to be measured with confidence for a higher fraction of SHIELD galaxies using our new technique (VLA Large Program 20a-330; PI Cannon).

\subsection{Application to the SHIELD~II galaxies}
The PV slices used to derive velocities and extents for the SHIELD~II galaxies are described in Section \ref{sec:gas} and presented in Appendix \ref{app:HI_atlas}. The offset binning used for the derivation of the maximum velocity extent is 12\arcsec, slightly smaller than the WSRT minor axis \hi\ beam size. The velocity binning used for deriving the spatial extent is 8 \kms, which is approximately the intrinsic velocity dispersion. AGC\,223254 was excluded a priori from the derivation of velocity and spatial extents as both its velocity field and PV diagram indicate a disordered system.  Three SHIELD~II galaxies (AGC~198691, AGC~229379, AGC~238890) were not fit as they have insufficient S/N in at least two bins for each slicing direction. We rejected the fits for seven SHIELD~II galaxies (AGC~102728, AGC~123352, AGC~198507, AGC~198508, AGC~200232, AGC~205590, AGC~223231) as the derived extent is not resolved along the PV slice direction. It is worthwhile to note that some of these latter systems are close to being resolved and appear to have well-behaved fits; with higher quality, namely higher spatial resolution data, we expect to be able to robustly derive velocity and spatial extents from the PV diagrams.

\begin{figure}
\centering
\includegraphics[width=0.95\linewidth]{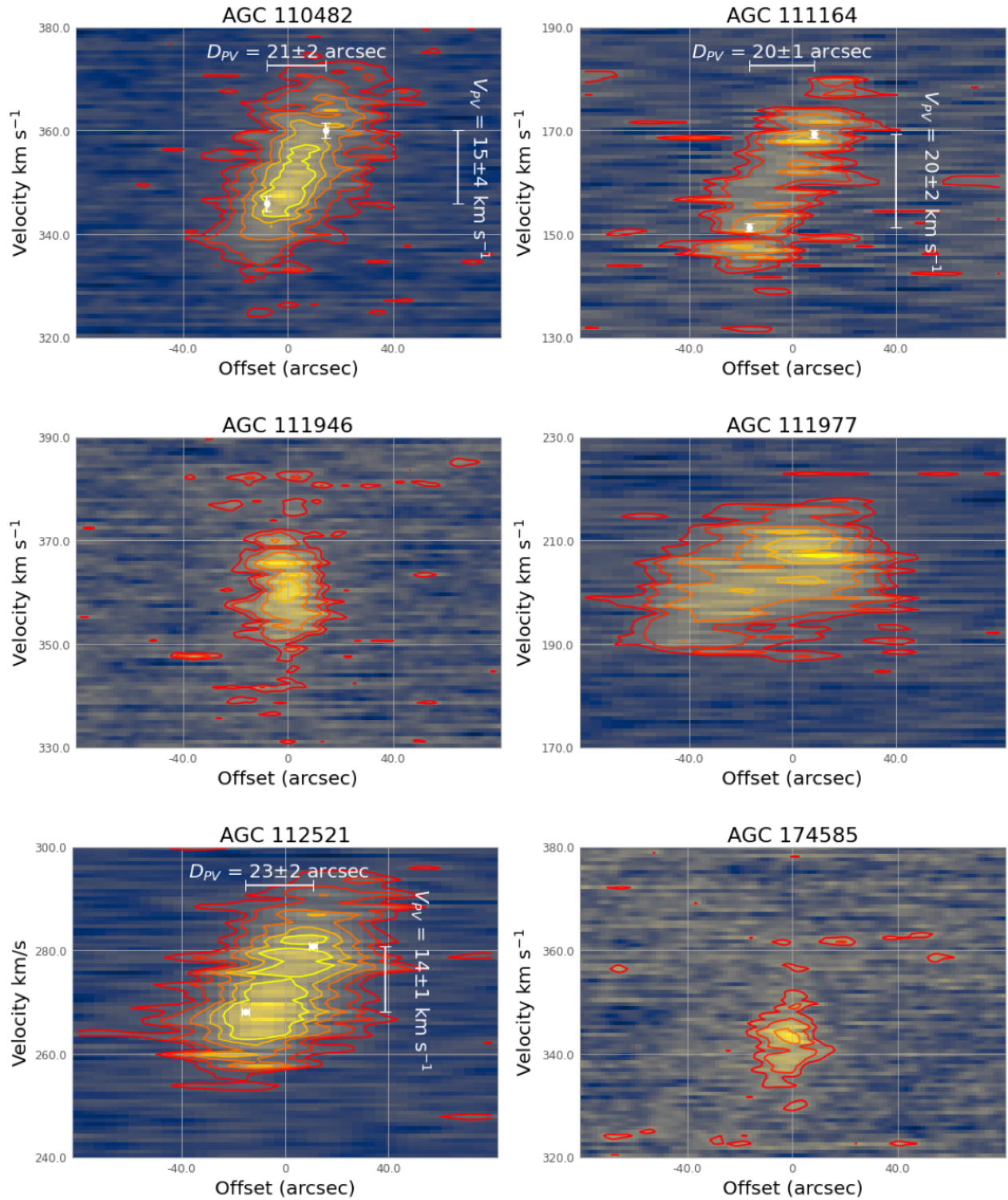}
\quad
\caption{PV diagrams for six of the twelve SHIELD~I galaxies, derived as described in \ref{app:shieldi_pv}. The newly derived velocity and spatial extents are shown for the galaxies which meet our criteria for robustly derived values.}
\label{fig:shieldi_pv_vel1}
\end{figure}

\begin{figure}
\centering
\quad
\includegraphics[width=0.95\linewidth]{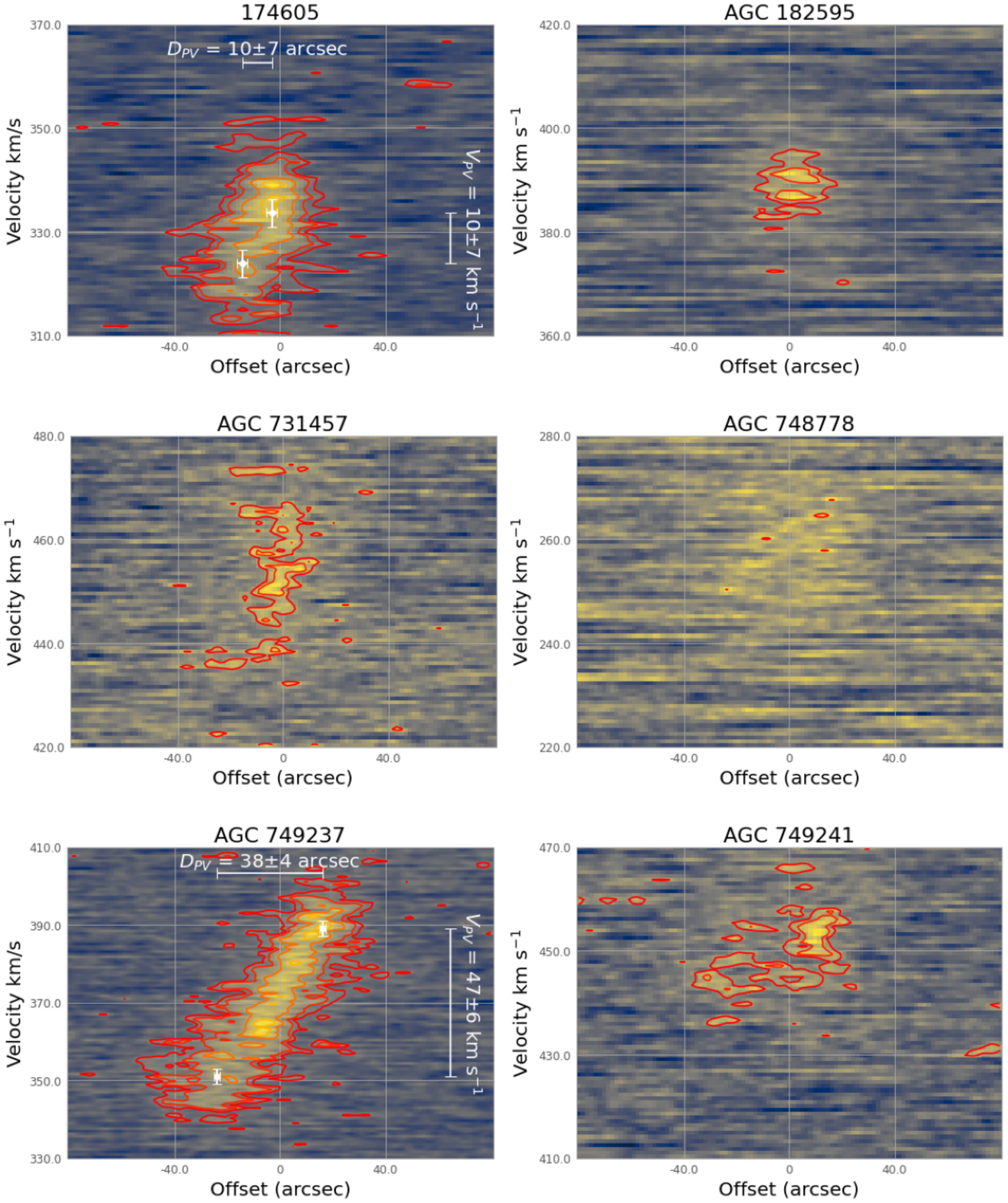}
\caption{PV diagrams for the remaining six SHIELD~I galaxies, derived as described in \ref{app:shieldi_pv}. The newly derived velocity and spatial extents are shown for the galaxies which meet our criteria for robustly derived values.}
\label{fig:shieldi_pv_vel2}
\end{figure}

\subsection{Application to the SHIELD~I galaxies}\label{app:shieldi_pv}
New PV slices were created for the SHIELD~I galaxies from the data presented in {McNichols et al. (2016)}\nocite{McNichols2016}.  Shown in Figures \ref{fig:shieldi_pv_vel1} \& \ref{fig:shieldi_pv_vel2}, these new PV slices have the same center and angle as those given in {McNichols et al. (2016)}, but the width is 50\arcsec, to be analogous to the PV slices derived for SHIELD~II (see Section \ref{sec:gas}.) 

The offset binning used for the derivation of the maximum velocity extent is 12\arcsec, which is roughly the typical SHIELD~I beam size and is consistent with the binning use for the SHIELD~II galaxies. The velocity binning used for deriving the spatial extent is 8 \kms, which is approximately the intrinsic velocity dispersion. When the derived velocity and spatial extents are robustly derived, we overplot these values on the PV diagrams in Figures~\ref{fig:shieldi_pv_vel1} \& \ref{fig:shieldi_pv_vel2}. Four SHIELD~I galaxies (AGC~182595, AGC~731457, AGC~748778,  AGC~749241) were not fit as there was insufficient S/N in our required minimum of two bins for each slicing direction. We rejected the fits for three SHIELD~I galaxies (AGC~111946, AGC~111977, AGC~174585) as the measured spatial extents were not resolved. 

\subsection{Comparison to ALFALFA velocity widths}
\begin{figure}
\centering
\includegraphics[width=0.5\linewidth]{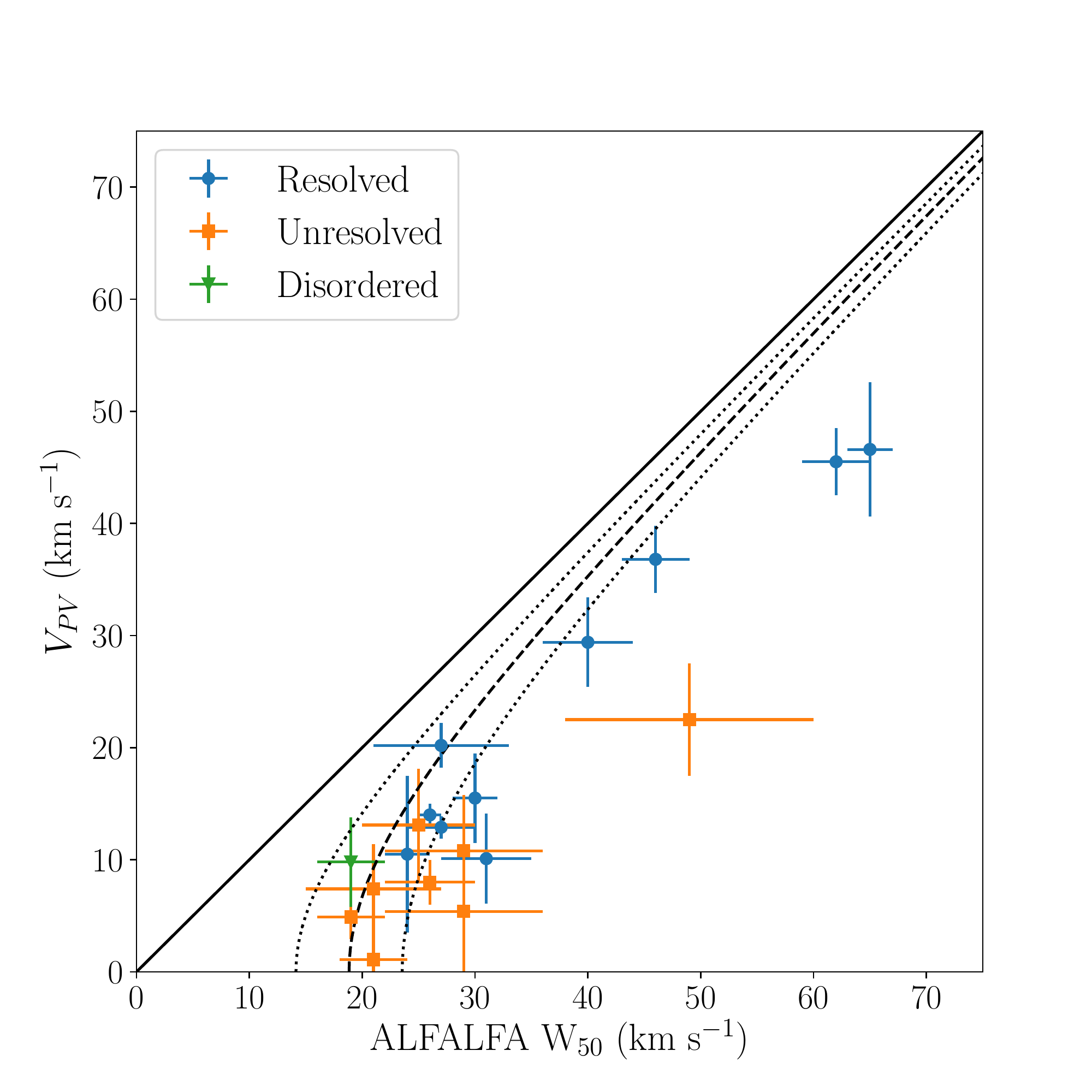}
\caption{The velocity extents, V$_{\rm PV}$, derived here compared to the velocity widths measured from single-dish observations, ALFALFA W$_{50}$. The solid line indicates the one-to-one relation, the dashed line indicates the relation accounting for an intrinsic velocity dispersion value of 8 \kms, and the lower and upper bounding dotted lines show the range assuming velocity dispersion values of 6 to 10 \kms\ respectively. See text for details.}
\label{fig:vel_w50}
\end{figure}

As a verification of this new approach, we undertook a comparison of our measured velocity extents, V$_{\rm PV}$, to the \hi\ velocity widths from the  ALFALFA data, W$_{50}$. Figure \ref{fig:vel_w50} shows this comparison, with sources color-coded by whether their measured spatial extent is resolved or unresolved; the disorder system AGC\,223254 is noted separately. While we do not report the velocities fit for the unresolved cases, we included them here for completeness. A one-to-one comparison is shown as a solid line; it is evident that velocity extents measured with our new methodology are systematically smaller than the ALFALFA W$_{50}$ values. This is to be expected given that the measured velocity widths include both the rotational velocity motion and the velocity dispersion of the gas. To account for this, we calculated the relationship between the measured velocity width, the velocity extent, and the velocity dispersion assuming that these values are all well-represented by Gaussians in the low-mass dwarf regime where both W$_{50}$ and V$_{\rm PV}$ measure the full-width at half-maximum of their respective Gaussians.
Specifically:
\begin{equation}
\sigma_{W_{50}}^2 = \sigma_{V_{PV}}^2 + \sigma_{disp}^2
\end{equation}

\noindent This relationship is represented as a dashed line in Figure \ref{fig:vel_w50} for a typical velocity dispersion of 8 \kms; the lower and upper bounding dotted lines are for $\sigma_{disp}$ values of 6 and 10 \kms, respectively. The fiducial line for a dispersion of 8 \kms\ nicely provides an upper limit to our measured values, indicating our measured velocity extents are consistent with the single-dish W$_{50}$ values. As our velocity values are measured at a limited spatial extent whereas the $W_{50}$ values include all the emission from the galaxies, it is not surprising that in many cases our values are below the line. This is especially worth keeping in mind for the galaxies where the spatial extents are smaller than the effective beam (i.e., the unresolved cases shown in orange). 

\end{document}